\documentclass{jfm}

\usepackage{graphicx}
\usepackage{gnuplottex}
\usepackage{tikz}
\usepackage{newtxtext}
\usepackage{newtxmath}
\usepackage{natbib}
\usepackage{hyperref}
\hypersetup{
    colorlinks = true,
    urlcolor   = blue,
    citecolor  = black,
}

\newcommand{\RomanNumeralCaps}[1]
\linenumbers

\newcommand{\bea}{\begin{equation}\begin{aligned}}
\newcommand{\eea}{\end{aligned}\end{equation}}

\newcommand{\dd}[2]{\frac{\partial #1}{\partial #2}}

\newcommand{\ave}[1]{\langle #1 \rangle}

\shorttitle{Vortex topology in the lee of a 6:1 prolate spheroid}
\title{Vortex topology in the lee of a 6:1 prolate spheroid}

\author[Marc Plasseraud and Krishnan Mahesh]
{Marc Plasseraud$^1$, Krishnan Mahesh$^1$
\thanks{Email address for correspondence: krmahesh@umich.edu}}

\affiliation{
$^1$Naval Architecture and Marine Engineering, University of Michigan, Ann Arbor, MI, 48109, USA}

\begin{document}
\maketitle

\begin{abstract}
A large scale parametric study of the flow over the prolate spheroid is presented to understand the effect of Reynolds number and angle of attack on the separation, the wake formation and the loads. Large--Eddy Simulation is performed for six Reynolds numbers ranging from $Re = 0.15\times 10^6$ to $4 \times 10^6$ and for eight angles of attack ranging from $\alpha = 10^\circ$ to $\alpha = 90^\circ$.
For all the cases considered, the boundary layer separates symmetrically and forms a recirculation region. Several distinct flow topologies are observed that can be grouped into three categories: proto--vortex, coherent vortex and recirculating wake. In the proto--vortex state, the recirculation does not have a distinct center of rotation, instead, a two--layer detached flow structure is formed. In the coherent vortex state, the separated shear layer rolls into a three--dimensional vortex that is aligned with the axis of the spheroid. This vortex has a clear center of rotation corresponding to a minimum of pressure and transforms the azimuthal momentum from the separated shear layer into axial momentum. In the recirculating wake regime, the recirculation is incoherent and the primary separation forms a dissipative shear layer that is convected in the direction of the free--stream. This symmetric pair of shear layers bounds a low--momentum recirculating cavity on the leeward side of the spheroid. The properties of these states are not constant, but evolve along the axis of the spheroid and are dictated by the characteristics of the boundary layer at separation. The variation of the flow with Reynolds number and angle of attack is described, and its connection to the loads on the spheroid are discussed.
\end{abstract}

\clearpage

\section{Introduction}
The phenomena of flow separation and vortex formation affect a wide variety of external flows, including marine and aerial vehicles. Whether boundary layer separation is detrimental (airfoil stall) or desired (delta wing), these effects dominate the loads. Thus, a correct understanding of the physics of separation and recirculation is critical to predict vehicle performance and maneuverability. To that end, past work has studied geometries, including spheres \citep{taneda1956, johnson1999}, cylinders \citep{williamson1995}, elliptic cones \citep{crabbe1965}, delta wings \citep{brown1954, lee1990}, and SUBOFF \citep{morse2021, kumar2017}. 

The 6:1 prolate spheroid is one such canonical streamlined body. Despite its geometrical simplicity, the spheroid generates a wide array of complex flow features such as streamline curvature, cross--flow transition, three--dimensional separations, and large recirculating flow structures. This complexity makes it a particularly attractive canonical model for evaluating numerical models, as performed in \cite{constantinescu2002, rhee2002, scott2004, wikstrom2004, xiao2007, fureby2009}.

When the spheroid is inclined in incidence with the flow, a three-dimensional boundary layer \citep{cebeci1987} forms on the windward side. The three--dimensionality is a consequence of streamline curvature and is characterized by a secondary flow, as described experimentally by \cite{chesnakas1994}. Multiple pathways of turbulent transition exist in the boundary layer, including Tollmien--Schlichting, separation line, attachment line, and cross--flow instabilities \citep{rubino2021}. The interaction between these modes makes the boundary layer challenging to resolve adequately, and predicting transition on the spheroid is still an area of active research. The importance of predicting transition of the attached flow is notable because the state of the boundary layer strongly influences the location of separation, which, in turn, leads to changes in the topology of the recirculation and the loads \citep{constantinescu2002, hedin2001}. The spheroid is often artificially tripped in experiments and simulations. The purpose of the trip is to bypass natural transition and to generate a turbulent boundary layer, thereby ensuring repeatability and behavior representative of a full--scale model. There are indications that the presence of a trip does not guarantee that the boundary layer becomes turbulent, particularly at higher incidence \citep{plasseraud2023}. 
The transition of the boundary layer from laminar to turbulent at separation leads to the existence of a critical Reynolds number between $Re = 2 \times 10^6$ and $Re = 3 \times 10^6$ \citep{ahn1992} for an angle of attack $\alpha \leq 20^\circ$. This transition is accompanied by changes in the location of separation, the size and vorticity of the recirculation, and the suction exerted on the spheroid. In addition, the primary vortex is attached and coherent at low angles of attack ($10^\circ$, $20^\circ$), while a wake forms at $90^\circ$, without a visible coherent vortical structure \citep{ElKhoury2010}.

The recirculation has been extensively studied at $10^\circ$ and $20^\circ$ in the subcritical \citep{guo2023}, critical \citep{ahn1992} and super--critical regime \citep{chesnakas1994, chesnakas1997}. The latter study offers particularly detailed measurements of velocity profiles, wall flow angle, turbulent quantities, and velocity perturbation correlations in the boundary layer and around the primary vortex. Fewer studies have looked at higher incidences. \cite{jiang2014} performed DNS of the flow for an incidence of $45^\circ$ at $Re \leq 1000$, and observed an unusual port/starboard asymmetry of the vortex pair despite the symmetry of the problem. \cite{ElKhoury2010, ElKhoury2012} performed DNS at $Re = 1e5$ in the laminar regime at $\alpha = 90^\circ$. The choice of this incidence had not been investigated prior to these publications, and it reveals interesting similarities to the flow around a cylinder.

The origin of the loads on the spheroid is a topic of particular importance and has been studied both numerically and experimentally. \cite{du2023} characterized the loads on the prolate spheroid in terms of the Josephson--Anderson relation \citep{josephson1965, anderson1966} which connects the vorticity flux to the drag on the body.
\cite{fu1994} specifically studied the effect of $Re/\alpha$ on the spheroid flow for small to moderate incidences around the critical Reynolds number. Their experimental study mainly investigated the origin of the force, its connection to the circulation in the primary vortex, and its evolution with varying $Re/\alpha$. They note that the circulation of the vortex increases with increasing angle of attack, while increasing $Re$ moves the separation leeward.
More generally, the suction force generated by the recirculation can be approached in terms of the vortex impulse, as formulated by \cite{truesdell1954, saffman1995} and applied by \cite{wu2007}.

\cite{plasseraud2023} performed wall--resolved, trip--resolved LES on the prolate spheroid at $10^\circ$ and $20^\circ$,  for a Reynolds number $Re = 4.2 \times 10^6$. They focused on the windward, attached boundary layer and recorded the velocity components, the turbulent stresses, and the production of turbulent kinetic energy along two attached streamlines. They concluded that the state of the attached boundary layer was influenced by the streamline curvature effects, which in turn affected the separation and the loads on the spheroid. The purpose of the present study is to relate the nature of separation and changes in the topology of the recirculation to the loads on the spheroid. Toward this goal, a large number of simulations (48) have been performed on a range of Reynolds numbers and angles of incidence varying from $Re = 0.15M$ to $Re = 4M$ and an angle of attack ranging from $\alpha = 10^\circ$ $\alpha = 90^\circ$. This study provides, to the best of our knowledge, the most expansive set of numerical data on the spheroid.

The paper is organized as follows: sections \ref{sec: numerical method} and \ref{sec: methodology} describe the LES solver used and the methodology used to predict the location of separation, the vortex boundary and the boundary layer thickness; section \ref{sec: grid convergence} details the grid convergence study; section \ref{sec: results and discussion} discusses the results and section \ref{sec: conclusion} concludes the paper. 



\section{Numerical method}
\label{sec: numerical method}
The overbar $\overline{(\cdot)}$ denotes spatial filtering, while the bracket $\langle \cdot \rangle$ is used to indicate time averaging. 
The incompressible, spatially filtered Navier-Stokes equation are solved in a Large-Eddy Simulation formulation:
\begin{eqnarray}
\frac{\partial \overline{u}_i}{\partial t} + \frac{\partial}{\partial x_j}(\overline{u}_i \overline{u}_j) & = &
-\frac{\partial \overline{p}}{\partial x_i}
+ \nu \frac{\partial^2 \overline{u}_i}{\partial x_j \partial x_j}
- \frac{\partial \tau_{ij}}{\partial x_j}, \\
\frac{\partial \overline{u}_i}{\partial x_i} & = & 0,
\end{eqnarray}
where $u_i$ is the velocity, $p$ is the pressure and $\nu$ is the kinematic viscosity. The Sub-Grid Stress (SGS) $\tau_{ij} = \overline{u_iu_j}-\overline{u}_i\overline{u}_j$ is modeled with the dynamic Smagorinsky model \citep{germano1991,lilly1992}.
A finite volume, second--order centered spatial discretization is used where the filtered velocity components and pressure are stored at the cell-centroids while the face-normal velocities are estimated at the face centers. The equations are marched in time with a second--order Crank-Nicolson scheme.
The algorithm has shown good performance for multiple complex flows, such as propeller flows \citep{verma2012, kroll2022, leasca2025} and flows over hulls \citep{kumar2017,morse2021}. The kinetic--energy conservation property of the method \citep{mahesh2004} makes it suitable for high Reynolds number flows such as the one presented in this paper.
The method was extended by \cite{horne2019a, horne2019b} to allow for overlapping (overset) grids and six-degree of freedom movement. Although the present geometry is stationary, overset grids are used for better grid efficiency by circumventing the need for a one-to-one match between the near-wall region and the far-field. In addition, overset grids provide more flexibility in the grid generation process as they allow for the refinement of specific areas of the flow such as the vortex region in the current study.

\section{Methodology}
\label{sec: methodology}
\subsection{Notations}
The angle of attack is denoted by $\alpha$ and $Re$ is the Reynolds number based on the major length $L$ of the spheroid and freestream velocity $U_0$. The azimuthal angle with respect to the spheroid is written as $\phi$, where $\phi = 0^\circ$ corresponds to the windward (i.e. pressure side) meridian and $\phi = 180^\circ$ is the leeward (i.e. suction side) meridian. On the $\phi \in [0^\circ, 180^\circ]$ side, the primary and tertiary vortices (if they are present) rotate counterclockwise with respect to the axis of the spheroid, while the secondary vortex rotates clockwise. The direction of rotation of all three types of vortices is opposite on the other side ($\phi \in [180^\circ, 360^\circ]$). $\vec{u}$ and $\vec{\omega}$ are the velocity and vorticity vector, normalized by $U_0$ and $U_0/L$ respectively, in the body--oriented reference frame. Following these conventions, the helicity density $\vec{u} \cdot \vec{\omega} /(\|\vec{u}\|\|\vec{\omega}\|)$ is negative in the primary vortex and positive in the secondary vortex for $\phi \in [0^\circ, 180^\circ]$. If the primary vortex is detected, it has a radius $r_0$, a circulation $\Gamma_v = \int_v \omega_x dv$, an area $A_v$, and an azimuthal angle $\theta$.

\subsection{Coordinate system}

\begin{figure}
\begin{center}
\begin{tikzpicture}
\draw[blue, ultra thick, ->, rotate=20] (-5,0.6) -- (-4,0.6);
\draw[blue, ultra thick, ->, rotate=20] (-5,0.4) -- (-4,0.4) node[anchor=west]{$U_0$};
\draw[blue, ultra thick, ->, rotate=20] (-5,0.2) -- (-4,0.2);
\draw[black, thick] (-4.75,-1.55) -- (-4.2,-1.55);
\draw[black, thick] (-4.2, -1.55) node[anchor=north]{$\alpha$} arc (0:20:0.55);
\draw[fill=blue!20!white, draw=black, thick] (-1.8,0.3) -- (2.5,0.6) arc (-90:90:0.3) -- (-1.8,0.3);
\draw[fill=blue!5!white, draw=black, ultra thick] (0,0) ellipse (3cm and 0.5cm);
\draw[black, ultra thick] (0,-0.5) arc (-90:90:0.5);
\draw[red, ultra thick, ->] (0,-0.5) arc (-90:10:0.5) node[anchor=west]{$\phi$};
\draw[fill=blue!30!white, draw=black, thick] (-1.5,0.2) -- (3,0.4) arc (-90:90:0.3) -- (-1.5,0.2);
\draw[red, ultra thick, ->] (3,0.4) arc (-90:20:0.3) node[anchor=west]{$\theta$};
\draw[black, ultra thick, <->] (-3,-0.8) -- (3,-0.8);
\draw (0.0,-1.0) node{$L$};
\draw[black, ultra thick, <->] (-3.8,-0.5) -- (-3.8,0.5);
\draw (-4.0,0.0) node{$D$};
\draw[black, ultra thick, <->] (3.8,0.4) -- (3.8,1);
\draw (4.3,0.65) node{$2r_0$};
\draw[red, ultra thick, ->] (-3,0) -- (-2,0) node[anchor=west]{$x$};
\draw[red, ultra thick, ->] (-3,0) -- (-3,1) node[anchor=west]{$y$};
\end{tikzpicture}
\caption{Schematic of the body--oriented coordinate systems.}
\label{fig: coordinate system}
\end{center}
\end{figure}
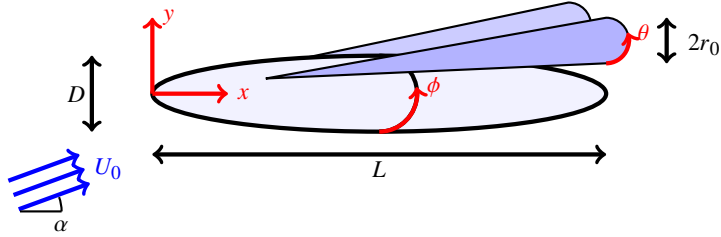

An angle of attack $\alpha$ is considered with respect to the major axis of the spheroid. Two coordinate systems are used and displayed in figure \ref{fig: coordinate system}. The first is Cartesian and originates at the nose of the spheroid. The $x$ axis is aligned with the major axis of the spheroid and is referred to as the `axial' coordinate. $x/L = 0$ is at the nose and $x/L = 1$ is on the tail end of the spheroid. The $y$ axis also rotates with the spheroid; its coordinate is negative on the windward side and positive on the leeward side such that $y$ aligns with the freestream velocity vector when $\alpha = 90^\circ$. The $z$ axis corresponds to the spanwise direction. The meridian plane will be referred to as the plane for which $z = 0$.
The second coordinate system is also body--oriented and cylindrical. The azimuthal direction $\phi$ is zero on the windward meridian and $180^\circ$ on the leeward meridian. The radial direction $r$ is orthogonal to the $x$ and $\phi$ axes; it is zero on the wall of the spheroid and increases with distance from the wall. The body--oriented reference frame is used unless otherwise specified.

\subsection{Statistics measurement}
The transient and development of the vortex has been observed to last approximately one length flow--through $t \approx L/U_0$ with small variations depending on the incidence. The flow is run for two spheroid flow through to wash the transient, then averaged for a duration of three flow--throughs. Time--averaged variables are denoted as $\ave{.}$.

\subsection{Location of separation}
\begin{figure}
\begin{centering}
\includegraphics[trim={10 170 10 50}, clip, width = 80 mm]{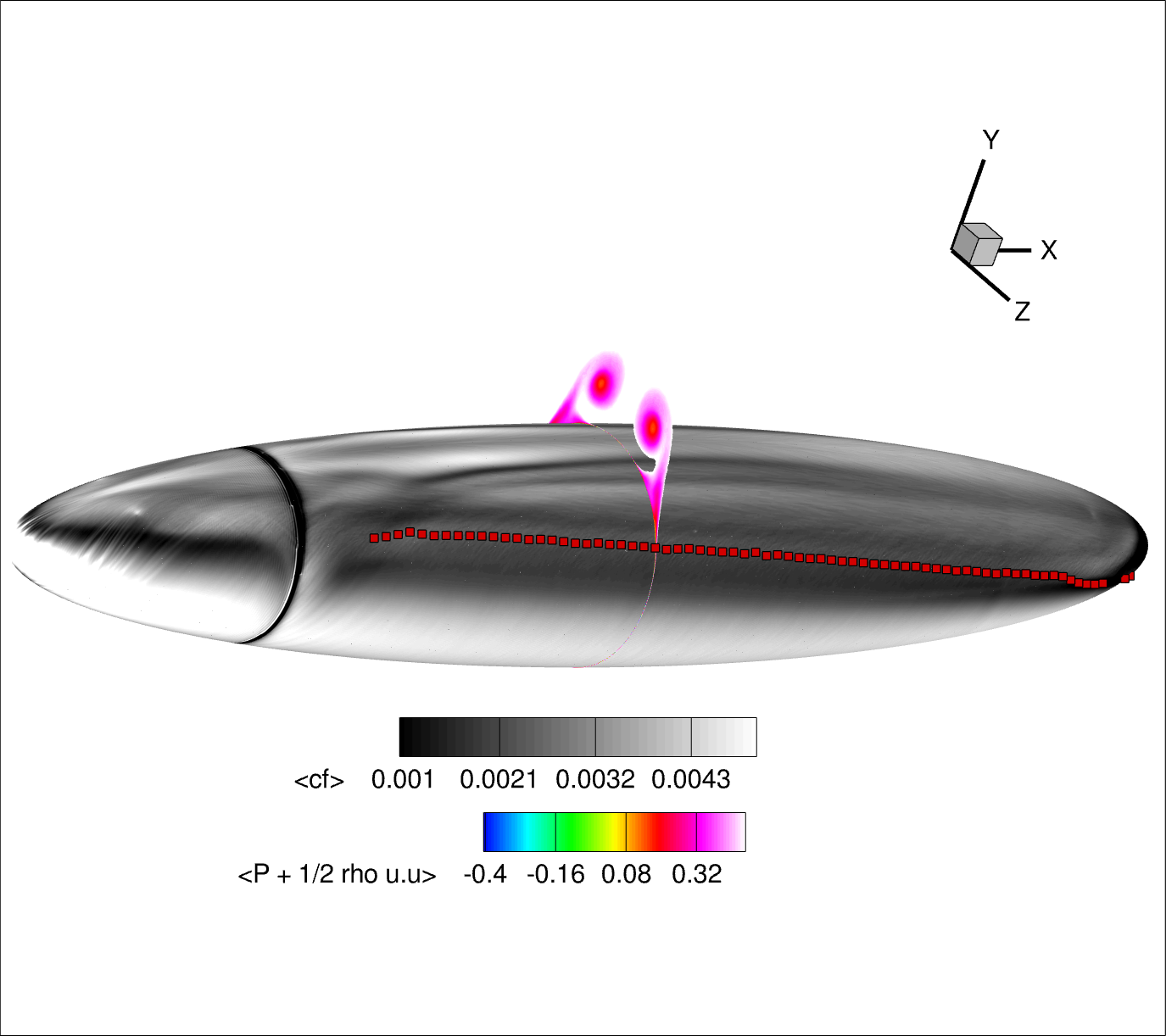}
\put(-220, 110){a)}
\includegraphics[trim={10 30 10 50}, clip, width = 40 mm]{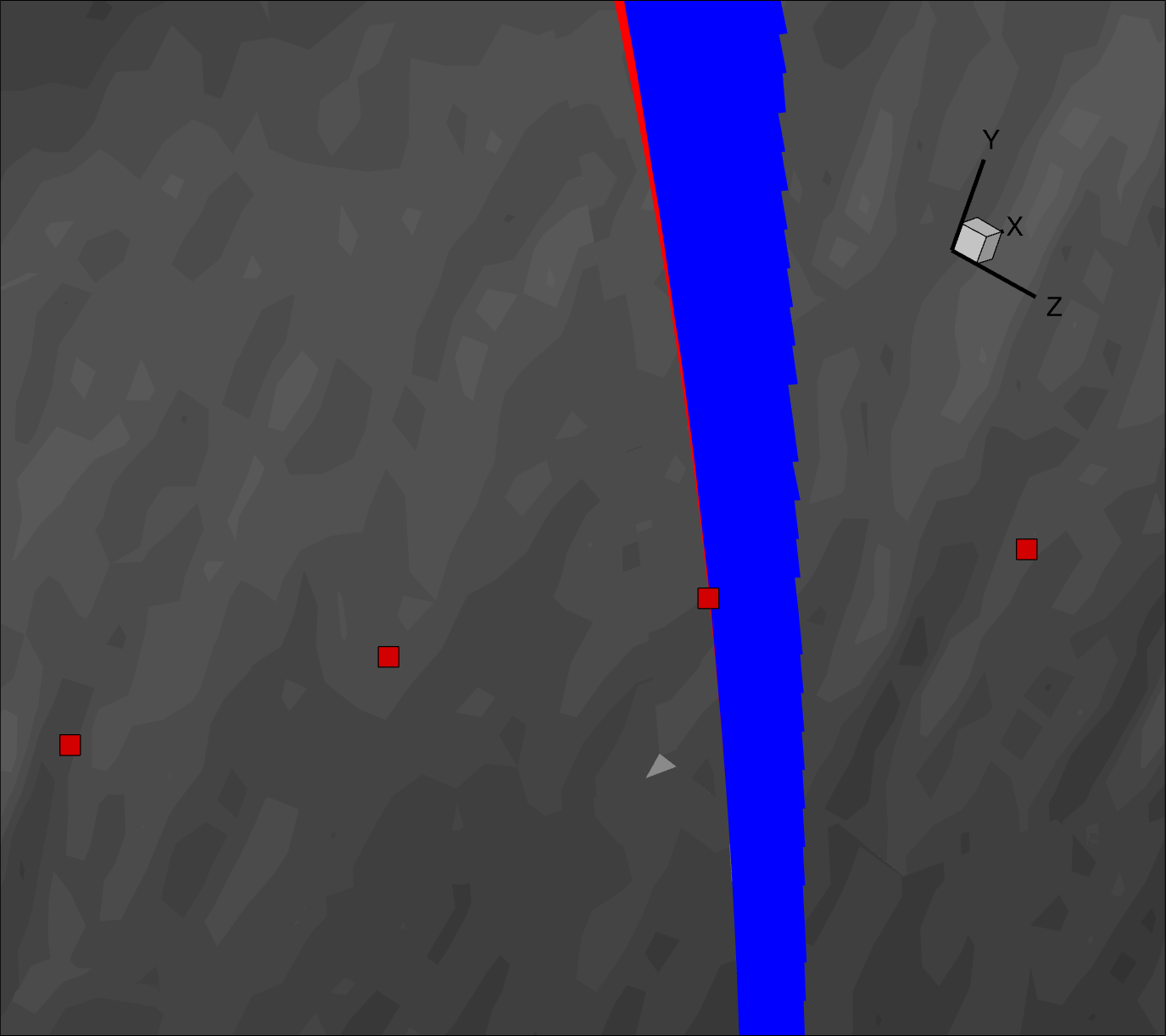}
\put(-110, 110){b)}
\caption{a): Time--averaged skin--friction coefficient $\ave{c_f}$ (spheroid wall), time--averaged stagnation pressure $P_s$ (transverse slice), root of helicity density (red symbols); b): scaled--up view of the same case showing the root of helicity density and sign of secondary velocity (red is negative, blue is positive) at $\alpha = 60^\circ$ and $Re = 4M$.}
\label{fig: AoA60_Re4M separation}
\end{centering}
\end{figure}

Past experiments and simulations have used the minimum magnitude of skin friction as an indicator of the location of primary separation. Although this metric is convenient and satisfactory at low angles of attack, it is an imprecise metric \citep{chesnakas1996} and can, at higher angles of attack, lead to false positives. Another approach is to locate the line of separation as the line of convergence of friction lines. This was done by \cite{kim2003} and \cite{morrison2003} to locate the separation and reattachment lines on the spheroid. Although this method yields good results for visual identification, the line of convergence is difficult to locate numerically. Instead, \cite{moffatt1992} have suggested that the helicity density defined as $h = \vec{u} \cdot \vec{\omega}$ switches sign across a three--dimensional separation. This method was also suggested by \cite{levy1990} to visualize three--dimensional separations and recirculation. \cite{chesnakas1996} observed this fact in the case of the spheroid at $\alpha = 10^\circ, 20^\circ$. The helicity density metric is used in the current study, since it was observed to be a reliable indicator of the separation locations for each of the cases studied.  Figure \ref{fig: AoA60_Re4M separation} shows the time--averaged skin--friction coefficient $\ave{c_f}$ on the wall of the spheroid and the points at which the helicity density switches sign (red symbols) at $\alpha = 60^\circ$, $Re = 4M$. In this case, the separation does not visually match a minimum of $\ave{c_f}$, in fact, there are two such minima around the separation. The root of helicity density, on the other hand, is close to the intercept of the wall and the direction of the separated shear layer; it is also close to the point where the secondary velocity switches sign. Note that even though the sign of secondary velocity is a good indicator of separation at high angle of attack where the separation is more two--dimensional, it is not a good metric at low angle of attack where flow separation can occur without sign reversal.
The symbol $\phi_s$ is used to indicate the azimuthal location of the separation throughout this study.

\subsection{Boundary layer thickness}
The definition of boundary layer thickness in flows with pressure gradient and curved wall is not trivial. Indeed, the metric should identify the region where the vorticity is negligible and the velocity is close to the freestream value, however vorticity decays asymptotically and the freestream velocity is non-uniform in the presence of pressure gradient. \cite{griffin2021} proposed to use the reconstructed velocity $u_p$ calculated from the freestream stagnation pressure $P_s^\infty = P^\infty + \frac{1}{2} \rho \vec{u}^\infty \cdot \vec{u}^\infty$ with the local pressure value such that $u_p = \sqrt{2(P_s^\infty - P)}$. The $\delta_{99}$ thickness could then be found as the point where the relative difference between the local value of velocity and the reconstructed velocity is less than $0.01$. This definition is consistent with the thickness of a flat plate boundary layer and offers a robust alternative for complex flows. 

\subsection{Identification of recirculation region}
For the same reasons, the domain of recirculation is challenging to assess and no mathematical method was found in the existing literature. However, a similar methodology as \cite{griffin2021} can be used to identify the recirculation region based on a threshold of stagnation pressure. Two options were considered to determine the stagnation pressure threshold to be consistent with the $\delta_{99}$ definition. First, the threshold could be taken as the stagnation pressure value at $r = \delta_{99}$. However, this is potentially sensitive to uncertainty in the measurement of boundary layer thickness. Second, since we have $\vec{u} \cdot \vec{u}/u_p^2 = \vec{u} \cdot \vec{u}/ (2 (P_s^\infty - P)) = 0.99^2$, the following threshold on $P_s$ could be taken:
\[
P_s = P (1.0 - 0.99^2) + 0.99^2 P_s^\infty \approx 0.49 P_s^\infty
\]

Thus, for any transverse location $x/L$ where separation occurs, the separated region is defined as the area where $P_s < 0.49 P_s^\infty$ and $r > \delta_{99}$. This methodology is easy to assess and leads to a definition consistent with that of the boundary layer thickness.

\subsection{Identification of primary vortex boundary}
Identifying the boundary of the primary vortex is difficult for similar reasons as the measurement of the boundary layer and the determination of the recirculation region. To be viable for the present study, a method must be fast enough to run on a large set of data, be robust to be applied to a wide variety of vortex topologies, be physically consistent, and provide additional information such as center of the vortex and reliable circulation estimates. Two common methods were initially considered:

The first method was to define the boundary of the vortex as $\lambda_2 = 0$, where $\lambda_2$ is the second invariant of velocity, as proposed by \cite{jeong1995}. Although the $\lambda_2$ method is a commonly used way to visualize vortices, the 0-boundary is very sensitive to noise and therefore unreliable to use for integral calculations.

The second method considered was to impose a threshold on vorticity and the center of the vortex. Assuming a Gaussian profile of vorticity, the vortex boundary can be taken as the line where $\omega_x = \omega_0/e$, where $\omega_0$ is the vortex center vorticity and $e$ is Euler's number. Although simple, this method gave poor results on the current data set because not all vortices had a Gaussian profile.

To address these limitations, \cite{plasseraud2024snh} devised a novel method, which defines the vortex as the largest closed iso--surface of stagnation pressure. This method was found to perform remarkably well on the wide range of flow that was studied and is used to analyse the results in the current study.


\section{Grid convergence}
\label{sec: grid convergence}
\begin{figure}
\begin{centering}
\includegraphics[trim={20 30 350 30}, clip, width = 40 mm]{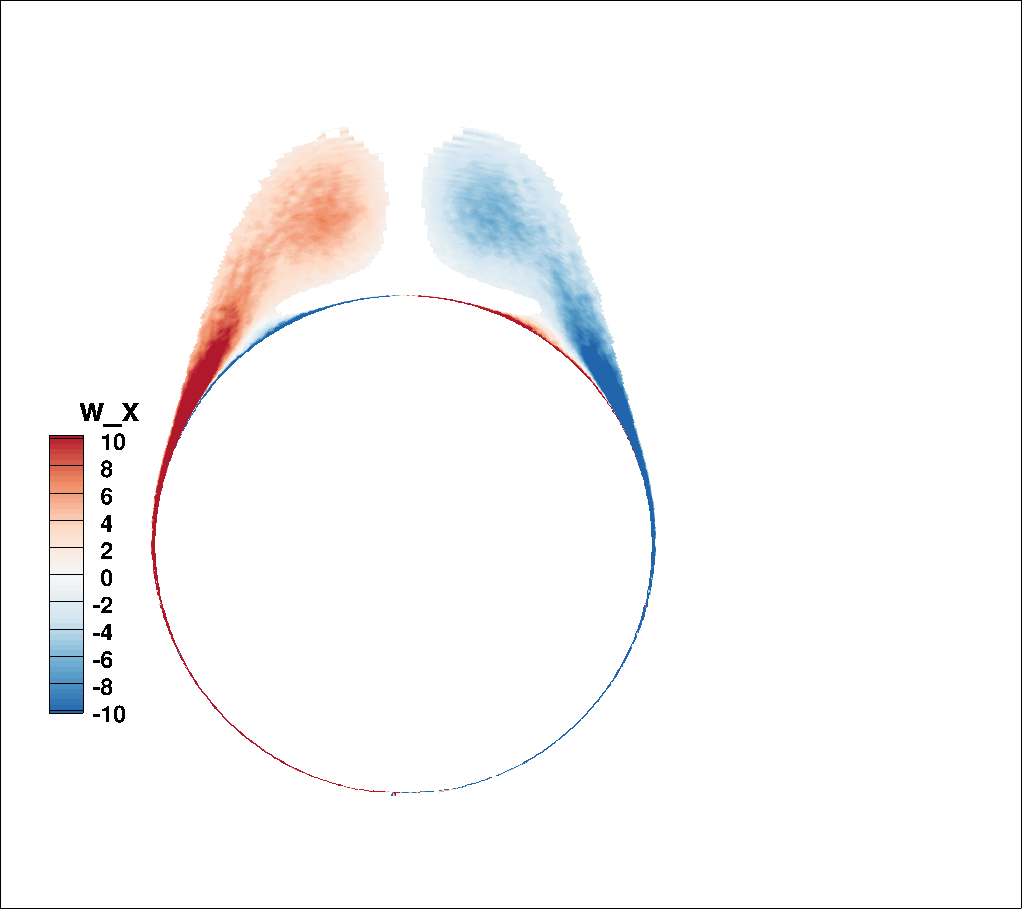}
\put(-90,120){(a)}
\includegraphics[trim={20 30 350 30}, clip, width = 40 mm]{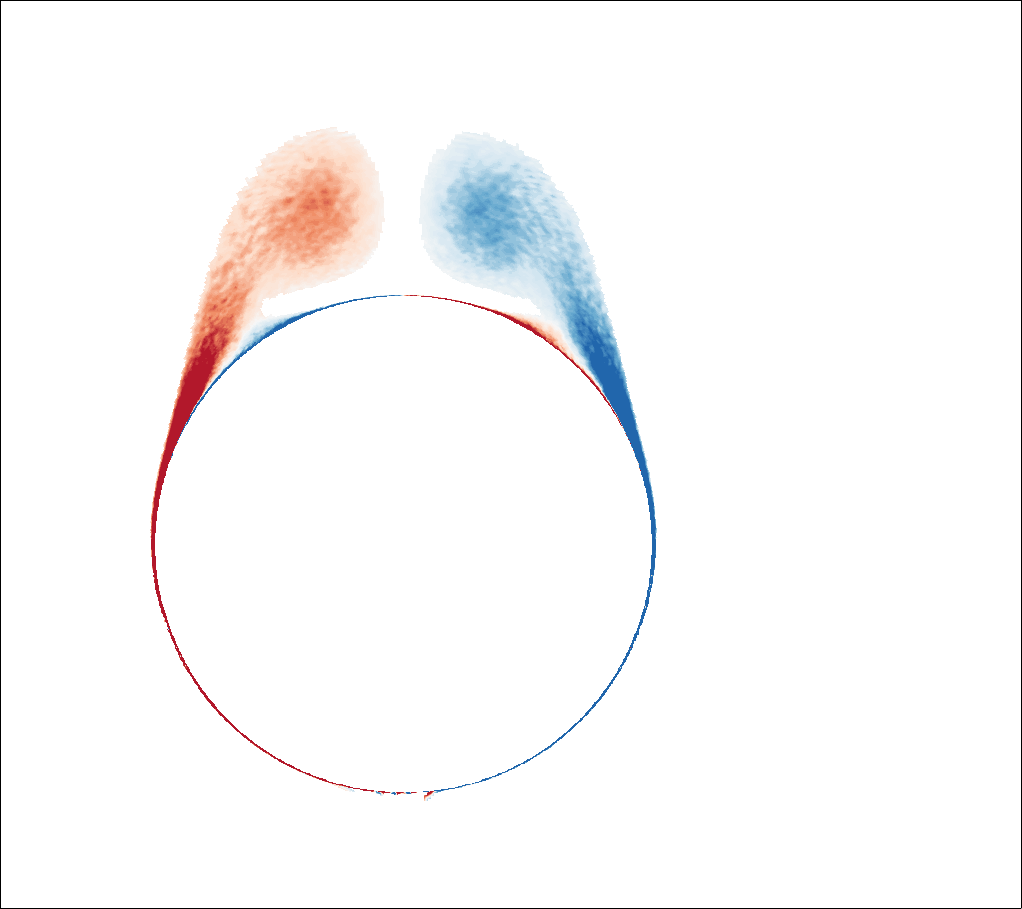}
\put(-90,120){(b)}
\includegraphics[trim={20 30 350 30}, clip, width = 40 mm]{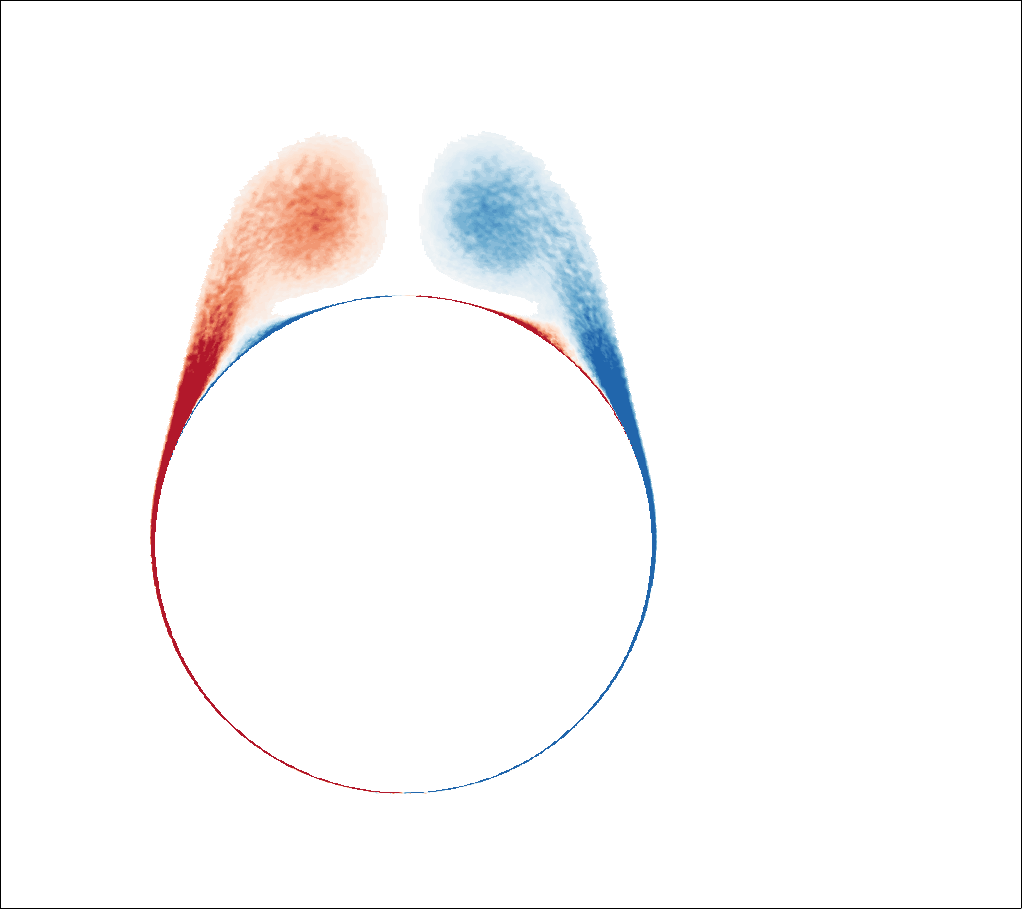}
\put(-90,120){(c)}
\caption{Time--averaged axial vorticity in a transverse section at $x/L = 0.8$, $\alpha = 20^\circ$, $Re = 4M$ for the (a) coarse grid, (b) medium grid and (c) fine grid.}
\label{fig: grid convergence wx}
\end{centering}
\end{figure}

\begin{figure}
\begin{centering}
\includegraphics[trim={90 90 10 600}, clip, width = 100 mm]{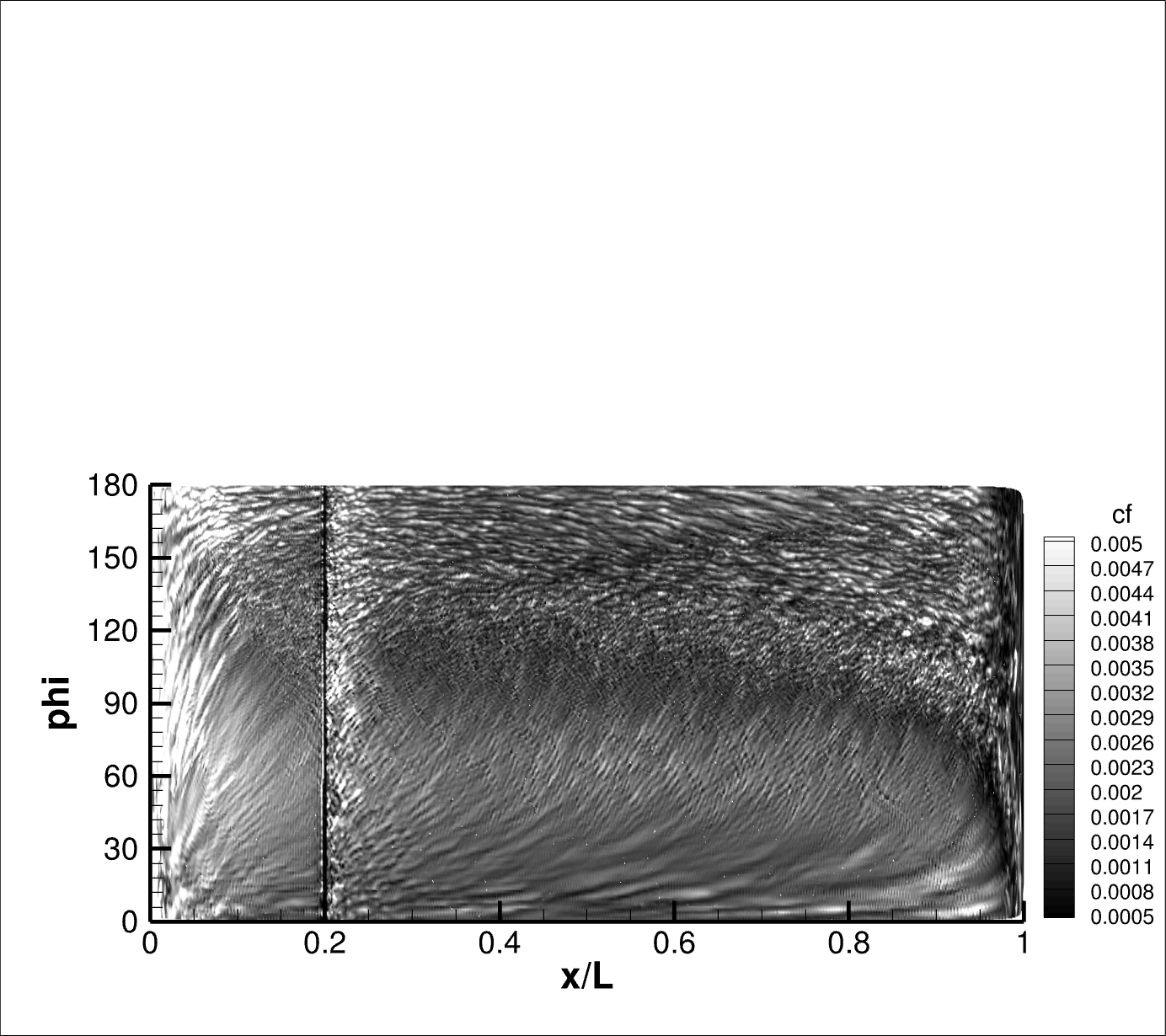}
\put(-280,120){(a)}
\put(-290,60){\rotatebox{90}{$\phi$}}

\includegraphics[trim={90 90 10 600}, clip, width = 100 mm]{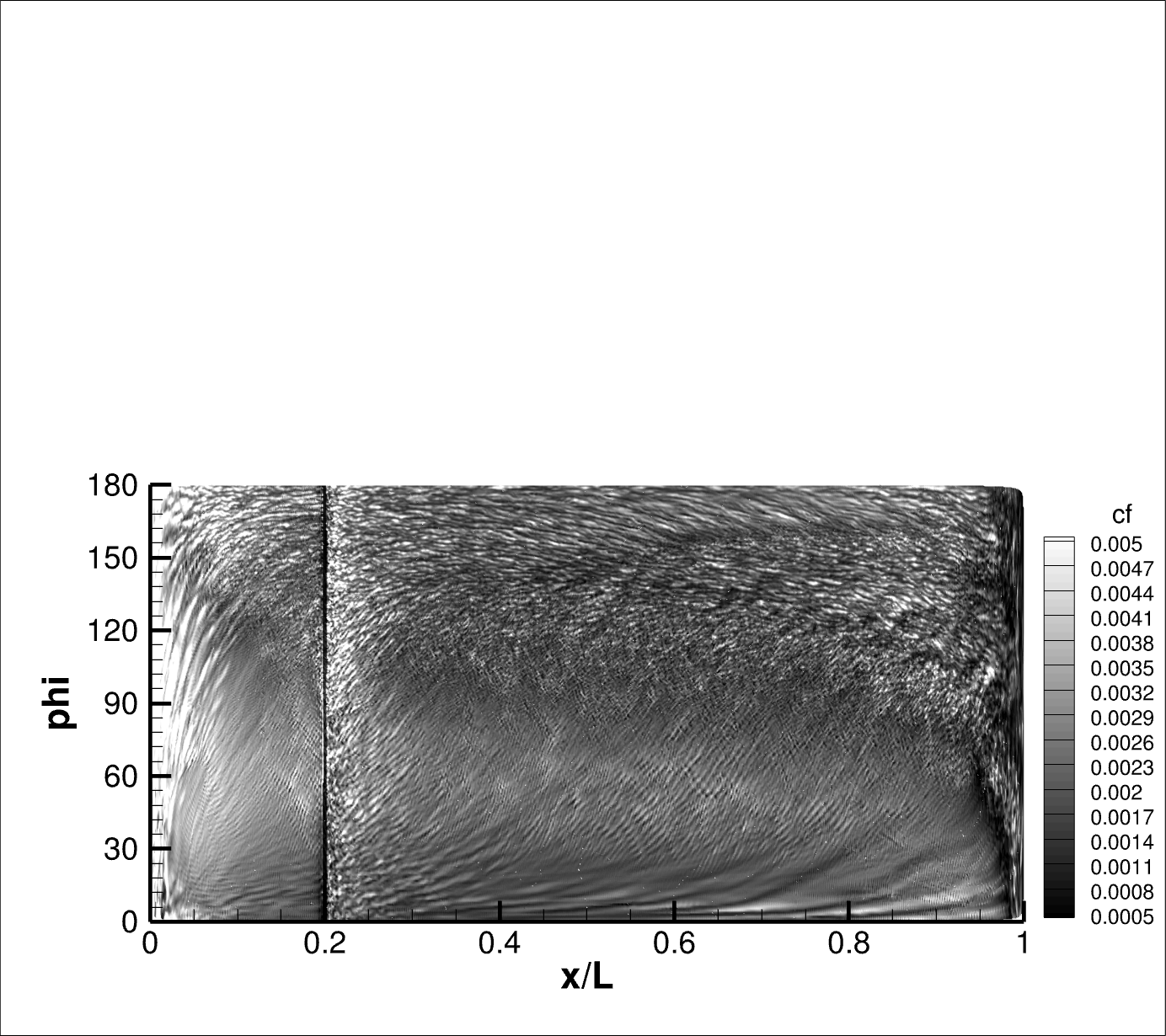}
\put(-280,120){(b)}
\put(-290,60){\rotatebox{90}{$\phi$}}

\includegraphics[trim={90 90 10 600}, clip, width = 100 mm]{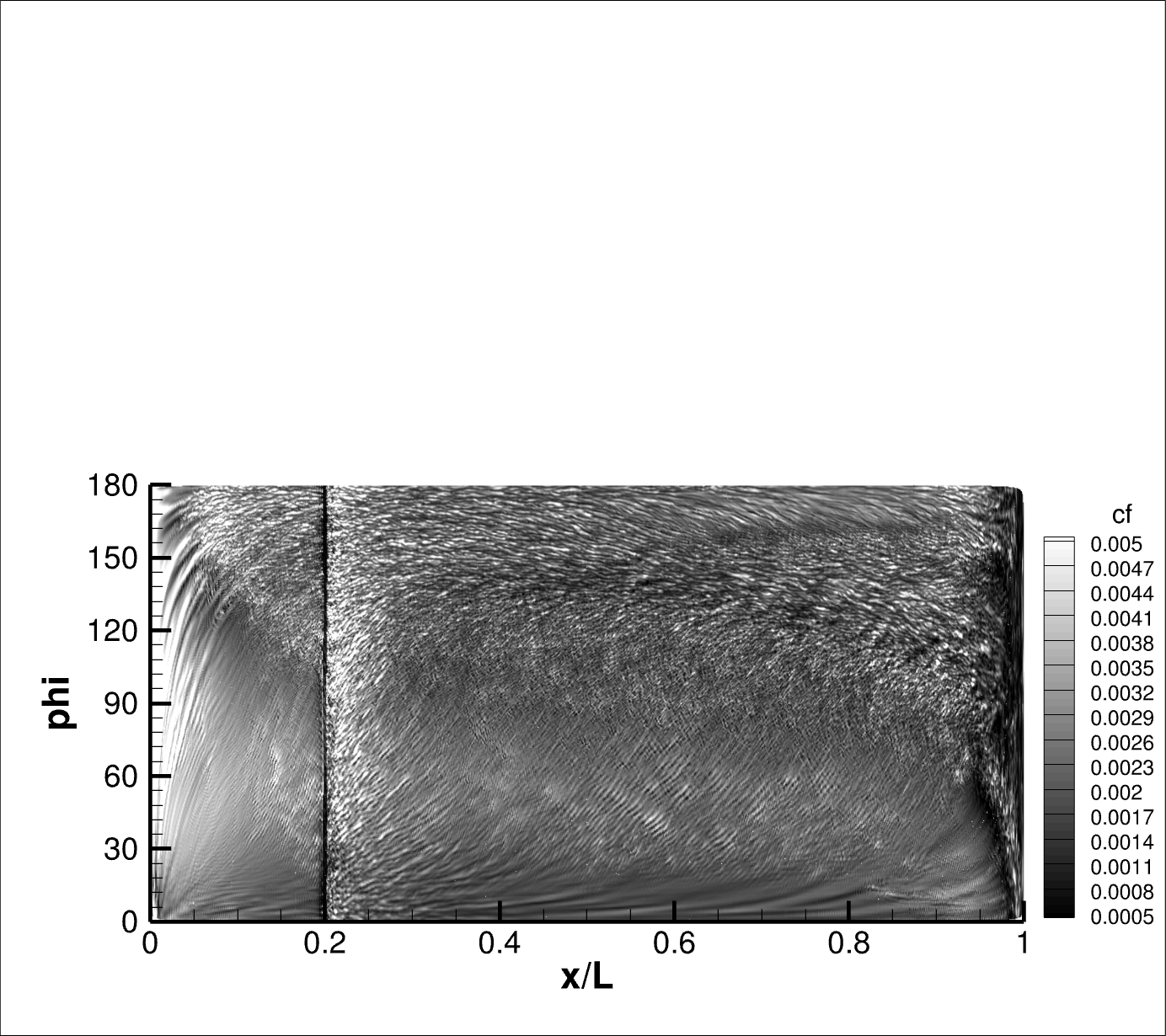}
\put(-280,120){(c)}
\put(-290,60){\rotatebox{90}{$\phi$}}
\put(-150,-10){$x/L$}

\caption{Skin friction coefficient versus $x/L$ and $\phi$ at $\alpha = 20^\circ$, $Re = 4M$ for the (a) coarse grid, (b) medium grid and (c) fine grid.}
\label{fig: grid convergence cf}
\end{centering}
\end{figure}
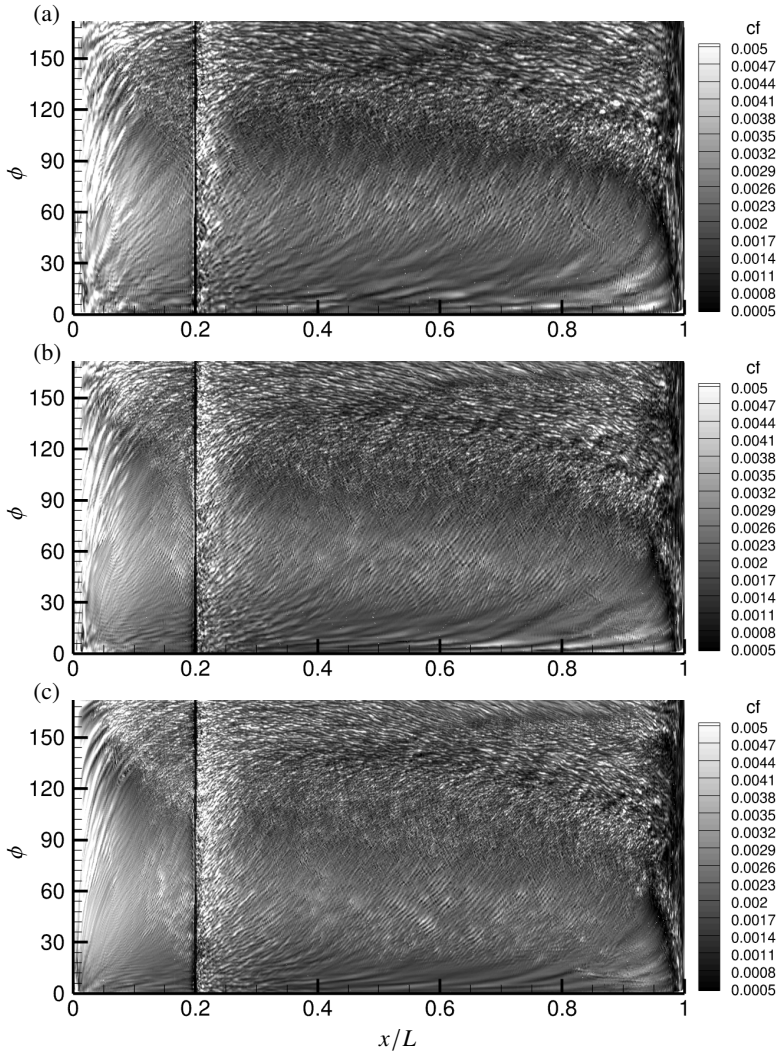

Three grid sizes are considered: 94M, 218M and 470M. All three grids have the same topology and only differ in the number of cells. The grids are composed of four different overset blocks used to refined the near--wall and the leeward regions. The convergence study is achieved by comparing the normal force and the time--averaged flow field at $\alpha = 20^\circ$ and $Re = 4M$. Table \ref{tab: grid convergence fy} shows the time--averaged normal force for the three grids. It is larger for the coarse grid and at close levels for the medium and fine grids. This suggests that the medium grid is sufficient to capture the physics of the flow. Figure \ref{fig: grid convergence wx} shows the axial vorticity for the three grids at $x/L = 0.8$. All three grids show a very similar flow composed of a primary separation leading to a primary vortex and a secondary, weaker separation. The vorticity in the center of the vortex increases with increasing refinement, albeit to a lower extent from the medium to the fine grid. Figure \ref{fig: grid convergence cf} shows the skin friction coefficient $c_f$ on the wall of the spheroid for the three grids. For all three grids, the primary and secondary separations are visible as minima of $c_f$ and a convergence of friction lines. In addition, perturbations are visible in the skin friction for all three grids. The perturbations have higher frequencies for the medium and fine grids compared to the coarse grid. The fine grid shows a slightly better definition of both separation lines and is the one used for this study.

\begin{table}
\begin{center}
\begin{tabular}{c c c c}
 Grid size & 94M & 218M & 470M \\
 \hline\hline
 Fy & 0.412779 & 0.418765 & 0.419225
\end{tabular}
\caption{Normal force on the three grids considered}
\label{tab: grid convergence fy}
\end{center}
\end{table}

\section{Results and discussion}
\label{sec: results and discussion}
\subsection{Overview}
\label{sec: overview}
\begin{figure}
\begin{centering}
\includegraphics[trim={10 10 50 10}, clip, width = 160 mm]{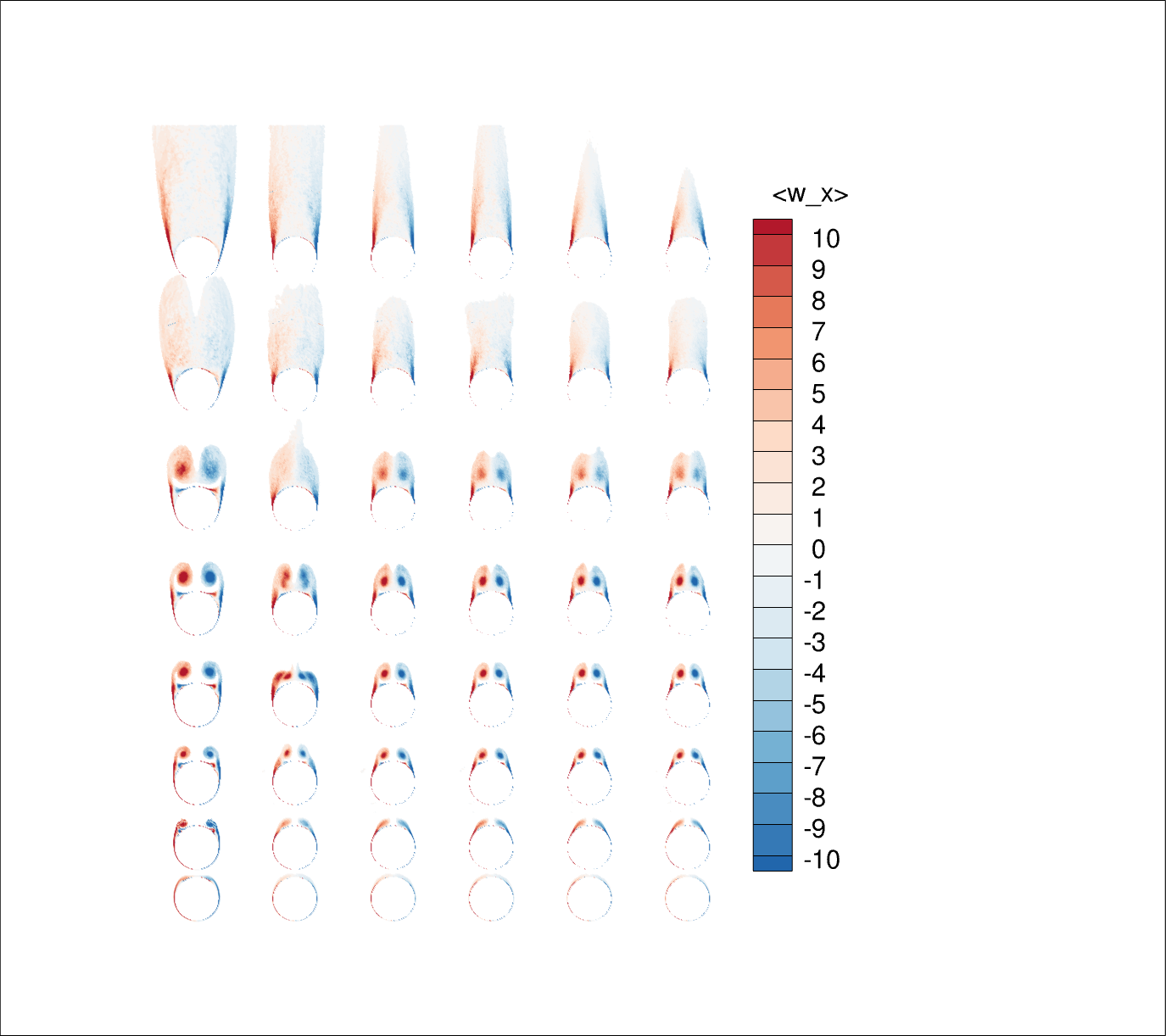}
\put(-400,20){$Re = 150k$}
\put(-355,20){$Re = 1M$}
\put(-316,20){$Re = 1.5M$}
\put(-272,20){$Re = 2M$}
\put(-233,20){$Re = 3M$}
\put(-192,20){$Re = 4M$}
\put(-430, 310){$\alpha = 90^\circ$}
\put(-430, 255){$\alpha = 70^\circ$}
\put(-430, 210){$\alpha = 60^\circ$}
\put(-430, 165){$\alpha = 50^\circ$}
\put(-430, 130){$\alpha = 40^\circ$}
\put(-430, 98){$\alpha = 30^\circ$}
\put(-430, 70){$\alpha = 20^\circ$}
\put(-430, 48){$\alpha = 10^\circ$}
\caption{Time--averaged axial vorticity in a transverse slice at $x/L = 0.6$ for all the cases, ordered from left to right by Reynolds number and from bottom to top by angle of attack.}
\label{fig: overview wx}
\end{centering}
\end{figure}
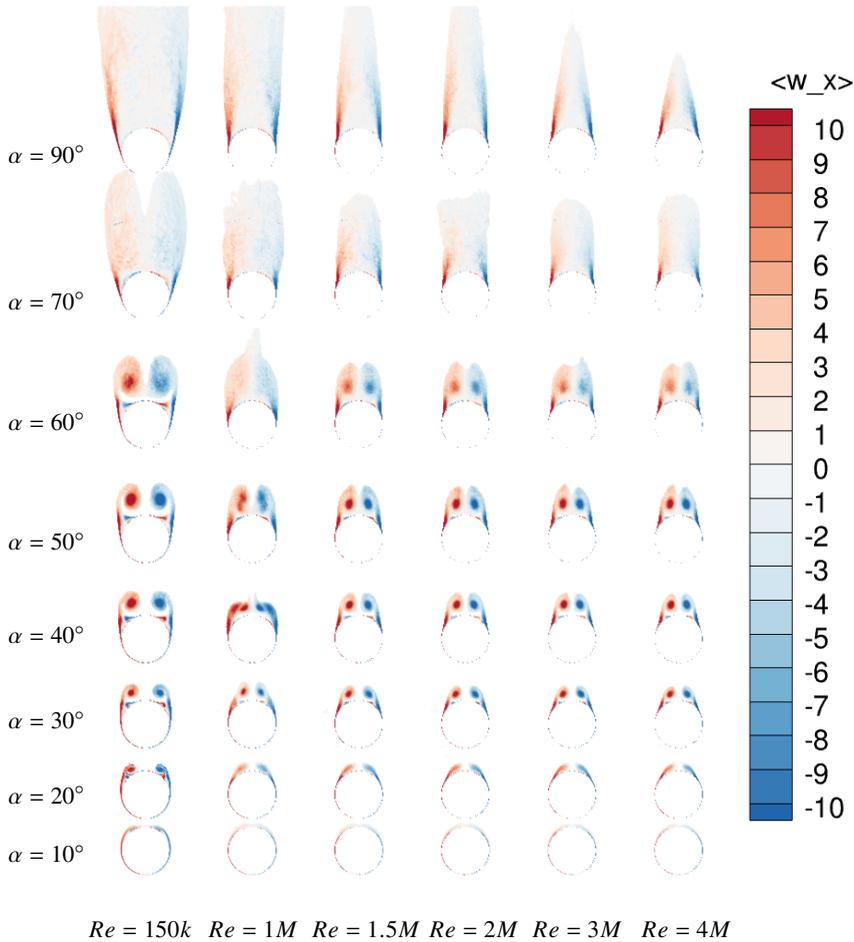

Figure \ref{fig: overview wx} gives an overview of the flows that were calculated. It shows the time--averaged axial vorticity at $x/L = 0.5$, sorted from left to right by increasing values of Reynolds number, and from bottom to top by increasing values of angle of attack. All the cases shown are symmetric and have at least one boundary layer separation, which leads to a recirculation. The topology of the recirculation varies and is affected by the angle of attack, the Reynolds number, and the $x/L$ location. The predominant effect of the angle of attack is to increase the size of the recirculation.  From $\alpha = 10^\circ$ to $50^\circ$, the recirculation consists of one to several pairs of vortices of increasing vorticity. At $\alpha = 70^\circ$, the vorticity decreases although the recirculation is still bounded. At $\alpha = 90^\circ$, the recirculation area is unbounded. The Reynolds number also alters the recirculation, albeit differently. The primary separation line moves leeward with increasing Reynolds number. This results in a closing of the separation angle and a shrinking of the primary separation. This effect is small at $\alpha = 10^\circ$ and $20^\circ$, although more pronounced from $30^\circ$ to $90^\circ$.  These physical phenomena are detailed in later sections.

\subsection{Separation}
\subsubsection{Separation line}
\label{sec:separation}

\begin{figure}
\begin{centering}
\includegraphics[width = 120 mm]{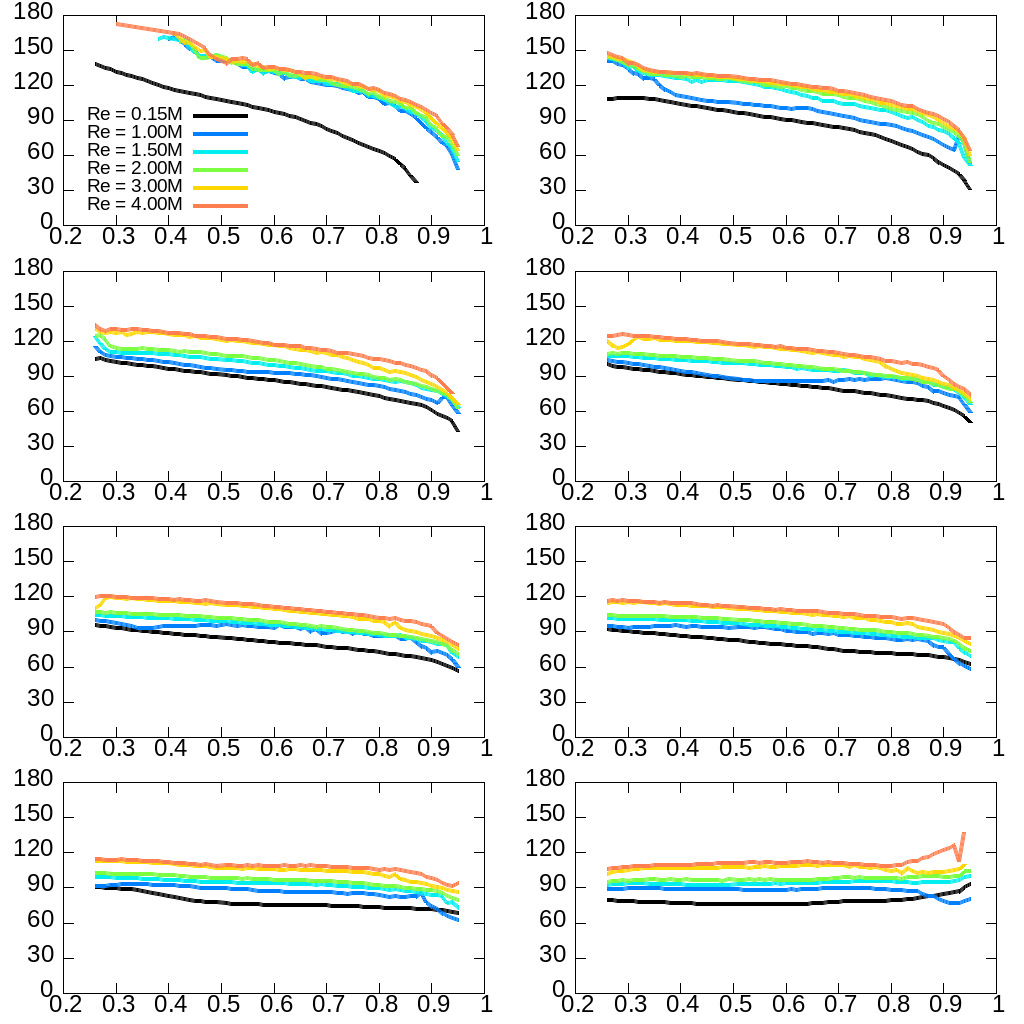}
\put(-260,-10){$x/L$}
\put(-90,-10){$x/L$}
\put(-350, 42){\rotatebox{90}{$\phi_s$}}
\put(-350, 126){\rotatebox{90}{$\phi_s$}}
\put(-350, 210){\rotatebox{90}{$\phi_s$}}
\put(-350, 294){\rotatebox{90}{$\phi_s$}}
\put(-195, 325){a)}
\put(-25, 325){b)}
\put(-195, 240){c)}
\put(-25, 240){d)}
\put(-195, 155){e)}
\put(-25, 155){f)}
\put(-195, 70){g)}
\put(-25, 70){h)}
\caption{Azimuthal location of primary separation versus x/L from a) to h): $\alpha = 10^\circ, 20^\circ, 30^\circ, 40^\circ, 50^\circ, 60^\circ, 70^\circ, 90^\circ$.}.
\label{fig: phi vs x}
\end{centering}
\end{figure}

For each case studied, the boundary layer detaches at least once and up to three times for the lower Reynolds number cases. 
Figure \ref{fig: phi vs x} shows the location of the primary separation for all the cases. For low angles of attack, no separation occurs until higher axial locations. The start of the primary separation moves toward lower axial locations as the angle of attack increases. In addition, the azimuthal position of the primary separation decreases with the axial position almost linearly between $x/L = 0.3$ and $x/L = 0.8$. 
This linear variation of the separation is also observed in experiments of \cite{fu1994} and \cite{barber1990}. In addition, \cite{fu1994} notes that increasing $Re$ leads to delayed separation in the sub--critical regime (defined by \cite{fu1994} as $Re < 1.3M$), as measured in the present data.

\begin{figure}
\begin{centering}
\includegraphics[width = 60 mm]{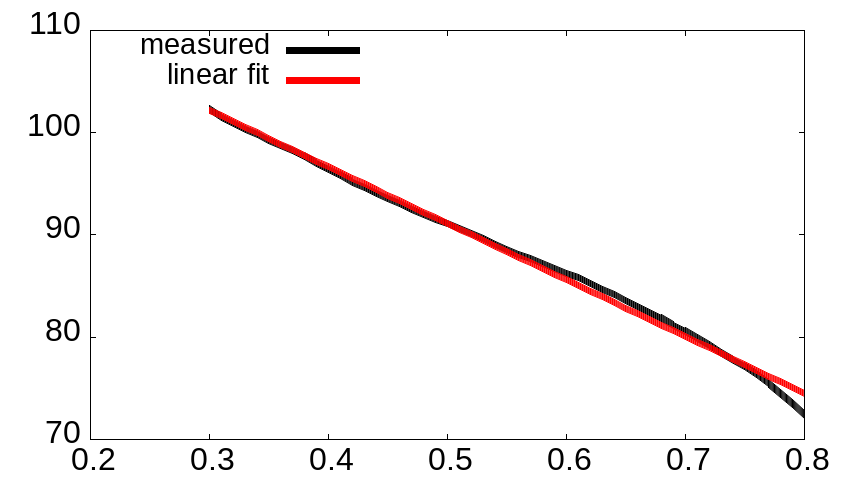}
\put(-90,-10){$x/L$}
\put(-175, 50){\rotatebox{90}{$\phi_s$}}
\caption{Separation line versus angle of attack at $\alpha = 30^\circ$, $Re = 0.15M$ (black) compared to a linear regression (red).}
\label{fig: phi separation vs x, AoA30 Re0p15M}
\end{centering}
\end{figure}

\begin{figure}
\begin{centering}
\includegraphics[width = 60 mm]{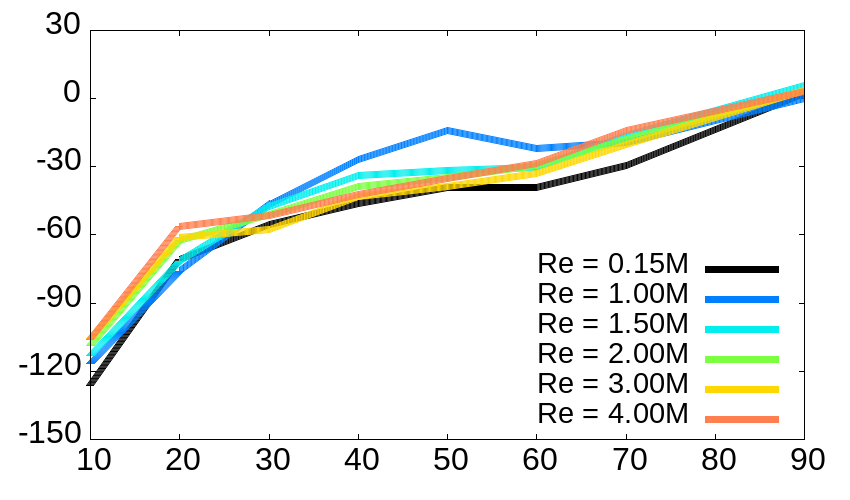}
\put(-85,-5){$\alpha$}
\put(-180, 45){\rotatebox{90}{$\dd{\phi_s}{x}$}}
\caption{Slope of separation line regression versus angle of attack. Each curve is a different Reynolds number.}
\label{fig: phi slope vs alpha}
\end{centering}
\end{figure}

Figure \ref{fig: phi separation vs x, AoA30 Re0p15M} shows the line of separation at $\alpha = 30^\circ$, $Re = 0.15M$. The separation curve matches well with a linear regression in the range $x/L \in [0.3, 0.8]$. The slope of the separation lines is calculated for all the cases and is shown in figure \ref{fig: phi slope vs alpha}.   The slope of the separation line varies little with the Reynolds number but increases linearly with the angle of attack for $\alpha > 10^\circ$. At $\alpha = 90^\circ$, the slope separation line is zero, i.e. the azimuth of separation is independent of $x/L$. This is expected because of the symmetry in $x$ at this angle, considering a negligible effect of the asymmetry introduced by the trip. The Reynolds number has an overall negligible effect on the slope of the separation.

\begin{figure}
\begin{centering}
\includegraphics[width = 60 mm]{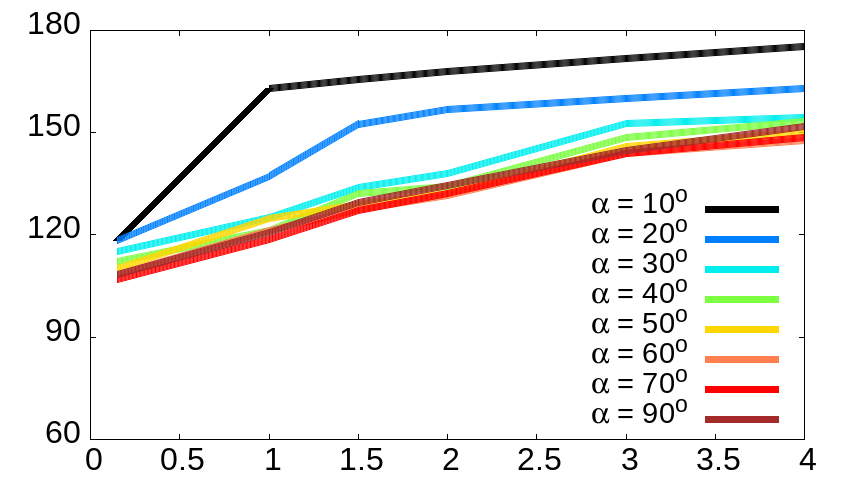}
\put(-85,-5){$Re$}
\put(-185, 45){\rotatebox{90}{$\overline{\phi}_s$}}
\caption{Average azimuth of separation line versus Reynolds number.}
\label{fig: phi centroid vs Re}
\end{centering}
\end{figure}

Figure \ref{fig: phi centroid vs Re} shows the average azimuth of separation versus the Reynolds number. The average is calculated between $x/L = 0.3$ and $x/L = 0.8$. For all angles of attack, the average increases with the Reynolds number until $Re = 3M$ and is constant for most incidences between $Re = 3M$ and $Re = 4M$. This is indicative of a delayed separation with increasing Reynolds number, and a saturation between $Re = 3M$ and $Re = 4M$.

\subsubsection{Boundary layer thickness at separation}
\begin{figure}
\begin{centering}
\includegraphics[width = 60 mm]{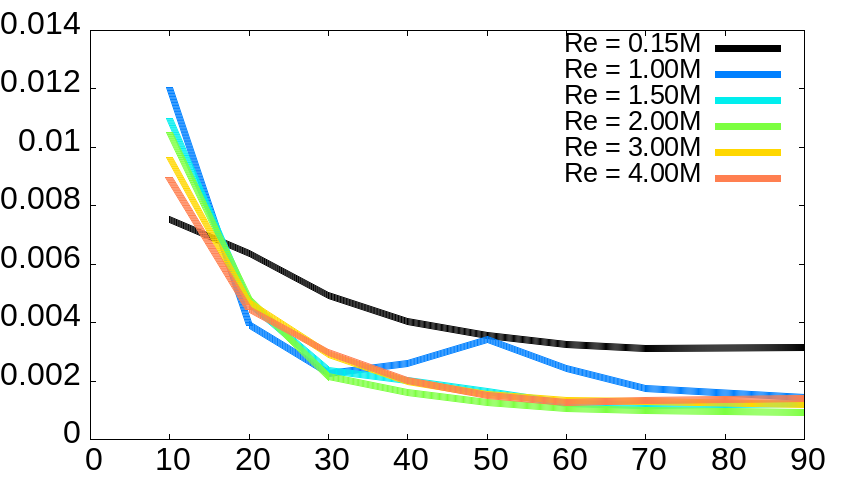}
\put(-80,-5){$\alpha$}
\put(-185, 45){\rotatebox{90}{$\delta_{99}$}}
\caption{$\delta_{99}$ averaged over the primary separation versus $\alpha$. Each curve is a different Reynolds number.}
\label{fig: d99 vs alpha}
\end{centering}
\end{figure}

Figure \ref{fig: d99 vs alpha} shows the boundary layer thickness $\delta_{99}$ versus $\alpha$ for all Reynolds numbers. The thickness decreases rapidly with increasing angle of attack until $\alpha = 50^\circ$, at which point it is constant. The cases with $Re = 0.15M$ and $Re = 1M$ are significantly different from the cases with $Re \ge 1.5M$. This can be explained by a change in the state of the upstream boundary layer from laminar to turbulent for $Re \approx 1.5M$.

\begin{figure}
\begin{centering}
\includegraphics[width = 60 mm]{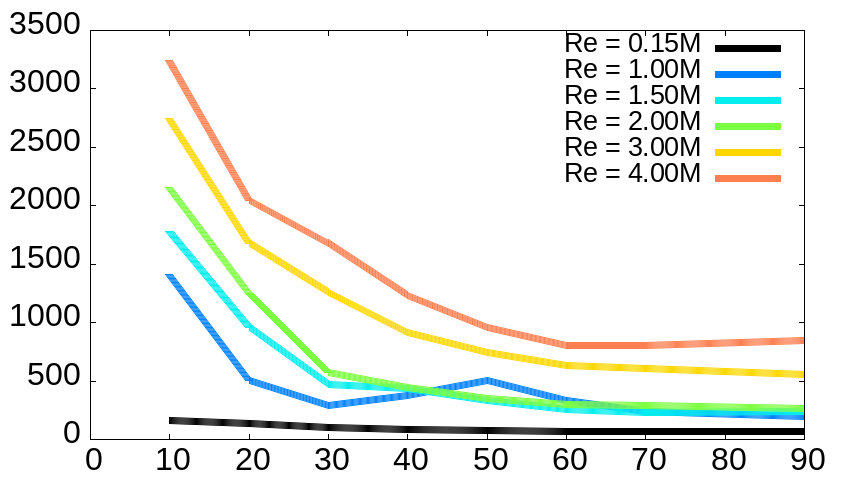}
\put(-80,-5){$\alpha$}
\put(-180, 45){\rotatebox{90}{$Re_{\theta}$}}
\caption{$Re_{\theta}$ averaged over the primary separation versus $\alpha$.}
\label{fig: Ret vs alpha}
\end{centering}
\end{figure}

Figure \ref{fig: Ret vs alpha} shows $Re_{\theta} = U_0 \theta/\nu$ versus $\alpha$ for the six Reynolds numbers considered. $Re_{\theta}$ follows a trend similar to $\delta_{99}$, with a strong decrease from $\alpha = 10^\circ$ to $50^\circ$, followed by a flattening of the curves. Apart from $Re = 1M$ , $\alpha \in [30^\circ, 70^\circ]$, which are close to the critical $Re$, $Re_{\theta}$ increases monotonically with the Reynolds number.

\subsubsection{Pressure gradient at separation}
\begin{figure}
\begin{centering}
\includegraphics[width = 60 mm]{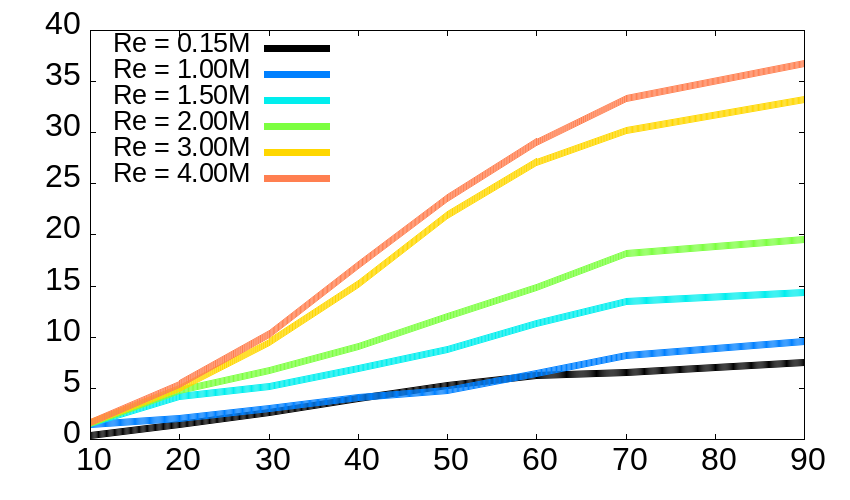}
\put(-80,-5){$\alpha$}
\put(-180, 40){\rotatebox{90}{$\|\dd{P}{x}\|$}}
\caption{Norm of the pressure gradient averaged over the primary separation versus $\alpha$.}
\label{fig: PG vs alpha}
\end{centering}
\end{figure}

Figure \ref{fig: PG vs alpha} shows the norm of the pressure gradient at the wall and averaged along the primary separation (later referred to as `separation--averaged'). The pressure gradient increases monotonically with $\alpha$ and linearly until $\alpha = 60^\circ$. This increase is understood as a consequence of the increase wall curvature: the elongation of the 6:1 prolate spheroid creates a low curvature in the axial direction and a high curvature in the azimuthal direction. When the angle of attack is small, the direction of the attached streamlines is predominantly along the axial direction, and the streamline curvature is small. When the angle of attack is large, the streamlines have a strong component in the azimuthal direction and the streamline curvature is large. This increase in streamline curvature with $\alpha$ is associated with an increase in pressure gradient. The Reynolds number also has a significant effect on the norm of the pressure gradient, with $\|\dd{P}{x}\|$ increasing with $Re$ for all angles of attack. This is understood to be a consequence of the boundary layer with higher momentum and higher $Re$, separating for higher pressure gradients.

\begin{figure}
\begin{centering}
\includegraphics[width = 70 mm, trim={1cm 14cm 1cm 1cm},clip]{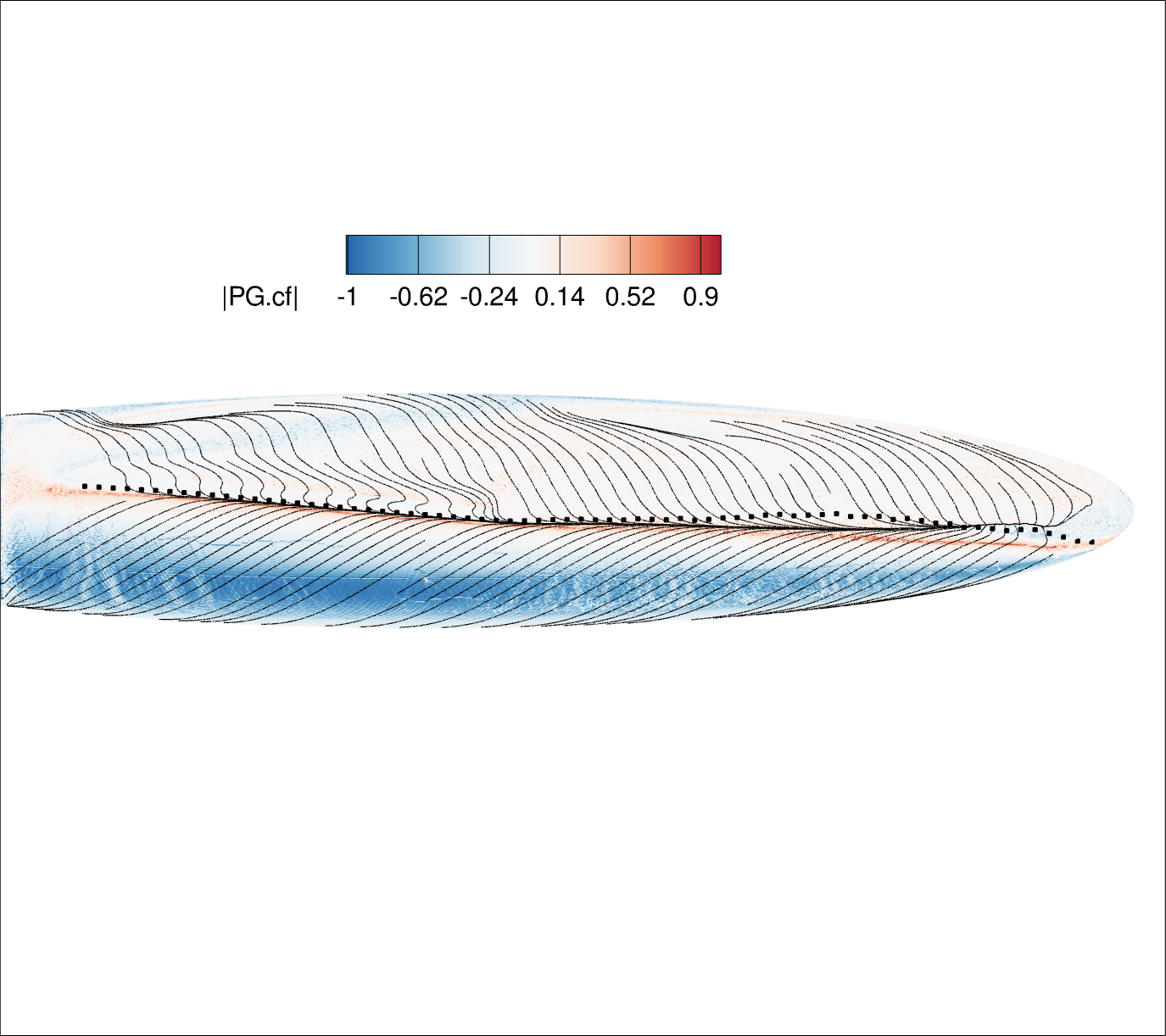}
\caption{Spheroid surface colored by normalized dot product of skin friction vector and pressure gradient at wall $\|\ave{c_f}\cdot \dd{P}{x_i}\|$ at $\alpha = 40^\circ$ $Re = 1M$, time--averaged friction lines, separation line based on helicity density root method (symbols).}
\label{fig: PG at AoA40 Re1M}
\end{centering}
\end{figure}

Figure \ref{fig: PG at AoA40 Re1M} shows the normalized dot product of the skin friction vector $c_f = \vec{\tau}_w/(1/2\rho U_{\infty}^2)$ (where $\vec{\tau}_w$ is the wall shear stress) with the pressure gradient $\|\ave{c_f}\cdot \dd{P}{x_i}\|$, the friction lines, and the root of helicity (the point at which helicity switches sign). The location of separation can be seen as a convergence of friction lines. Both the friction line method and the helicity root method predict a similar location of separation. On the windward side of the spheroid (bottom of figure \ref{fig: PG at AoA40 Re1M}), $\|\ave{c_f}\cdot \dd{P}{x_i}\| < 0$ indicating a favorable pressure gradient. On the leeward side (top of figure \ref{fig: PG at AoA40 Re1M}), the value is close to zero, indicating that the pressure gradient is orthogonal to the direction of the near--wall flow. In the upstream vicinity of separation,  $\|\ave{c_f}\cdot \dd{P}{x_i}\|$ is maximum and close to unity. This indicates that the pressure gradient becomes adverse and collinear with the direction of the flow at separation.

\subsubsection{Angle of separation}
\begin{figure}
\begin{centering}
\includegraphics[width = 60 mm]{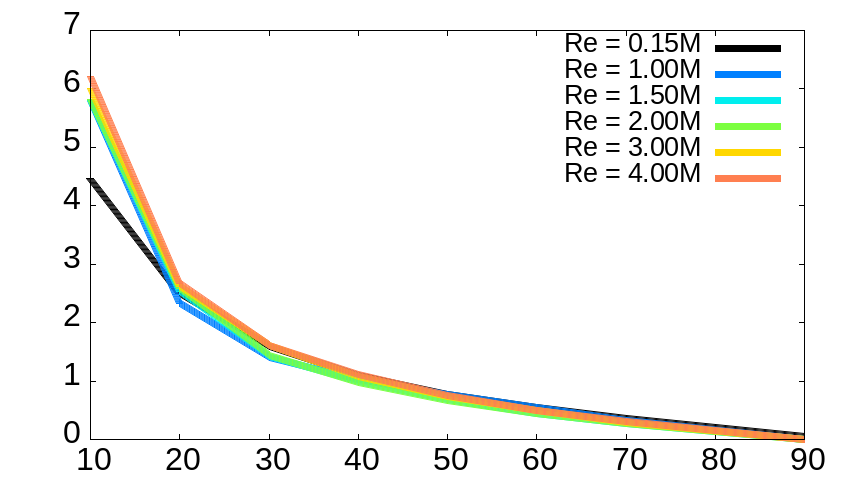}
\put(-85,-5){$\alpha$}
\put(-180, 35){\rotatebox{90}{$\frac{u}{\sqrt{v^2+w^2}}$}}
\caption{Averaged ratio of axial to azimuthal velocity at primary separation versus $\alpha$.}
\label{fig: sep angle vs alpha}
\end{centering}
\end{figure}

Figure \ref{fig: sep angle vs alpha} shows the separation--averaged ratio of axial to azimuthal velocity at primary separation versus $\alpha$. High values at low angles of attack indicate that the flow is mainly axial and that the separation is strongly three--dimensional. The decrease in this ratio with $\alpha$, to reach 0 at $\alpha = 90^\circ$ is interpreted as the separation becoming closer to a two-dimensional separation. This decreasing ratio at separation will later be associated with a decreasing swirl number in the vortex (see figure \ref{fig: swirl Re = 2M}). Apart from the $10^\circ$ case, the Reynolds number has a negligible effect on the value of this ratio.

\subsection{Recirculation}
\begin{figure}
\begin{centering}
\includegraphics[width=40mm,trim={0.5cm 1cm 1cm 8cm},clip]{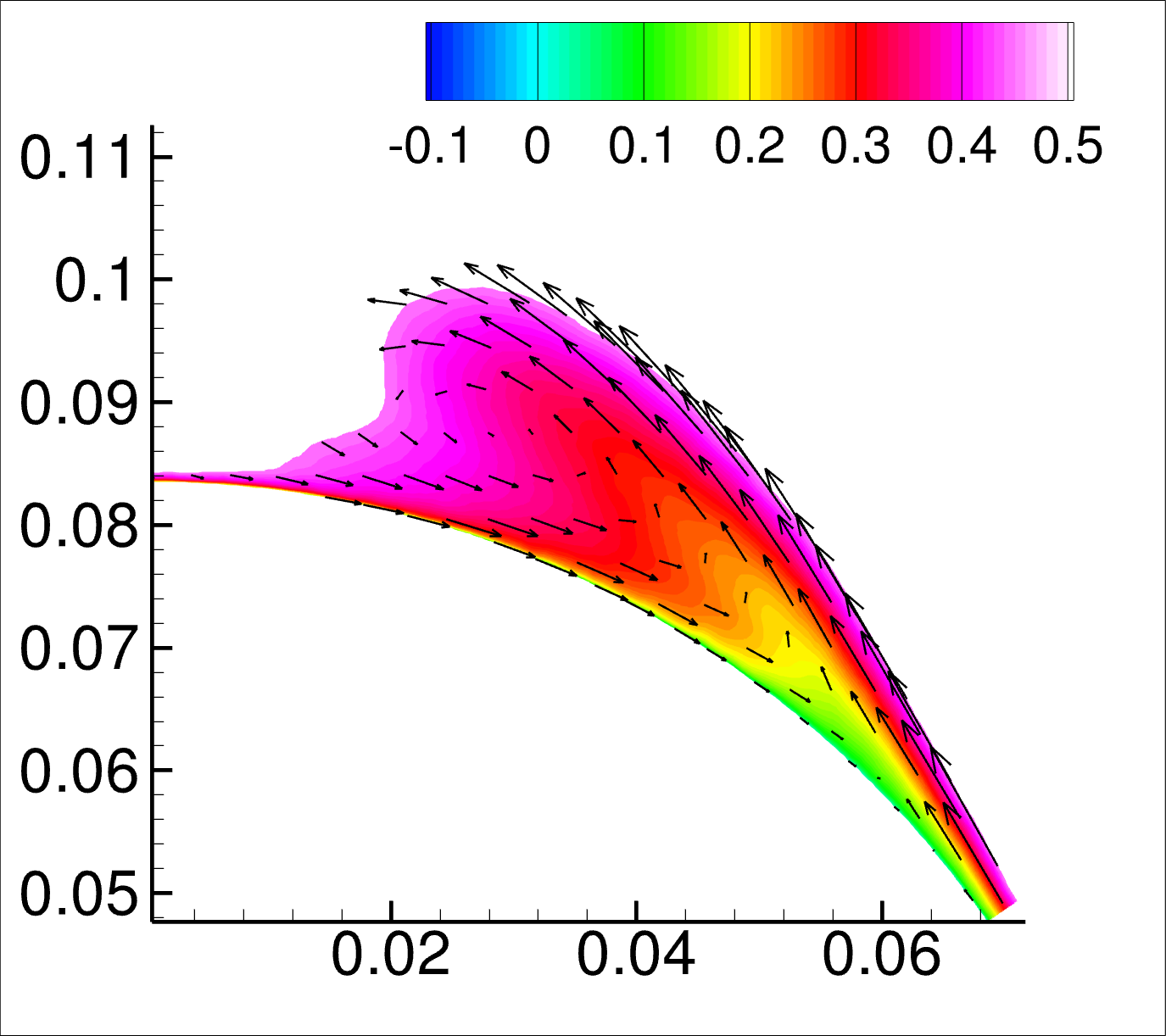}
\put(-95,20){(a)}
\includegraphics[width=40mm,trim={1cm 2.1cm 1cm 1cm},clip]{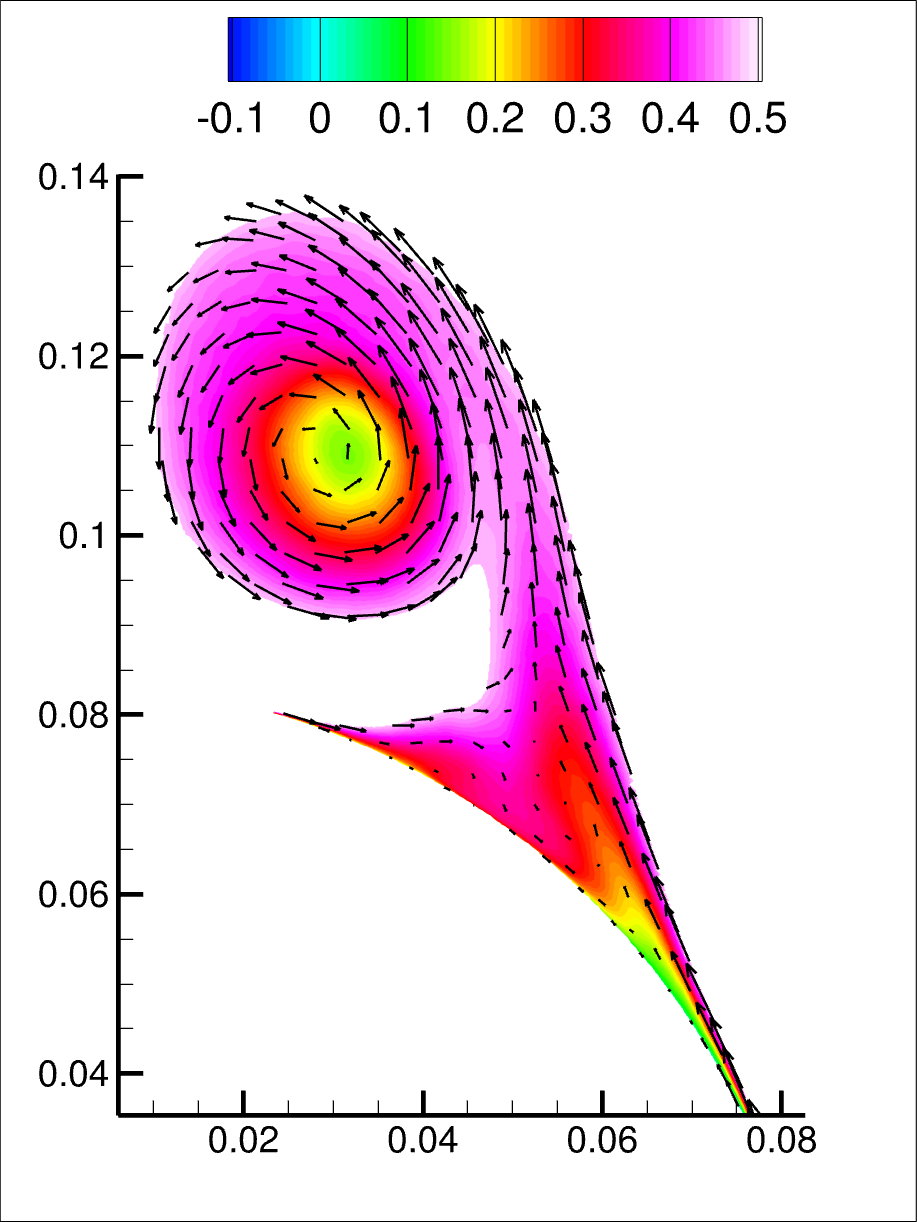}
\put(-95,20){(b)}
\put(-103,143){$P_s$}
\includegraphics[width=40mm,trim={1cm 2.2cm 1cm 5cm},clip]{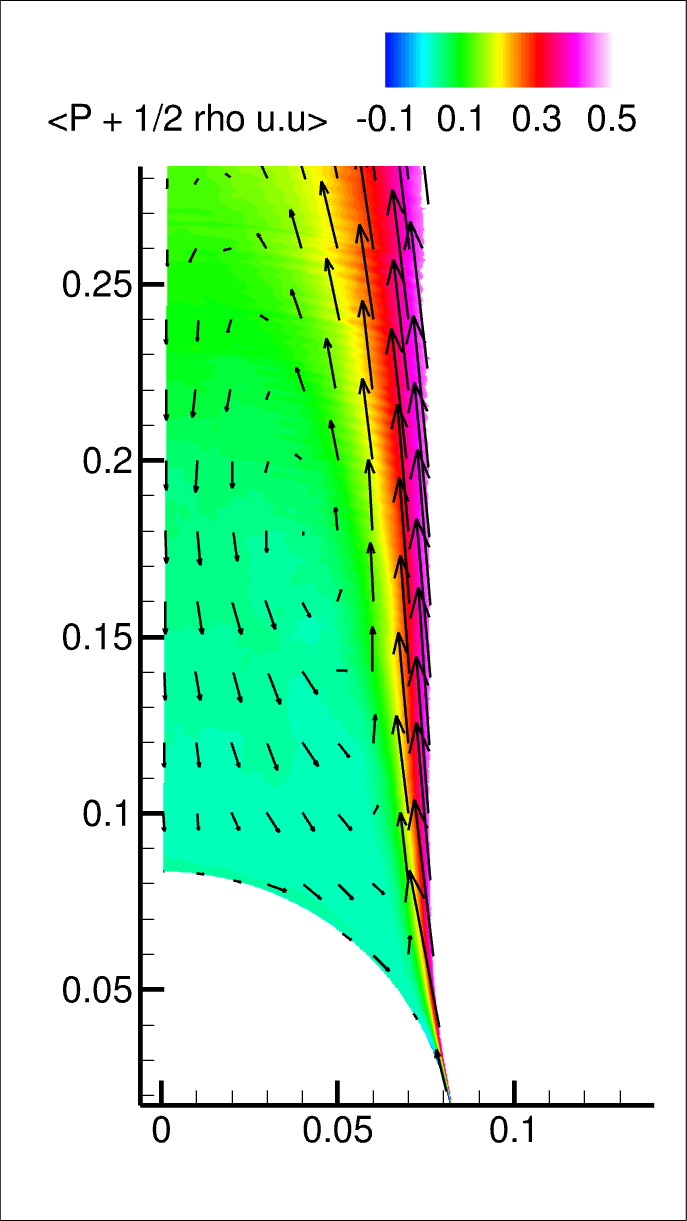}
\put(-85,20){(c)}
\caption{Time--averaged stagnation pressure in a transverse slice at (a): $(x/L, \alpha, Re) = (0.5, 20^\circ, 2M)$, showing the proto--vortex; (b) $(x/L, \alpha, Re) = (0.5, 40^\circ, 4M)$, showing the 3D coherent vortex; $(x/L, \alpha, Re) = (0.5, 90^\circ, 2M)$, showing the recirculating wake.}
\label{fig: states}
\end{centering}
\end{figure}

As discussed in section \ref{sec: overview}, a wide variety of recirculating flows are produced depending on the values of $Re$, $\alpha$ and the axial location $x/L$. These recirculations can be grouped into three states, as seen in figure \ref{fig: states}. Figure \ref{fig: states} (a) shows a "proto--vortex" state in which the separation is small and confined close to the wall. A weak recirculation is visible, although its center does not correspond to an extremum of pressure, stagnation pressure, or vorticity. Figure \ref{fig: states} (b) shows a coherent vortex state in which a vortex core is visible, bounded by a closed isoline of stagnation pressure. Similarly to the proto--vortex state, the boundary layer separates on the side of the spheroid, however, the separated shear layer travels farther from the wall and rolls up in a counter--rotating vortex pair. In addition, after attachment on the leeside, the boundary layer separates a second time below the vortex and forms a secondary vortex pair. An entrainment region is visible between the vortex pair, which pulls fluid from the freestream, below the primary vortex. Figure \ref{fig: states} (c) shows a recirculating wake without visible coherent vortex. In this state, the shear layer does not reattach and is advected in the freestream direction. The shear layer pair slowly diffuses vorticity with increasing distance from inception. A weak, low--velocity recirculation region is present between the symmetric shear layers.

In both the proto--vortex and coherent vortex states, the entirety of the separated layer is redirected in the axial direction. In the shedding recirculation state, 2D streamlines projected in a transverse plane and originating from the separation are unbounded. Thus, part of the flow originating from the separated boundary layer is not carried into the recirculation but is indefinitely advected along the freestream direction. The following sections describe these three states in more detail.

\subsection{Recirculation in proto--vortex case}

\begin{figure}
\begin{centering}
\includegraphics[width=60mm,trim={0.5cm 1cm 1cm 1cm},clip]{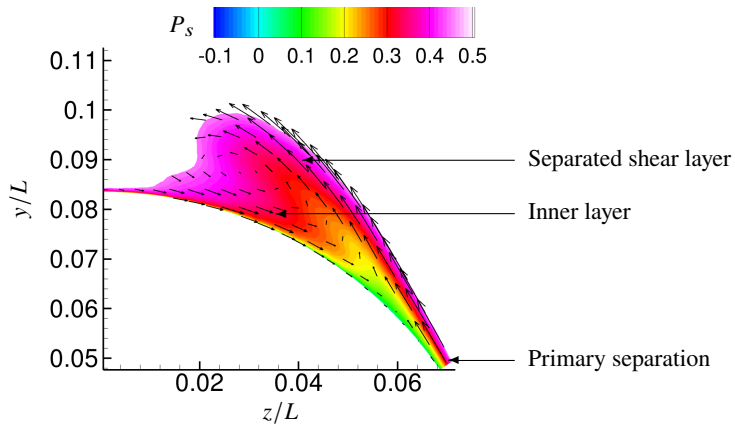}
\put(10,90){Separated shear layer}
\put(5,92){\vector(-1,0){80}}
\put(10,70){Inner layer}
\put(5,72){\vector(-1,0){90}}
\put(10,15){Primary separation}
\put(5,17){\vector(-1,0){25}}
\put(-185, 70){\rotatebox{90}{$y/L$}}
\put(-90, -5){$z/L$}
\put(-125, 140){$P_s$}
\caption{Time--averaged stagnation pressure in a transverse slice at $(x/L, \alpha, Re) = (0.5, 20^\circ, 2M)$, showing the proto--vortex.}
\label{fig: proto overview}
\end{centering}
\end{figure}

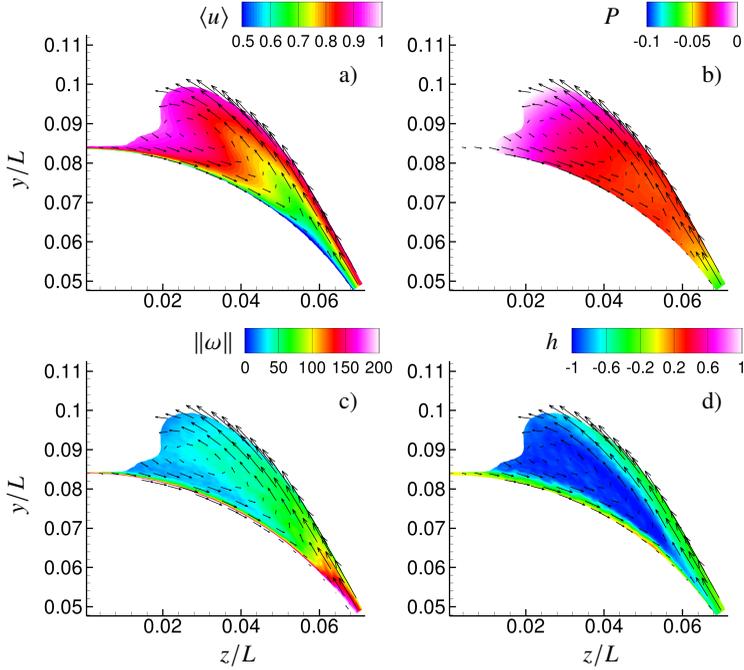
\begin{figure}
\begin{centering}
\includegraphics[width=48mm,trim={0.5cm 0.5cm 0.5cm 0.5cm},clip]{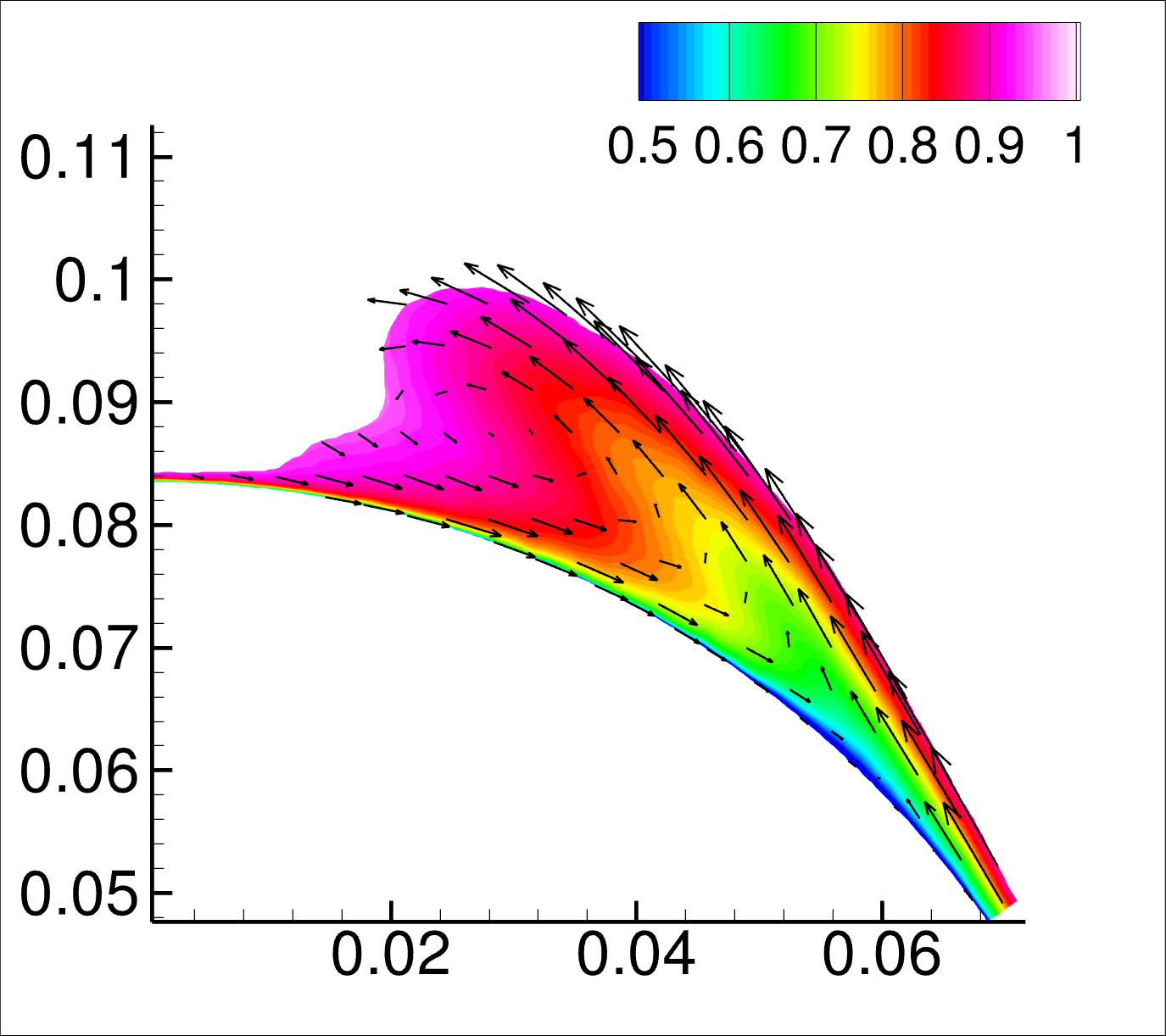}
\put(-25, 90){a)}
\put(-150, 50){\rotatebox{90}{$y/L$}}
\put(-78, 112){$\ave{u}$}
\includegraphics[width=48mm,trim={0.5cm 0.5cm 0.5cm 0.5cm},clip]{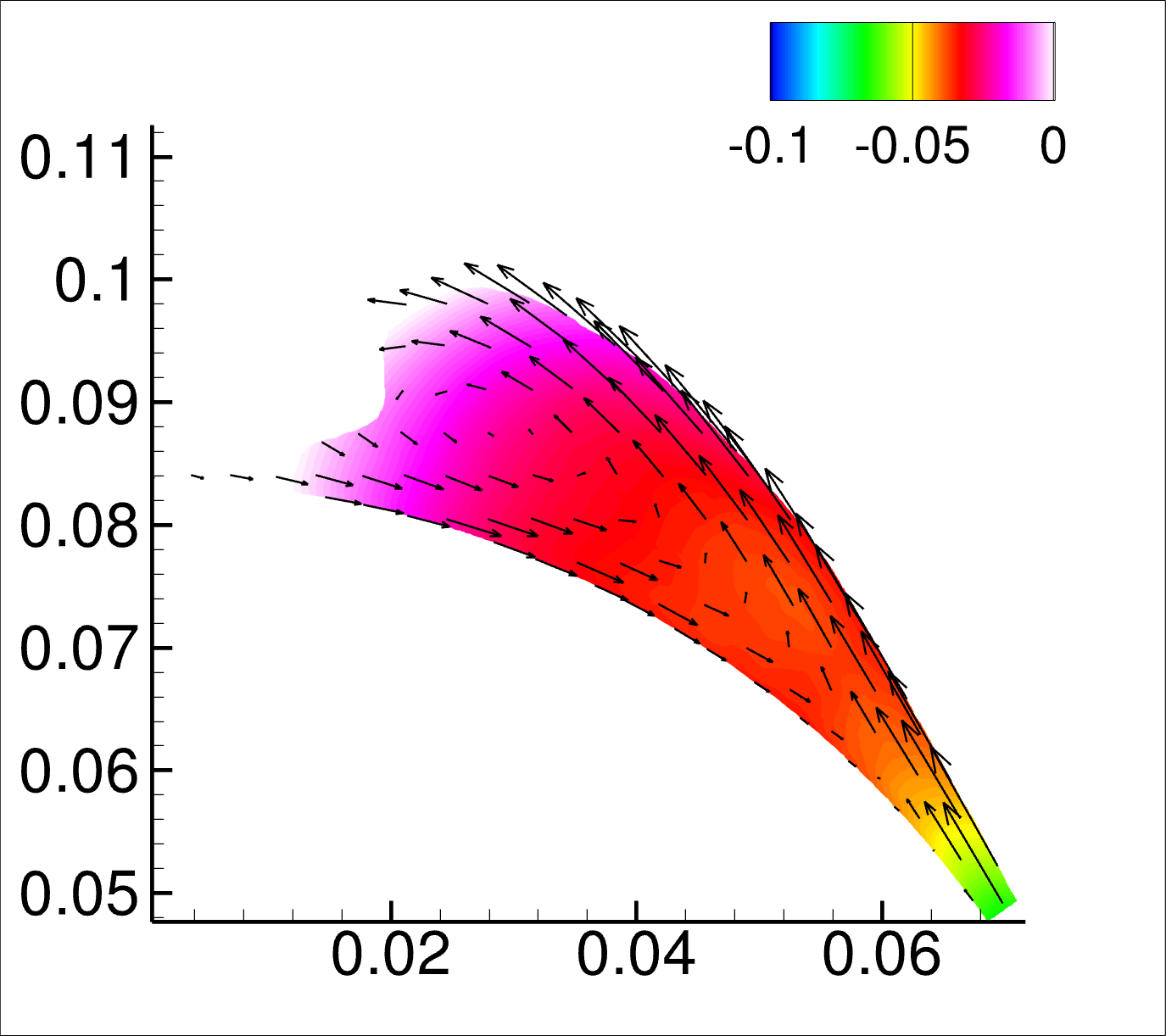}
\put(-25, 90){b)}
\put(-62, 112){$P$}

\includegraphics[width=48mm,trim={0.5cm 0.5cm 0.5cm 0.5cm},clip]{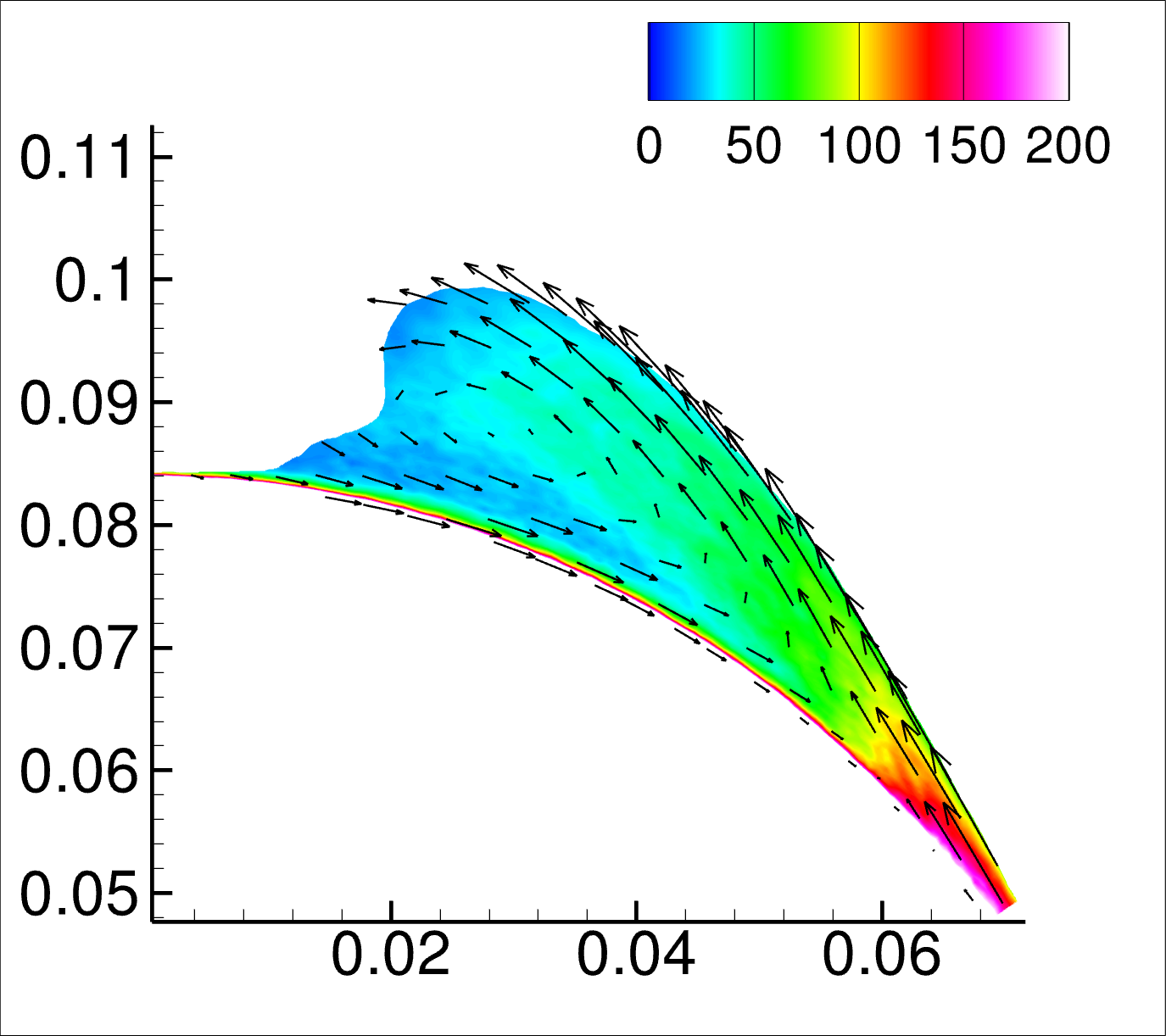}
\put(-25, 90){c)}
\put(-150, 50){\rotatebox{90}{$y/L$}}
\put(-70, -5){$z/L$}
\put(-80, 112){$\| \omega \|$}
\includegraphics[width=48mm,trim={0.5cm 0.5cm 0.5cm 0.5cm},clip]{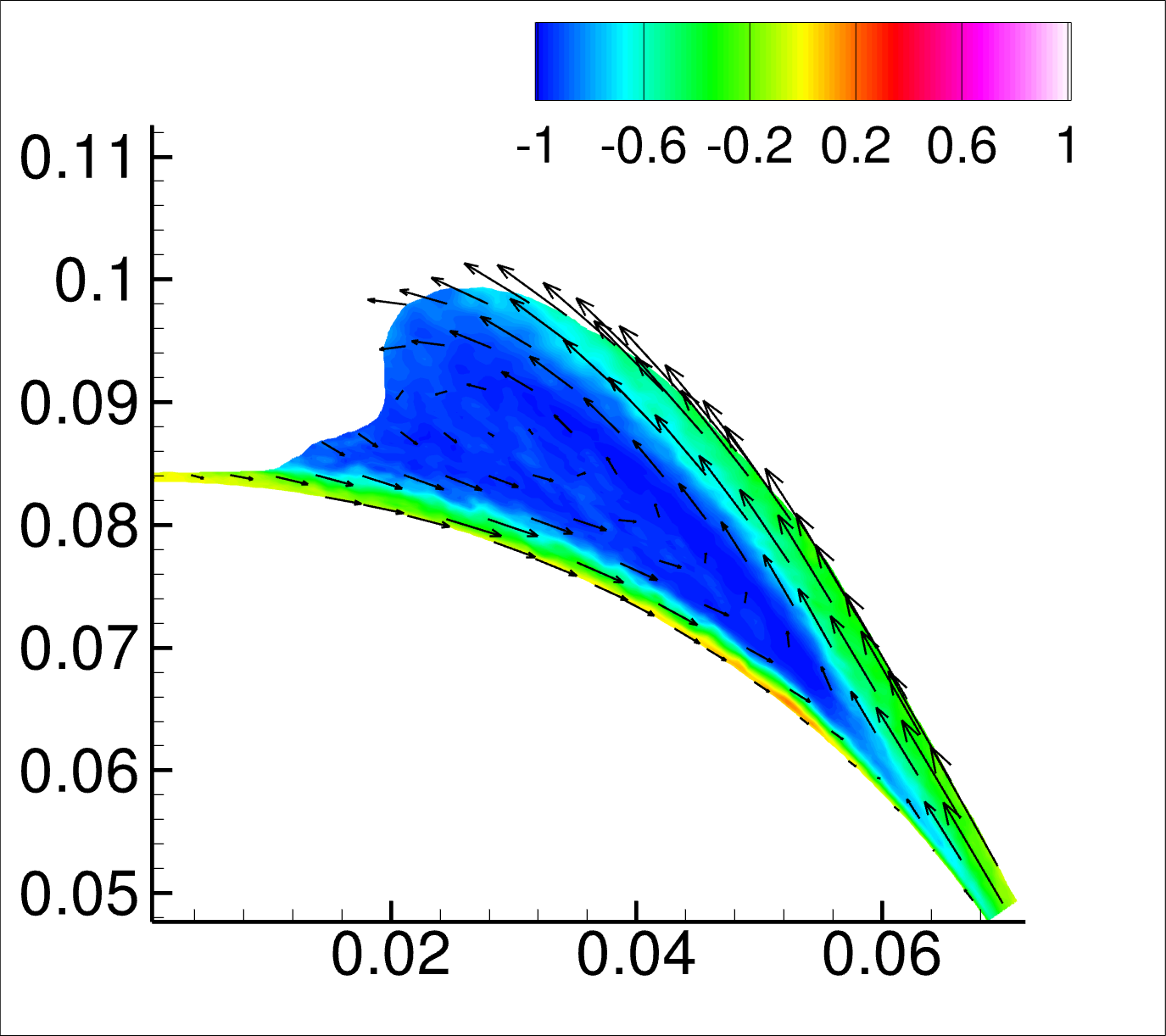}
\put(-25, 90){d)}
\put(-70, -5){$z/L$}
\put(-84, 112){$h$}
\caption{Time--averaged a) axial velocity, b) pressure, c) vorticity magnitude and d) normalized helicity density in a transverse slice at $(x/L, \alpha, Re) = (0.5, 20^\circ, 2M)$, showing the proto--vortex.}
\label{fig: proto fields}
\end{centering}
\end{figure}

Figure \ref{fig: proto overview} shows the time--averaged stagnation pressure at $(x/L, \alpha, Re) = (0.5, 20^\circ, 2M)$, for which a proto--vortex recirculation is formed. The flow can be divided into two layers. The outer layer is formed from the primary separation, moves in the leeward and axial directions, and has a higher vorticity. The inner layer is under the outer layer and travels in the windward direction. Figure \ref{fig: proto fields} shows the axial velocity, pressure, vorticity magnitude and normalized helicity density for the same parameters. The axial velocity, vorticity and helicity density also show the formation of a two--layer structure. In contrast to a coherent vortex, where the azimuthal velocity cancels at a point, the azimuthal velocity cancels along a boundary between the two layers. The magnitude of the helicity density is smaller at separation and in the outer layer; it is nearly unity in the recirculation region. This indicates that the vorticity is closely aligned with the direction of the streamlines. No localized minimum pressure or vorticity maximum is visible along the boundary. Instead, an adverse pressure gradient is visible in the outer layer and a favorable pressure gradient is observed in the inner layer.

\subsection{Recirculation in 3D vortex case}
\begin{figure}
\begin{centering}
\includegraphics[width=60mm,trim={1cm 1cm 1cm 1cm},clip]{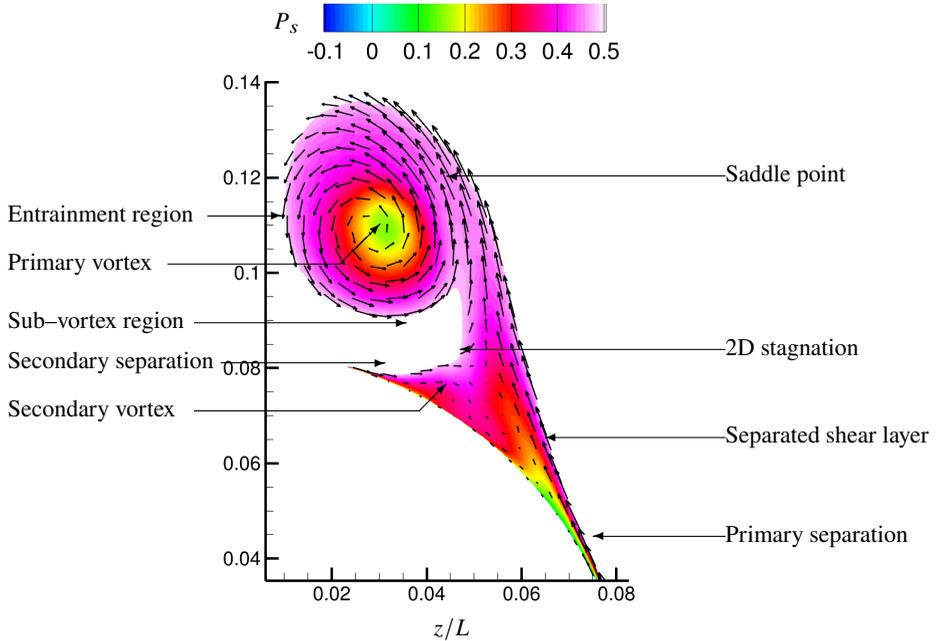}
\put(-250,150){Entrainment region}
\put(-175,152){\vector(1,0){28}}
\put(-250,132){Primary vortex}
\put(-190,134){\line(1,0){65}}
\put(-125,134){\vector(1,1){15}}
\put(-250,110){Sub--vortex region}
\put(-180,112){\vector(1,0){80}}
\put(-250,95){Secondary separation}
\put(-170,97){\vector(1,0){62}}
\put(-250,77){Secondary vortex}
\put(-180,79){\line(1,0){85}}
\put(-95,79){\vector(1,1){10}}
\put(20,30){Primary separation}
\put(20,32){\vector(-1,0){50}}
\put(20,67){Separated shear layer}
\put(20,69){\vector(-1,0){68}}
\put(20,100){2D stagnation}
\put(20,102){\vector(-1,0){100}}
\put(20,165){Saddle point}
\put(20,167){\vector(-1,0){105}}
\put(-90, -5){$z/L$}
\put(-150, 222){$P_s$}
\caption{Time--averaged stagnation pressure in a transverse slice at $x/L = 0.7$ for $\alpha = 40^\circ$, $Re = 2M$, showing the coherent 3D vortex.}
\label{fig: 3d vortex overview}
\end{centering}
\end{figure}

\begin{figure}
\begin{centering}
\includegraphics[width=48mm,trim={1cm 1cm 1cm 1cm},clip]{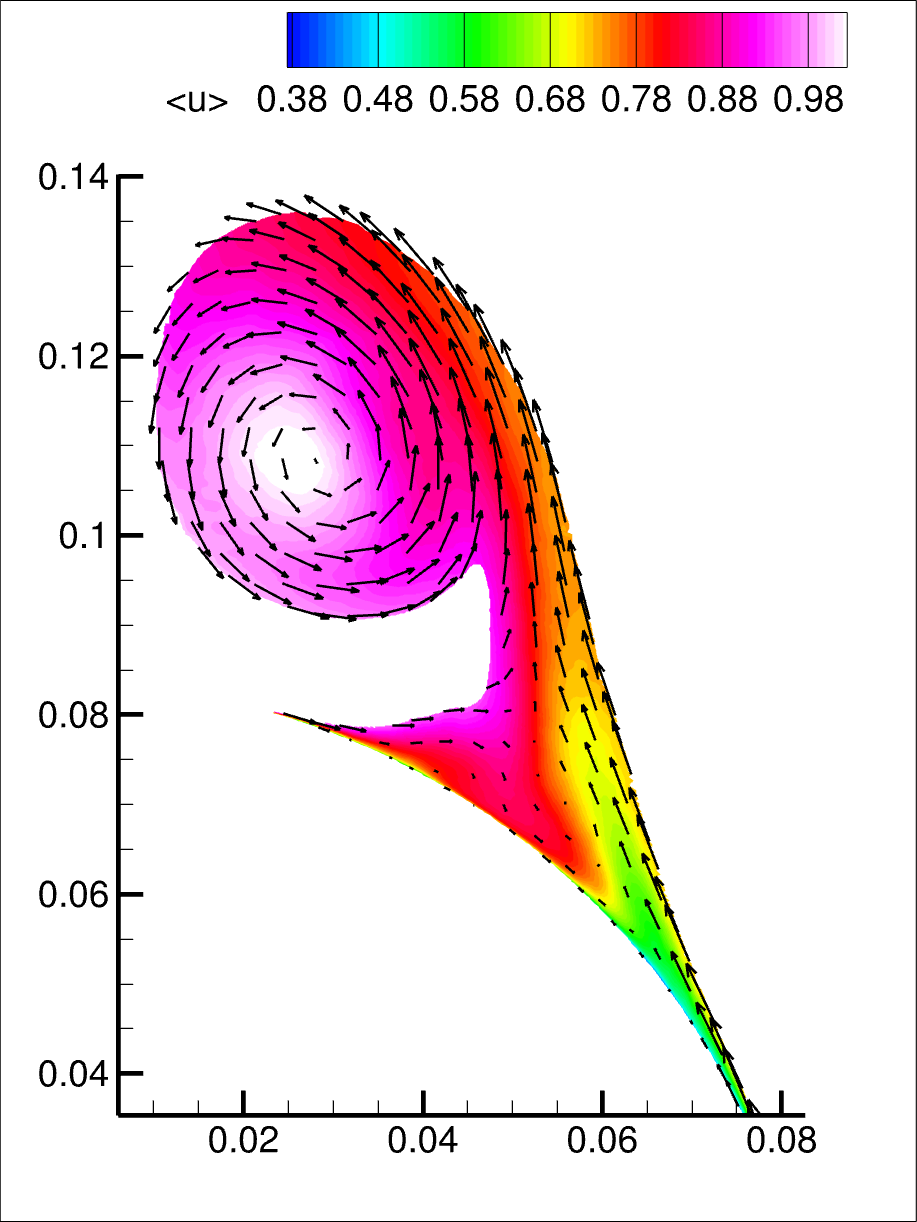}
\put(-30,150){a)}
\put(-145, 85){\rotatebox{90}{$y/L$}}
\includegraphics[width=48mm,trim={1cm 1cm 1cm 1cm},clip]{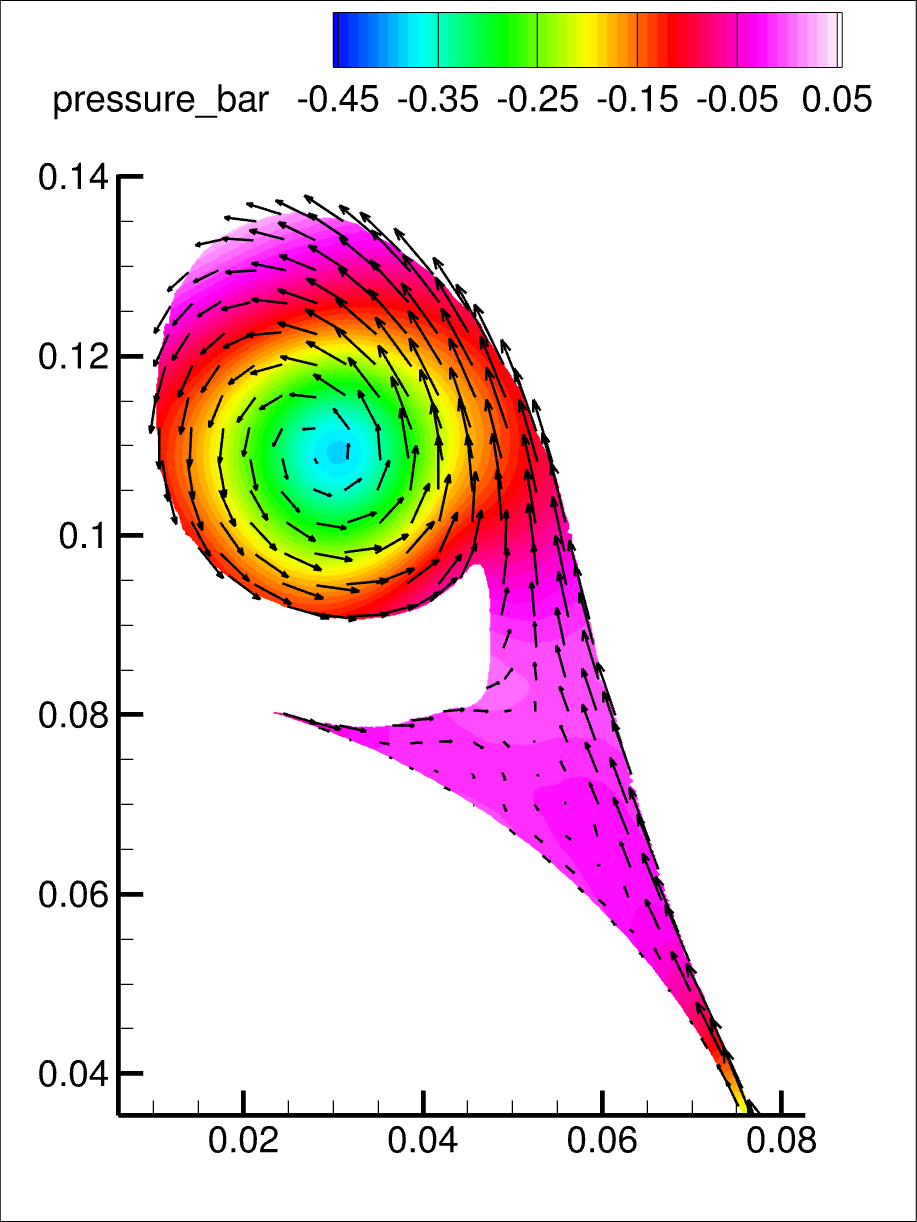}
\put(-30,150){b)}

\includegraphics[width=48mm,trim={1cm 1cm 1cm 1cm},clip]{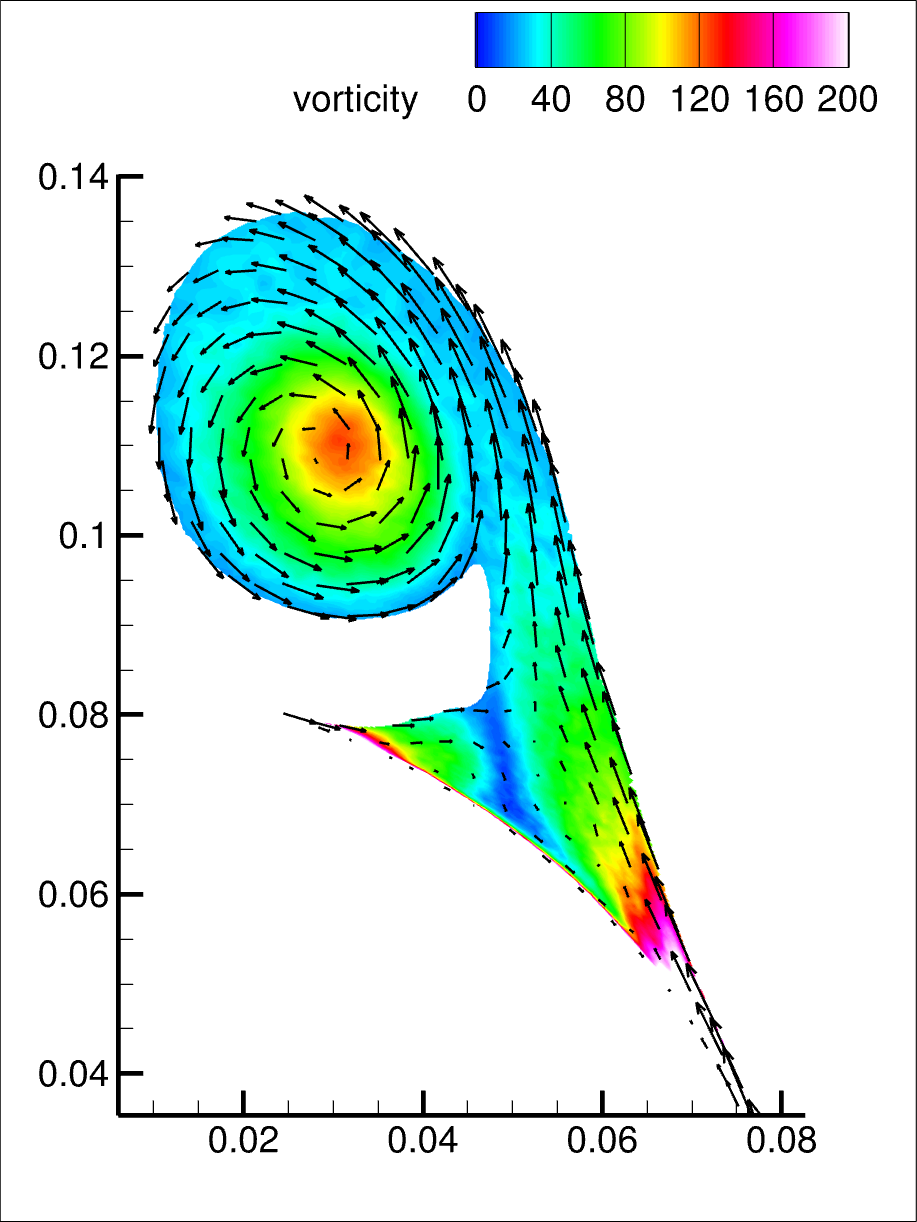}
\put(-30,150){c)}
\put(-145, 85){\rotatebox{90}{$y/L$}}
\put(-70, -5){$z/L$}
\includegraphics[width=48mm,trim={1cm 1cm 1cm 1cm},clip]{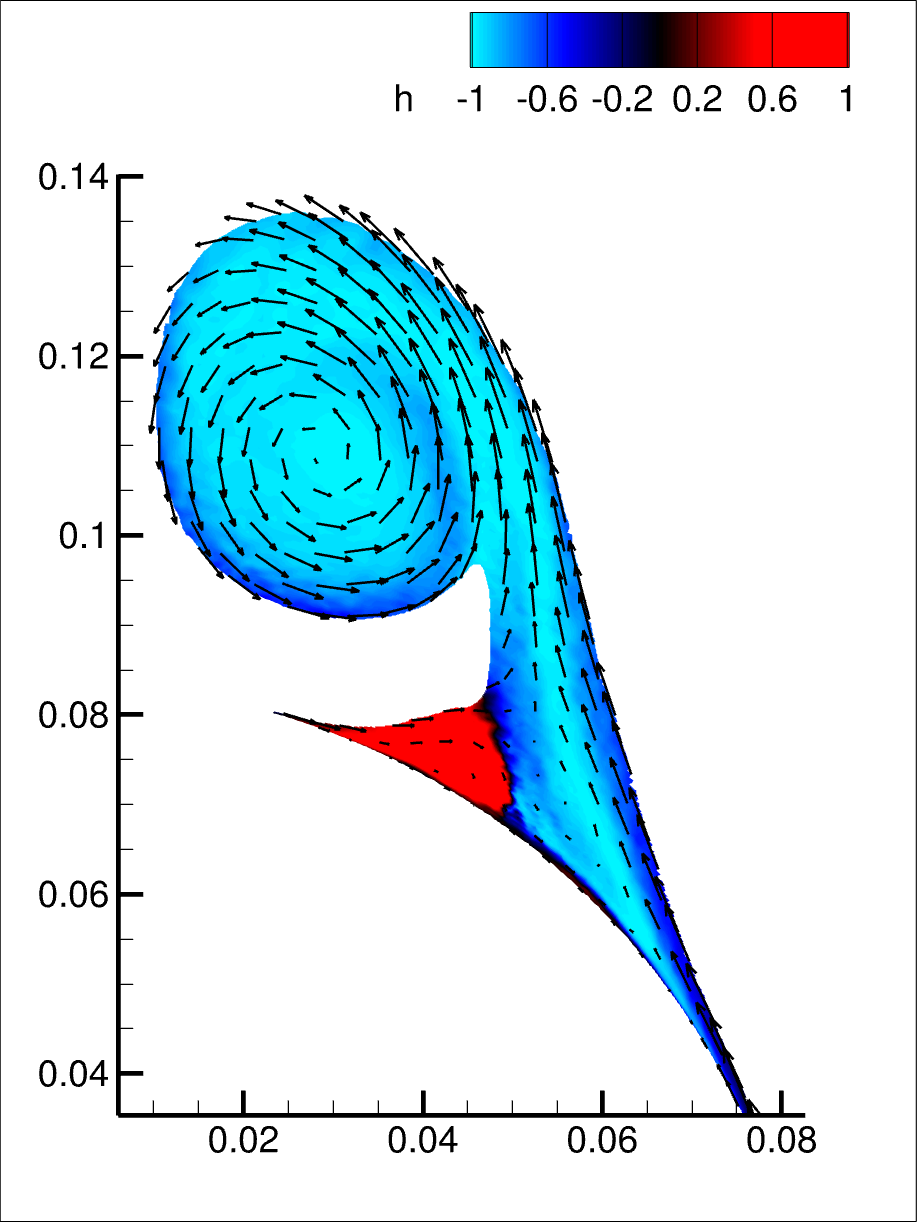}
\put(-30,150){d)}
\put(-70, -5){$z/L$}
\caption{Time--averaged a) axial velocity, b) pressure, c) vorticity magnitude and d) normalized helicity density in a transverse slice at $x/L = 0.7$ for $\alpha = 40^\circ$, $Re = 2M$.}
\label{fig: 3d vortex fields}
\end{centering}
\end{figure}

Figure \ref{fig: 3d vortex overview} shows time--averaged stagnation pressure for $(x/L, \alpha, Re) = (0.5, 40^\circ, 4M)$, where the recirculation forms a coherent 3D vortex.  After the primary separation, the separated boundary layer is convected along the y--direction, where it rolls up into a distinct vortical structure. This vortex entrains flow from the freestream and induces a secondary separation that leads to a secondary vortex rotating in the opposite direction. Figure \ref{fig: 3d vortex fields} shows the axial velocity $\ave{u}$, pressure, vorticity magnitude, and normalized helicity density for $(x/L, \alpha, Re) = (0.5, 40^\circ, 4M)$. The following subsections describe each area of the 3D vortex flow in more detail.

\subsubsection{Separation sheet}
\label{sec: 3d separation sheet}
Downstream of the primary separation, the separated shear layer travels tangentially to the wall, with an axial component inversely proportional to the incidence. The fields displayed in figure \ref{fig: 3d vortex fields} show that the shear layer is actually composed of two sub-layers, similarly to the proto--vortex state: the inner sub-layer has low secondary velocity, high axial velocity, while the outer sub-layer has high secondary velocity, low axial velocity. The inner layer also has low vorticity, suggesting that the fluid originates from the freestream entrained between the primary vortex pair. The outer sub-layer has high vorticity and high turbulent kinetic energy. It is also a region of production of TKE. This high production likely comes from the shear between the two sub-layers and the displacement of the shear layer in time. This is related to the large wall pressure and velocity fluctuations that \cite{goody1998} observed close to separation and into the separation layer, respectively.
In addition, the separated shear layer is a pressure maximum and has a helicity density close to unity, which indicates that the axis of vorticity is closely aligned with the streamlines. This observation is consistent with a vortex comprised of small helical vortices as seen in DNS in the laminar $Re$ regime \citep{jiang2016, strandenes2019}. This structure consisting of small vortices in the separation layer, helically wrapped to form a coherent vortex, is not unique to the spheroid and has been observed in the flow over a ducted propeller \citep{leasca2025}. The higher TKE in this location was also observed by \cite{chesnakas1996}.
The shear layer is then accelerated outside the vortex by a favorable pressure gradient, correlated with a decrease in vorticity and TKE. The shear layer then attaches to the primary vortex at a point that will be referred to as the saddle point. This name is used because $\dd{P_s}{\theta} = 0$, $\dd{P_s}{r} = 0$, $\frac{\partial^2 P_s}{\partial r^2} \cdot \frac{\partial^2 P_s}{\partial \theta^2} < 0$ where $(r,\theta)$ are the cylindrical coordinates with respect to the center of the vortex. This point lies on the boundary of the vortex and is used to assess the highest value of stagnation pressure inside it. 

\subsubsection{Primary vortex}
Figure \ref{fig: 3d vortex fields} shows that the primary vortex itself can be divided into two regions. The first outer region is asymmetric and originates from the roll-up of the separated shear layer as previously discussed. The inner region, or vortex core, is close to axisymmetric, has high vorticity, high TKE, low pressure, low radial velocity, close to uniform tangential velocity, and axial velocity that increases in the radial direction. 

\begin{figure}
\begin{centering}
\includegraphics[width = 120 mm]{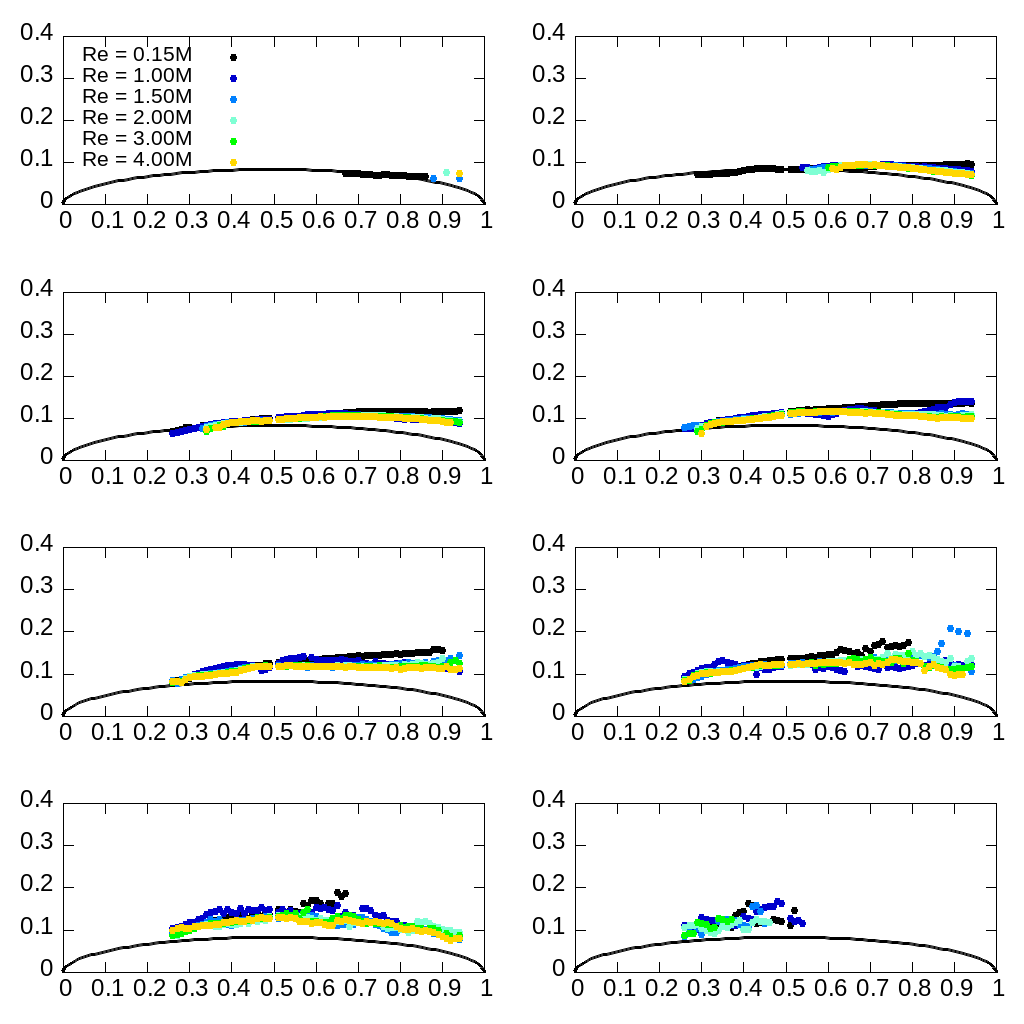}
\put(-260,-5){$x/L$}
\put(-90,-5){$x/L$}
\put(-350, 299){\rotatebox{90}{$y/L$}}
\put(-350, 214){\rotatebox{90}{$y/L$}}
\put(-350, 129){\rotatebox{90}{$y/L$}}
\put(-350, 44){\rotatebox{90}{$y/L$}}
\put(-195, 315){a)}
\put(-25, 315){b)}
\put(-195, 230){c)}
\put(-25, 230){d)}
\put(-195, 145){e)}
\put(-25, 145){f)}
\put(-195, 60){g)}
\put(-25, 60){h)}
\caption{Location of the primary vortex center vs $x/L$ from a) to h): $\alpha = 10^\circ, 20^\circ, 30^\circ, 40^\circ, 50^\circ, 60^\circ, 70^\circ, 90^\circ$.}
\label{fig: vortex position vs x}
\end{centering}
\end{figure}

Figure \ref{fig: vortex position vs x} shows the location of the center of the vortex when it is formed, for all cases. The vortex lines are almost horizontal, with a small positive slope. That slope is slightly higher on the first half of the spheroid compared to that on the second half. This is understood to be related to the vortex growth pushing the center point farther from the wall. The wall slopes downward for $x/L > 0.5$, which accommodates a larger space for the vortex to grow. Angle of attack and Reynolds number do not visibly affect the slope of the vortex line, although the vertical offsets of the lines increase with $\alpha$. This increasing offset is related to the vortex starting to develop at an increasingly lower axial position with increasing $\alpha$.

\begin{figure}
\begin{centering}
\includegraphics[width = 40 mm]{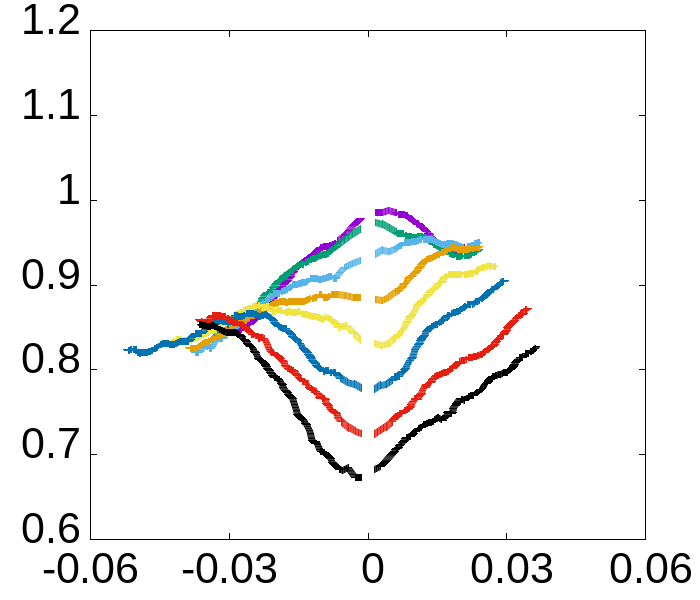}
\put(-55,-5){$r$}
\put(-122, 50){\rotatebox{90}{$u$}}
\put(-95, 85){a)}
\hspace{3mm}
\includegraphics[width = 40 mm]{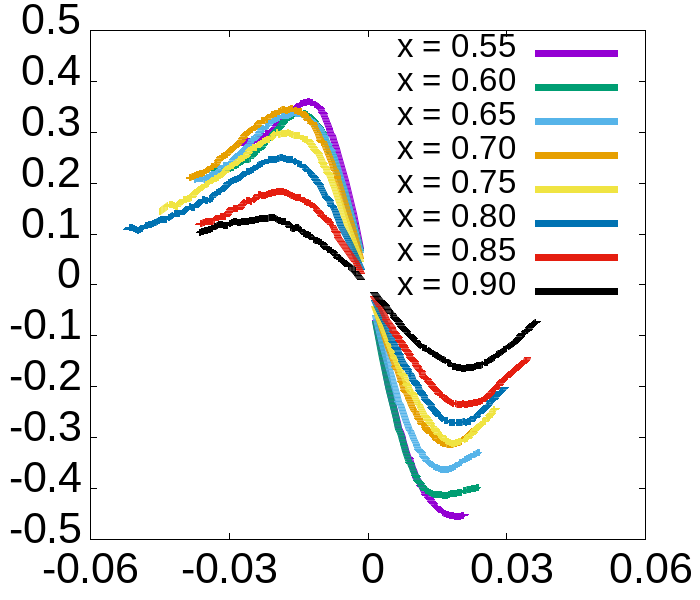}
\put(-55,-5){$r$}
\put(-122, 50){\rotatebox{90}{$u_{\theta}$}}
\put(-95, 85){b)}
\hspace{3mm}
\includegraphics[width = 40 mm]{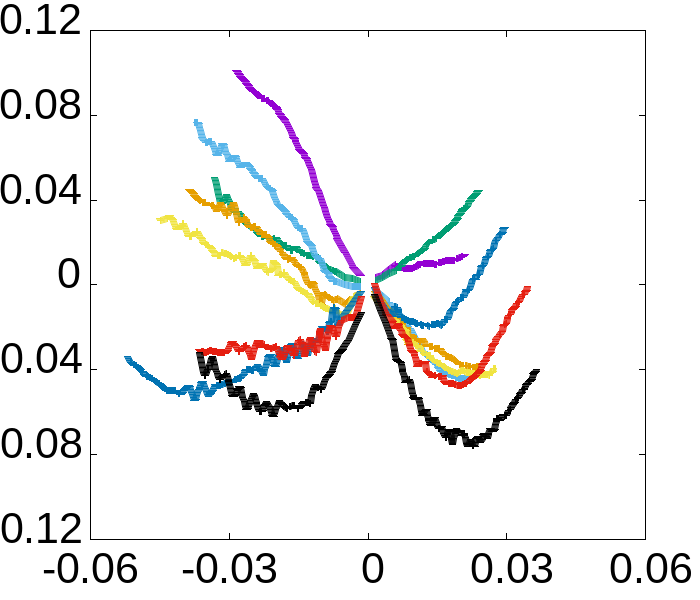}
\put(-55,-5){$r$}
\put(-122, 50){\rotatebox{90}{$u_r$}}
\put(-95, 85){c)}
\caption{Time--averaged a) axial, b) tangential, c) radial velocity along a $y$ aligned segment passing by the center of the primary vortex at $\alpha = 40^\circ$, $Re = 4M$. The different curves represent different axial locations from $x/L = 0.55$ to $x/L = 0.9$ by $\Delta x = 0.05 L$ increment.}
\label{fig: profiles AoA40 Re4M}
\end{centering}
\end{figure}

\begin{figure}
\begin{centering}
\includegraphics[width = 40 mm]{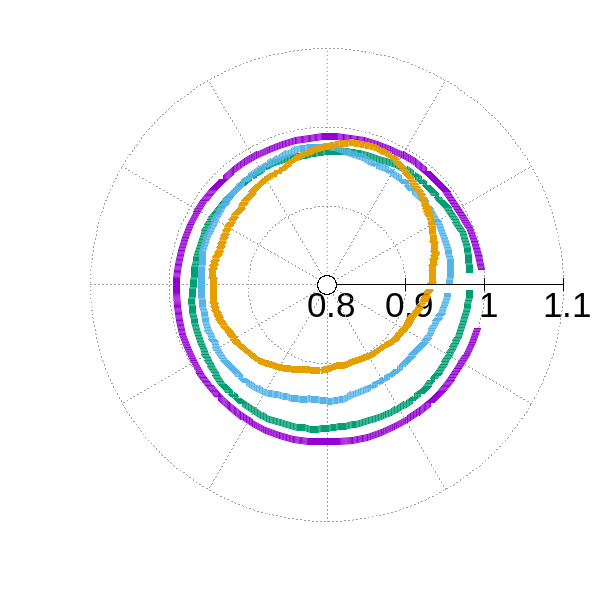}
\put(-95, 95){a)}
\hspace{3mm}
\includegraphics[width = 40 mm]{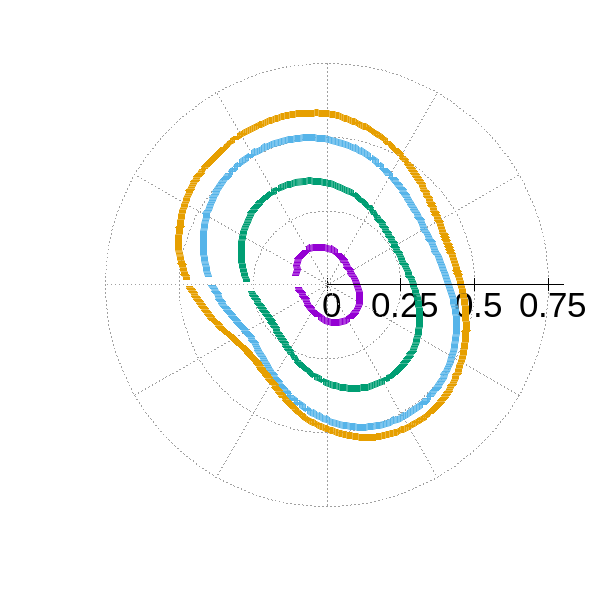}
\put(-95, 95){b)}
\hspace{3mm}
\includegraphics[width = 40 mm]{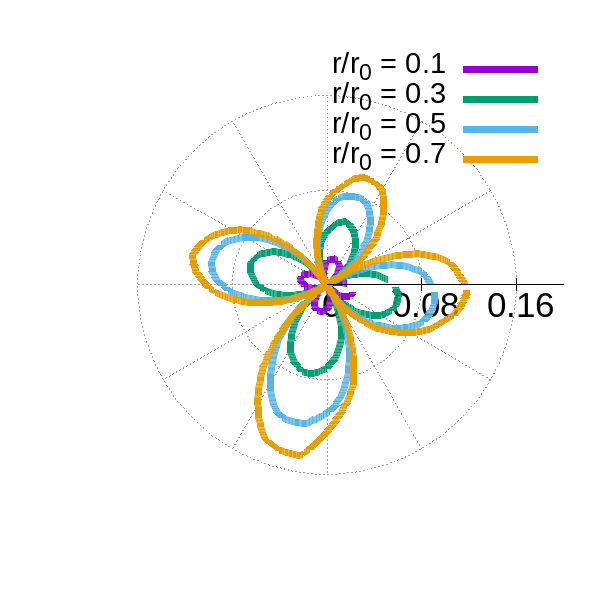}
\put(-95, 95){c)}
\caption{Time--averaged a) axial, b) tangential, c) radial velocity along circle of radius $r/r_0 = 0.1, 0.3, 0.5, 0.7$ at $\alpha = 40^\circ$ and $Re = 4M$.}
\label{fig: azim AoA40 Re4M}
\end{centering}
\end{figure}

\begin{figure}
\begin{centering}
\includegraphics[width = 60 mm]{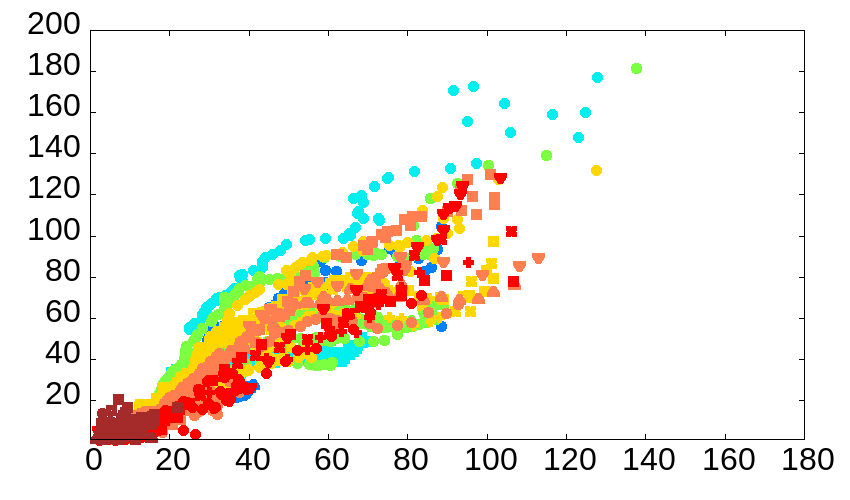}
\put(-90,-10){$\overline{\omega}_x^v$}
\put(-185, 40){\rotatebox{90}{$\dd{u_{\theta}} {r}_{|r=0}$}}
\caption{Slope at origin of $u_{\theta}$ versus average axial vorticity in the vortex $\overline{\omega}_x^v$.}
\label{fig: ut vs ave(wx)}
\end{centering}
\end{figure}

Figure \ref{fig: profiles AoA40 Re4M} shows the three components of velocity in cylindrical coordinate $(u_x, u_{\theta}, u_r)$ along two radial lines originating at the center of the vortex, at angles $\theta = \pi/6$ and $\theta = 7 \pi/6$, for $\alpha = 40^\circ$, $Re = 4M$ and from $x/L = 0.55$ to $x/L = 0.9$ by step of $0.05L$. At $x/L = 0.55$, the axial velocity is maximal in the center of the vortex and decreases linearly. The center velocity decreases with $x$ such that at $x/L = 0.9$, it is a local minimum, with $u_x$ increasing linearly with $r$. The aximuthal velocity increases linearly from the center to around half of the radius of the vortex, then decays until the border of the vortex. The slope and maximum velocity are maximal at $x/L = 0.55$ and decrease with axial distance. This is similar to a Burgers vortex $(u_x, u_{\theta}, u_r) = (-ar, \frac{\Gamma}{2\pi r}(1-exp(-\frac{ar^2}{2\nu})), 2az)$. Note that not all $(Re, \alpha)$ pairs resulted in a Burgers vortex, although most cases have a linear slope of $u_{\theta}$ at the origin of the vortex. This linear slope close to the center of the vortex is indicative of a solid--body rotation, as demonstrated by figure \ref{fig: ut vs ave(wx)} showing the slope at the origin of the vortex of azimuthal velocity versus the average axial vorticity.

\begin{figure}
\begin{centering}
\includegraphics[width = 60 mm]{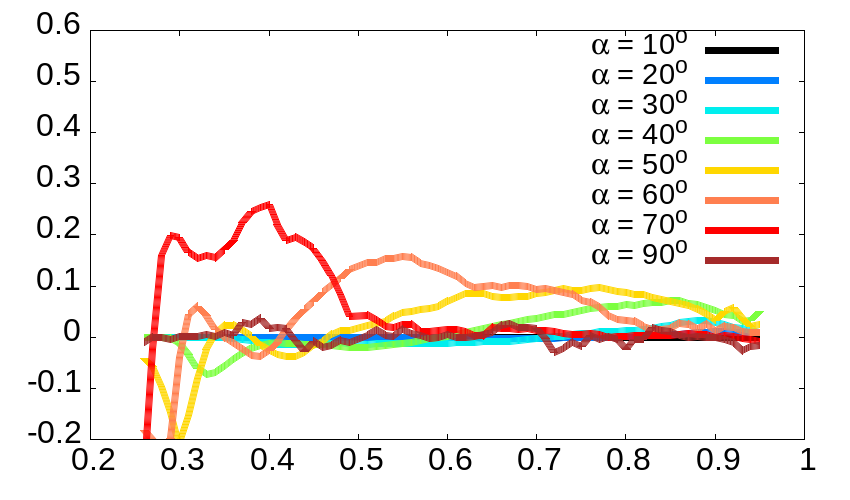}
\put(-90,-5){$x/L$}
\put(-185, 40){\rotatebox{90}{$\omega_j\dd{u_x}{x_j}$}}
\caption{Vortex stretching $\omega_j\dd{u_x}{x_j}$ vs $x/L$ for $Re = 2M$.}
\label{fig: vortex stretching vs x}
\end{centering}
\end{figure}

\begin{figure}
\begin{centering}
\includegraphics[width = 60 mm]{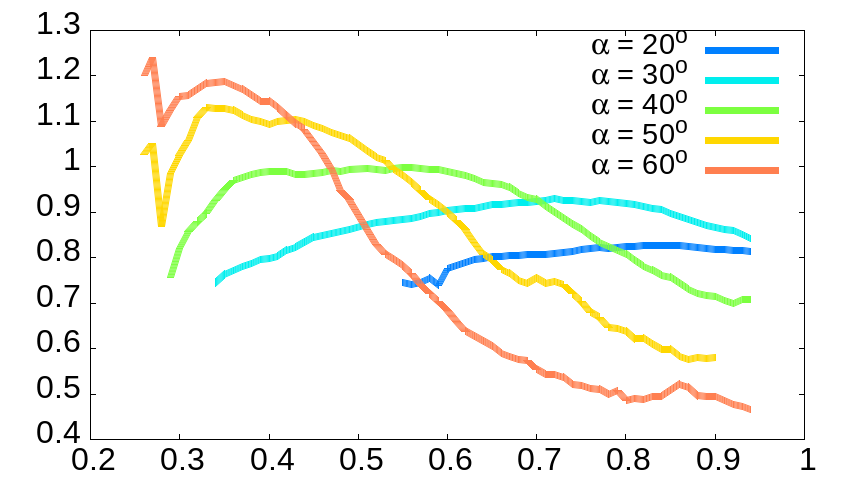}
\put(-90,-5){$x/L$}
\put(-175, 40){\rotatebox{90}{$u_{r=0}$}}
\caption{Vortex center velocity vs $x/L$ for $Re = 2M$.}
\label{fig: vortex center velocity vs x}
\end{centering}
\end{figure}

The radial velocity is positive at $x/L = 0.5$ and decreases with $x$ to be negative at $x/L = 0.9$. This indicates that initially the streamlines spiral into the vortex, the radial flow decreases with $x$ and close to the tail of the spheroid, the streamlines spiral out of the vortex. This observation, combined with the fact that the center velocity decreases with $x$, suggests that the vortex experiences a contraction. This is confirmed by figure \ref{fig: vortex stretching vs x}, showing that the vortex first experiences a stretching (negative values, with $\omega_x < 0$ in the primary vortex) followed by a contraction. The contraction starts earlier and is stronger with increasing $\alpha$. Figure \ref{fig: vortex center velocity vs x} shows the velocity in the center of the vortex for $Re = 2M$. The velocity first peaks and decreases until the tail of the spheroid. The location of the peak is earlier and the height of the peak is higher as $\alpha$ increases. The peak does not match the point where the vortex starts to compress, but the location where the compression is maximal.

\begin{figure}
\begin{centering}
\includegraphics[width = 120 mm]{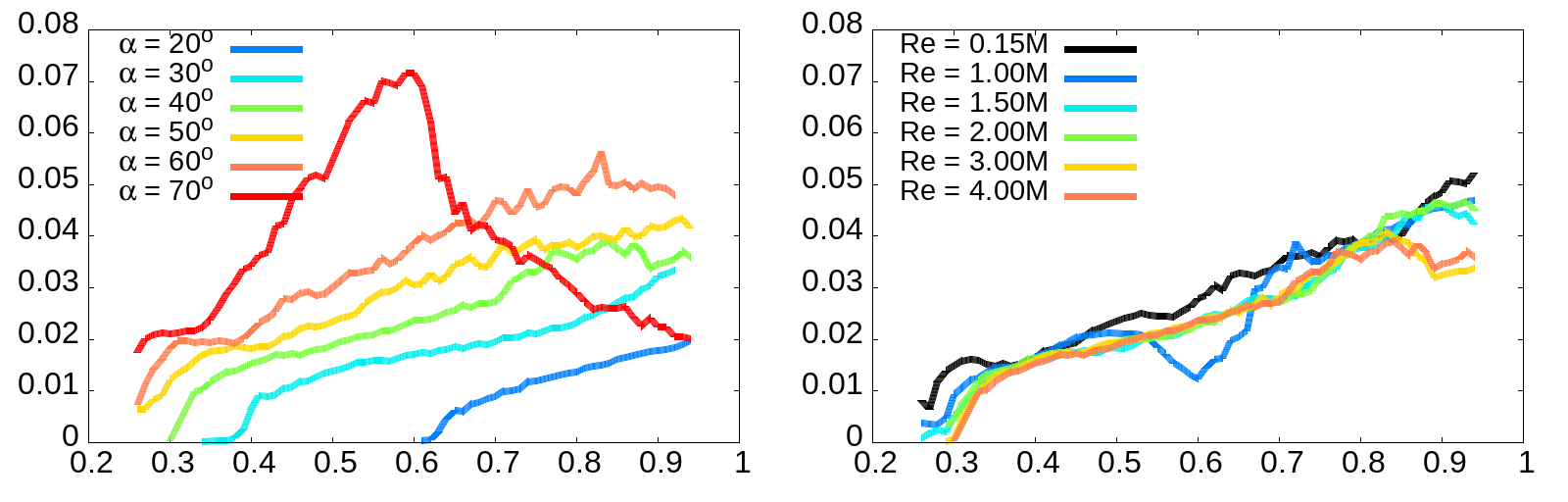}
\put(-260,-5){$x/L$}
\put(-90,-5){$x/L$}
\put(-195,90){a)}
\put(-25,90){b)}
\put(-350, 60){\rotatebox{90}{$r$}}
\put(-60,28){double vortex}
\put(-61,30){\vector(-1,0){15}}
\caption{Primary vortex radius $r$ vs $x/L$ for $\alpha \in [20^\circ, 70^\circ]$, $Re = 4M$ (a) and for $\alpha = 40^\circ$, $Re \in [0.15M, 4M]$ (b).}
\label{fig: vortex radius vs x}
\end{centering}
\end{figure}

\begin{figure}
\begin{centering}
\includegraphics[width = 80 mm]{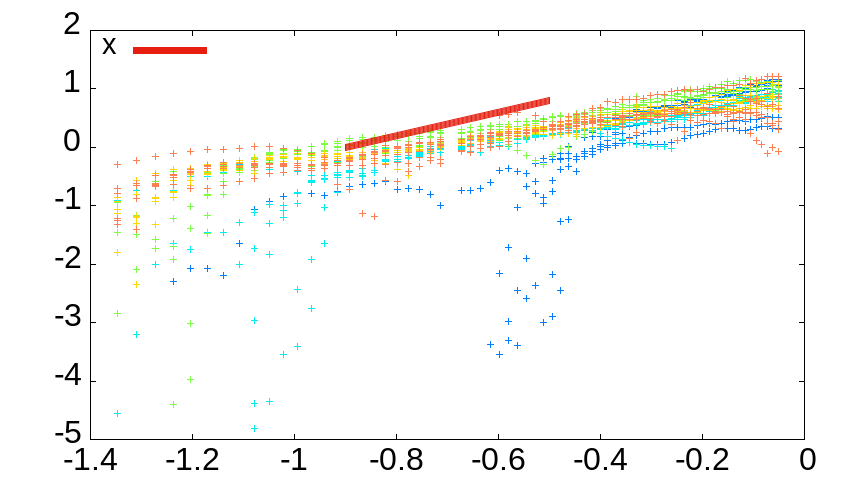}
\put(-120,-5){$log(x)$}
\put(-230, 50){\rotatebox{90}{$log(r)$}}
\caption{Logarithm of the primary vortex radius vs $log(x)$ for $\alpha \in [20^\circ, 60^\circ]$.}
\label{fig: log vortex radius vs x}
\end{centering}
\end{figure}

Figure \ref{fig: vortex radius vs x} shows the radius of the primary vortex along the spheroid for $\alpha \in [20^\circ, 70^\circ]$, $Re = 4M$ and for $\alpha = 40^\circ$, $Re \in [0.15M, 4M]$. Three regions are observed: 

(i) Early in the vortex formation, the radius increases linearly with a large slope ($ > 0.2 r/(x/L)$) for a short distance ($\approx 0.1 x/L$).

(ii) The large increase is followed by a slower increase ($\approx 0.04 r/(x/L)$) until the tail of the spheroid for $\alpha <= 30^\circ$ and until $x/L \approx 0.6$ at $\alpha = 50^\circ$ and $x/L \approx 0.4$ for $70^\circ$. The angle of attack does not have a strong effect on this slope.

(iii) For $\alpha = 70^\circ$, a rapid linear increase is followed by a decrease in radius at high $Re$.

\begin{figure}
\begin{centering}
\includegraphics[trim={35 350 10 10}, clip, width = 100 mm]{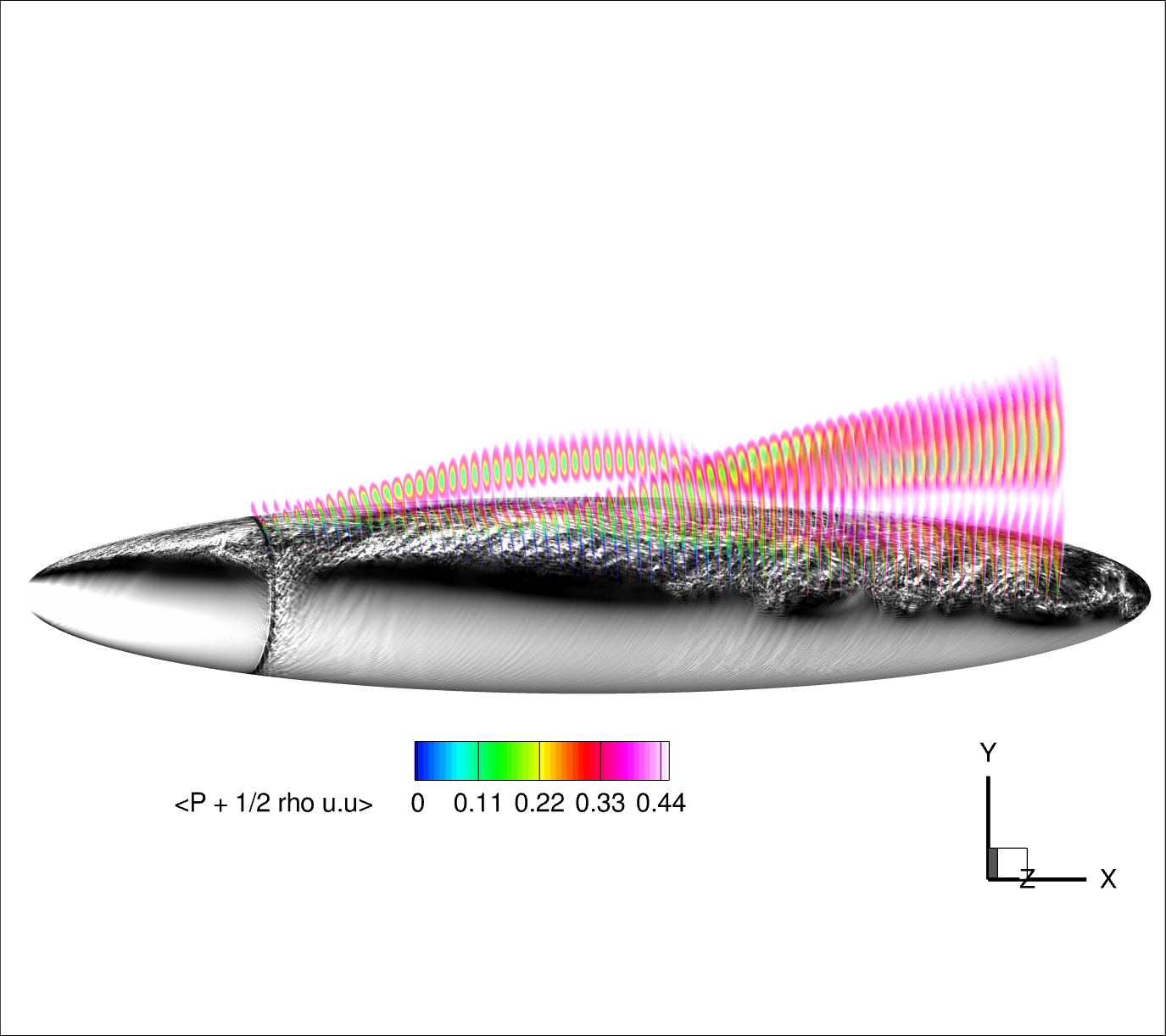}
\put(-280, 60){a)}

\includegraphics[trim={35 300 30 450}, clip, width = 100 mm]{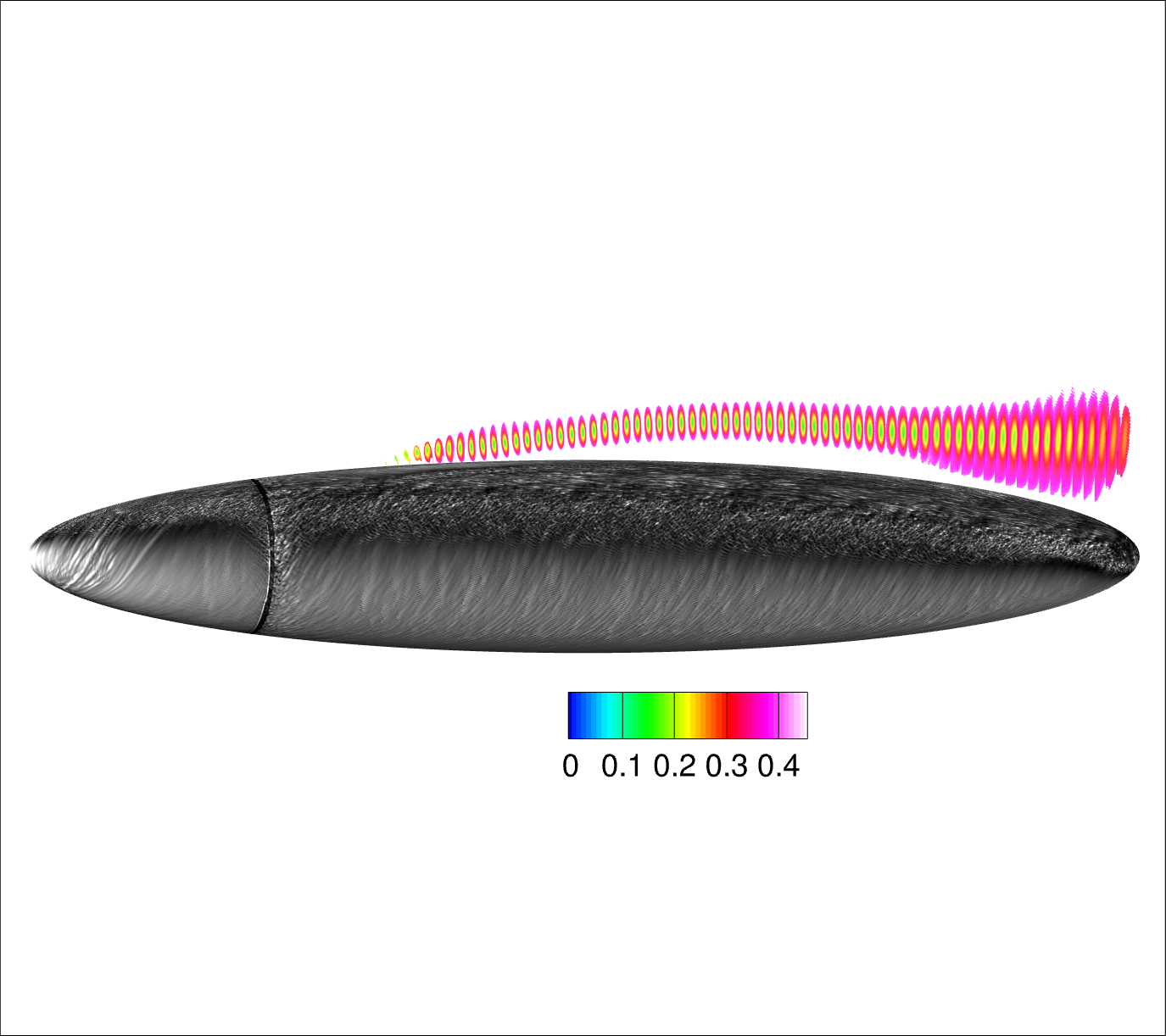}
\put(-280, 78){b)}
\put(-160, 13){$P_s$}
\caption{Visualization of the spheroid at a) $\alpha = 40^\circ$ and $Re = 1M$; b) $\alpha = 40^\circ$ and $Re = 4M$. The wall of the spheroid is shaded by instantaneous skin friction coefficient; one half of the flow ($\phi \in [0^\circ, 180^\circ]$) is shown in transverse slices spaced by $\Delta x = 0.01$ and shaded by time--averaged stagnation pressure.}
\label{fig: AoA40_Re1M}
\end{centering}
\end{figure}

Note that the case $(\alpha, Re) = (40^\circ, 1M)$ shows a deviation with the linear increase, which is due to the merger of two co--rotating primary vortices. This merger is visible in figure \ref{fig: AoA40_Re1M} a), representing the time--averaged stagnation pressure in the lee of the spheroid. For comparison, a more commonly observed single vortex formation is provided in figure \ref{fig: AoA40_Re1M} b), obtained for $(\alpha, Re) = (40^\circ, 4M)$. This also shows the linear increase of the vortex radius and the near--collinearity of the axis of the vortex with the axis of the spheroid, both of which have been previously observed in figures \ref{fig: vortex radius vs x} and \ref{fig: vortex position vs x}.

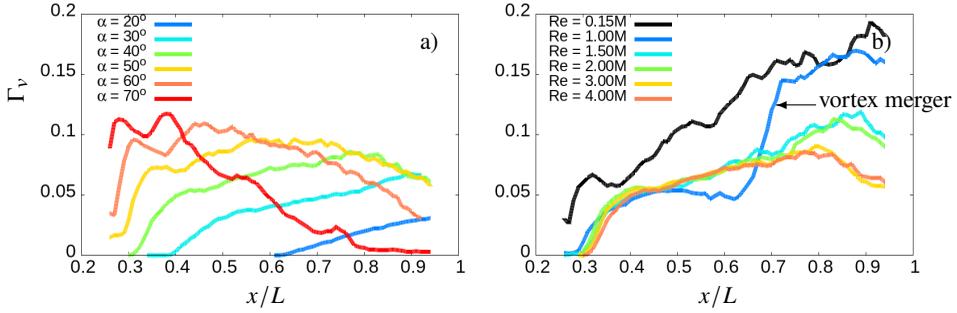
\begin{figure}
\begin{centering}
\includegraphics[width = 120 mm]{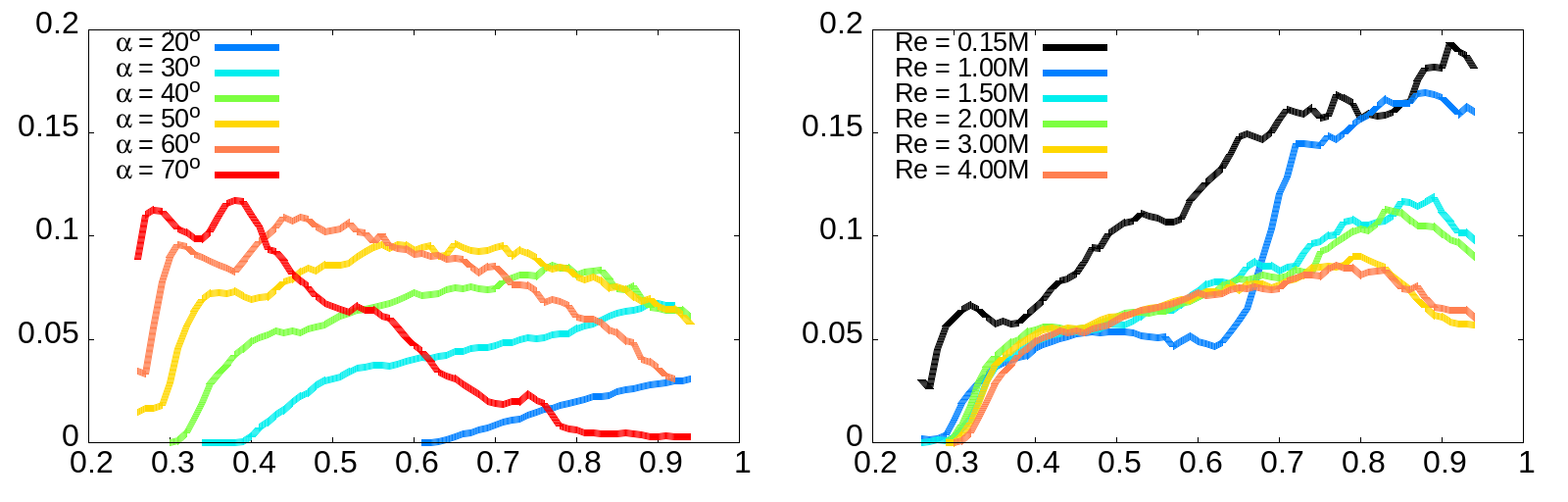}
\put(-260,-5){$x/L$}
\put(-90,-5){$x/L$}
\put(-195,90){a)}
\put(-25,90){b)}
\put(-350, 70){\rotatebox{90}{$\Gamma_v$}}
\put(-45,68){vortex merger}
\put(-46,69){\vector(-1,0){15}}
\caption{Primary vortex circulation vs $x/L$ for $\alpha \in [20^\circ, 70^\circ]$, $Re = 4M$ (a) and for $\alpha = 40^\circ$, $Re \in [0.15M, 4M]$ (b).}
\label{fig: vortex circulation vs x}
\end{centering}
\end{figure}

The evolution of the recirculation shown in figure \ref{fig: vortex circulation vs x} follows a similar evolution as the radius, except the increase is mostly linear for $\alpha \leq 30^\circ$ and a clearer positive correlation between the slope and the angle of attack. The circulation increases monotonically for $\alpha \leq 30^\circ$. For $\alpha > 30^\circ$, the circulation increases, peaks, and decreases until the tail of the spheroid. This peak occurs increasingly early with increasing incidence, such that at $\alpha = 70^\circ$, the circulation is highest at $x \approx 0.4L$ and decreases from that point until the tail of the spheroid. These three--part curves correspond to the three phases of the vortex mentioned in the discussion of figure \ref{fig: vortex radius vs x}: inception, development, and decay. The slope of increase in circulation is not strongly correlated with $\alpha$ or $Re$, although the slope of decay is higher for higher incidences. For all angles, the $Re = 0.15M$ cases are distinctly larger than the higher Reynolds numbers, although the behavior is similar. 

\begin{figure}
\begin{centering}
\includegraphics[width = 80 mm]{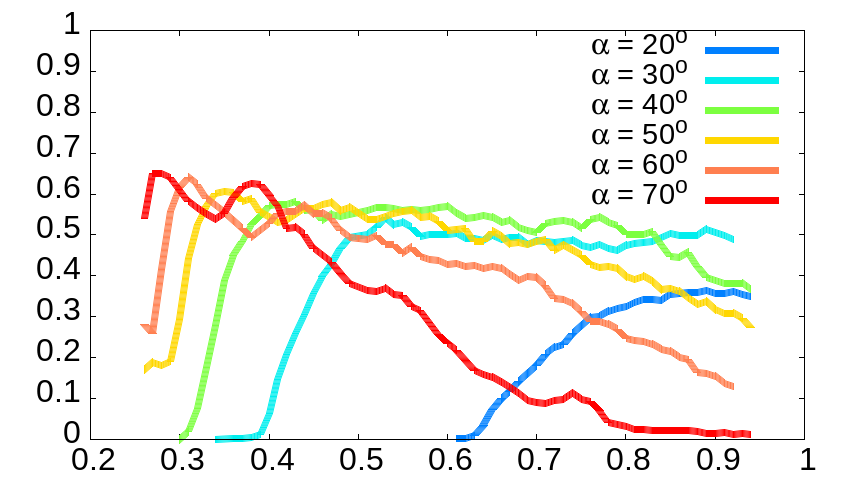}
\put(-120,-5){$x/L$}
\put(-235, 50){\rotatebox{90}{$\Gamma_v/\Gamma_t$}}
\caption{Ratio of primary vortex circulation over total circulation vs $x/L$.}
\label{fig: vortex circulation / total circulation vs x}
\end{centering}
\end{figure}

Figure \ref{fig: vortex circulation / total circulation vs x} shows the vortex circulation as a fraction of the total circulation. The ratio increases rapidly, then decreases slightly for most of the length and at intermediate $\alpha$, and decreases faster for $\alpha \geq 50^\circ$. The circulation in the primary vortex is mostly between $0.65$ and $0.5$ that of the total circulation. This is a similar value as measured by \cite{fu1994}. The rapid increase corresponds to the formation of the vortex; the slight decrease occurs when the main vortex development; the fast decrease corresponds to the weakening of the vortex where more vorticity is added to the recirculation while the vortex shrinks.

\begin{figure}
\begin{centering}
\includegraphics[width = 120 mm]{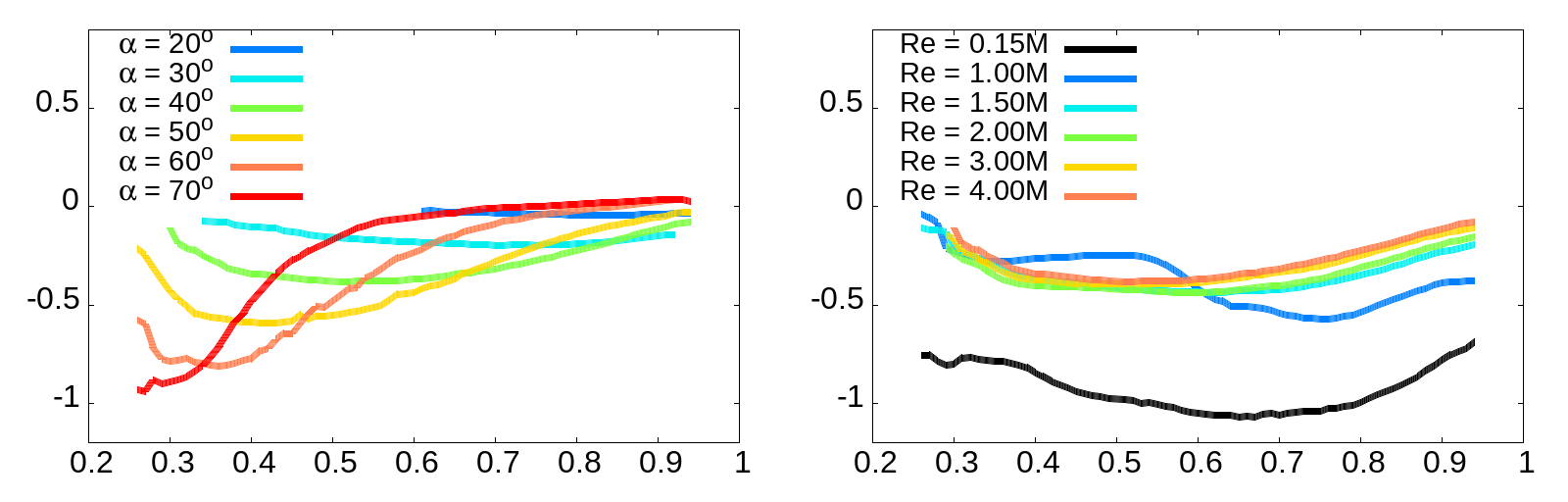}
\put(-260,-5){$x/L$}
\put(-90,-5){$x/L$}
\put(-195,90){a)}
\put(-25,90){b)}
\put(-345, 55){\rotatebox{90}{$P$}}
\caption{Pressure at the center of the primary vortex vs $x/L$ for $\alpha \in [20^\circ, 70^\circ]$, $Re = 4M$ (a) and for $\alpha = 40^\circ$, $Re \in [0.15M, 4M]$ (b).}
\label{fig: vortex pressure vs x}
\end{centering}
\end{figure}

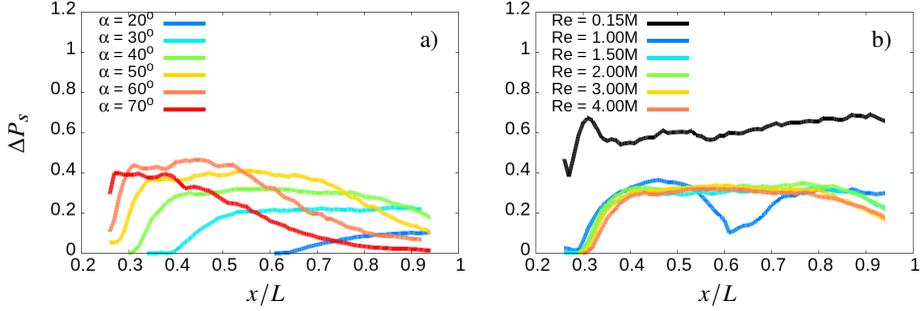
\begin{figure}
\begin{centering}
\includegraphics[width = 120 mm]{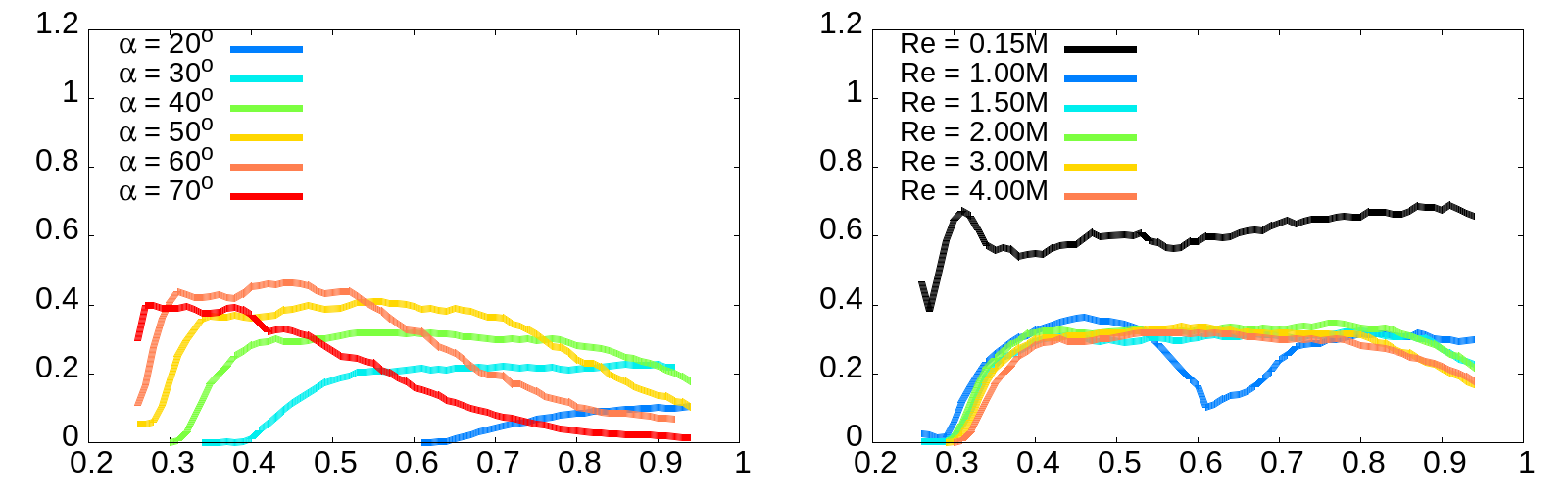}
\put(-260,-5){$x/L$}
\put(-90,-5){$x/L$}
\put(-195,90){a)}
\put(-25,90){b)}
\put(-350, 50){\rotatebox{90}{$\Delta P_s$}}
\caption{Stagnation pressure differential in the primary vortex vs $x/L$ for $\alpha \in [20^\circ, 70^\circ]$, $Re = 4M$ (a) and for $\alpha = 40^\circ$, $Re \in [0.15M, 4M]$ (b).}
\label{fig: vortex dPs vs x}
\end{centering}
\end{figure}

Figures \ref{fig: vortex pressure vs x} and \ref{fig: vortex dPs vs x} show the pressure in the center of the vortex and the stagnation pressure differential $\Delta P_s = P_s^{saddle} - P_s^0$ where $P_s^{saddle}$ is the stagnation pressure at the saddle point and $P_s^0$ is the stagnation pressure at the center. Both $P$ and $\Delta P_s$ can be split into three regions corresponding to the three stages of the vortex: At the inception of the vortex, the pressure drops and the stagnation pressure differential rises; during the main development, the pressure has small variations and $\Delta P_s$ remains remarkably constant; during the decay of the vortex, both the center pressure and $\Delta P_s$ converge to 0.

\begin{figure}
\begin{centering}
\includegraphics[width = 60 mm]{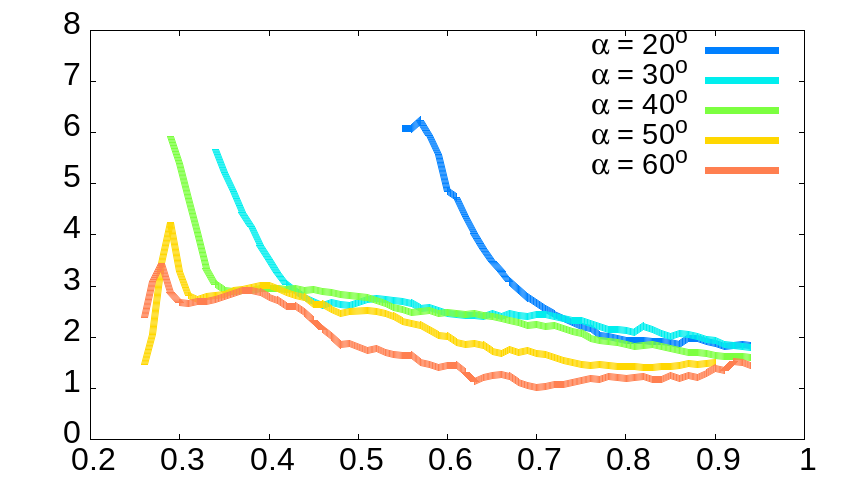}
\put(-90,-5){$x/L$}
\put(-175, 40){\rotatebox{90}{$log(S)$}}
\caption{Swirl number vs x for $\alpha \in [20^\circ, 60^\circ]$, $Re = 2M$.}
\label{fig: swirl Re = 2M}
\end{centering}
\end{figure}

Figure \ref{fig: swirl Re = 2M} shows the swirl $S$ in the vortex for $Re = 2M$, $\alpha \in [20^\circ, 60^\circ]$. The swirl number $S$ is defined as the ratio of azimuthal flux to axial flux \citep{reynolds1962}. The curves can be decomposed into three successive behaviors: 

(i) A fast initial decay at the formation of the vortex (3 orders of magnitude in $\Delta x = 0.05L$ at $\alpha = 40^\circ$). The start of the decay occurs closer to the nose for higher values of $\alpha$. These high swirl values are the consequence of high vorticity and small vortex radius at the inception of the vortex. The maximum values of swirl are overall higher for lower angles of attack.

(ii) A slower quadratic decay $S \propto x^{-2}$  (slightly faster decay for $\alpha \geq 50^\circ$), which is related to a faster increase in vortex area ($\sim x^2$) compared to its increase in circulation ($\sim x$).

(iii) For $\alpha \geq 50^\circ$: An almost constant swirl number $10 < S < 100$ from $x/L \approx 0.75$ to $x/L = 0.95$.

\subsubsection{Entrainment zone and sub--vortex region}
\label{sec: entrainment}

\begin{figure}
\begin{centering}
\includegraphics[width=40mm,trim={1cm 1cm 1cm 1cm},clip]{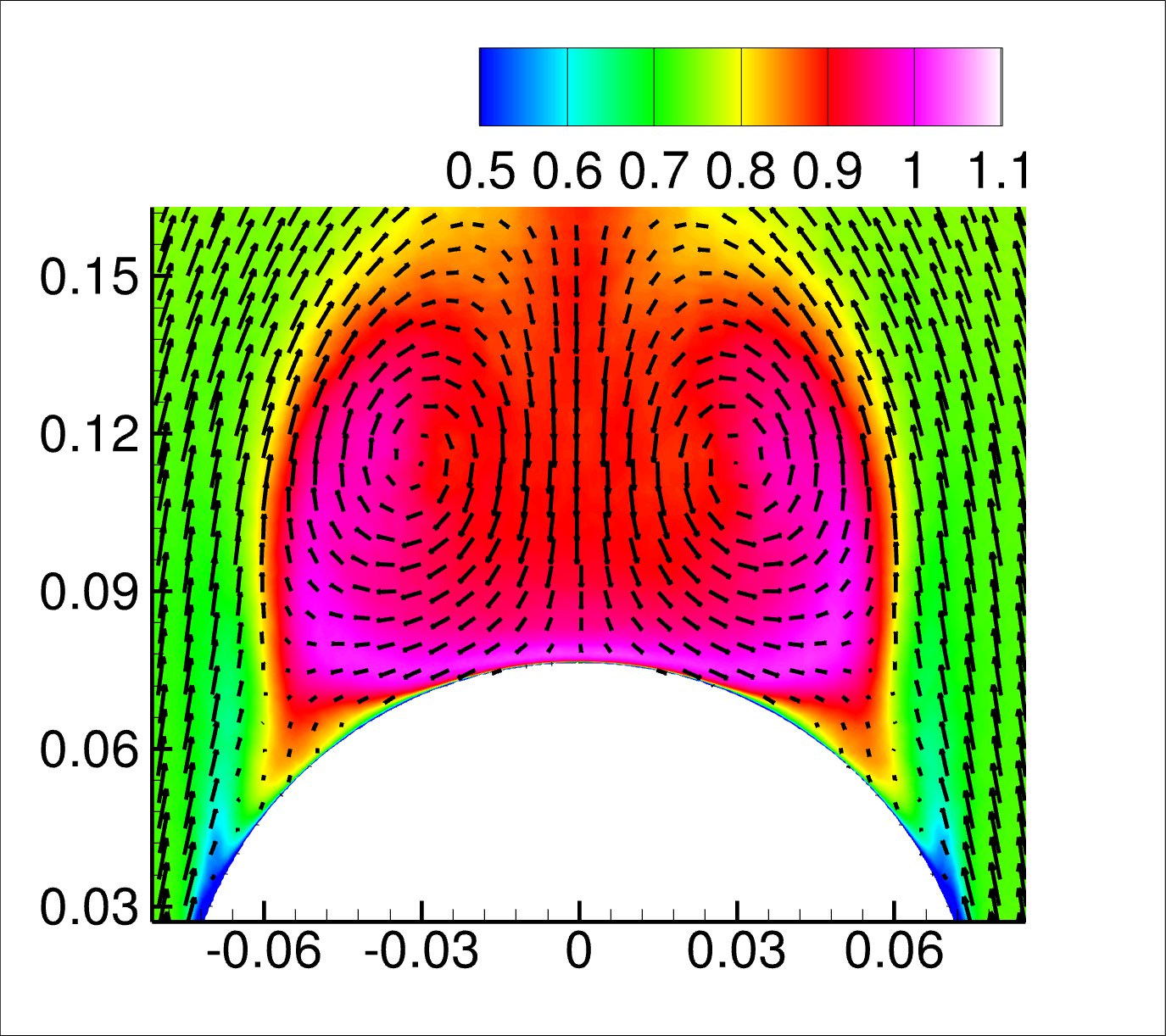}
\put(-115, 85){a)}
\put(-85, 92){$\ave{u}$}
\put(-125, 40){\rotatebox{90}{$y/L$}}
\put(-64, -5){$z/L$}
\includegraphics[width=40mm,trim={1cm 1cm 1cm 1cm},clip]{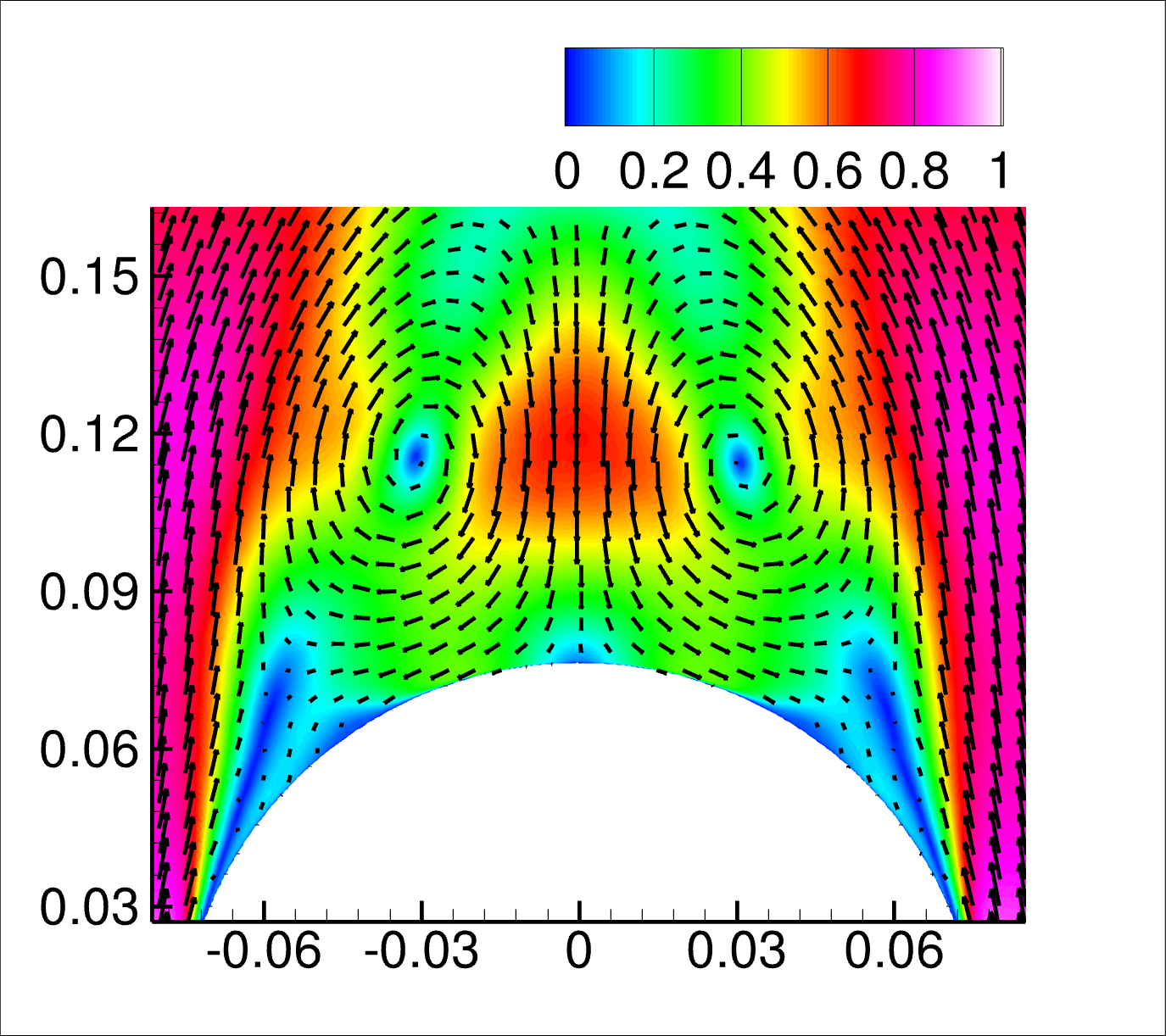}
\put(-115, 85){b)}
\put(-110, 92){$\sqrt{\ave{v}^2+\ave{w}^2}$}
\put(-64, -5){$z/L$}
\includegraphics[width=40mm,trim={1cm 1cm 1cm 1cm},clip]{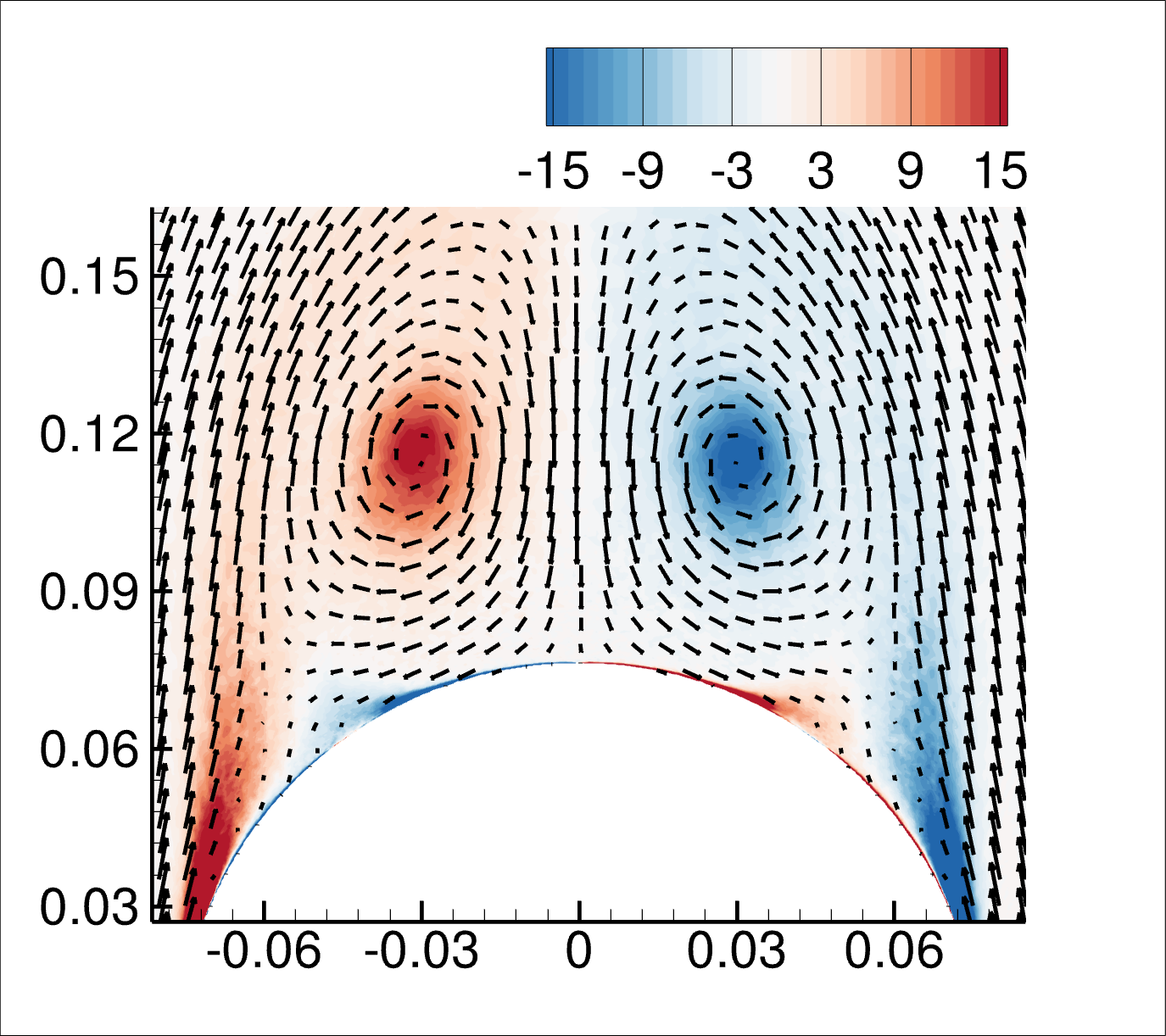}
\put(-115, 85){c)}
\put(-85, 92){$\ave{w_x}$}
\put(-64, -5){$z/L$}
\caption{Time--averaged a) axial velocity, b) secondary velocity and c) axial vorticity in the entrainment zone in a transverse slice at $x/L = 0.7$ for $\alpha = 40^\circ$, $Re = 2M$.}
\label{fig: entrainment}
\end{centering}
\end{figure}
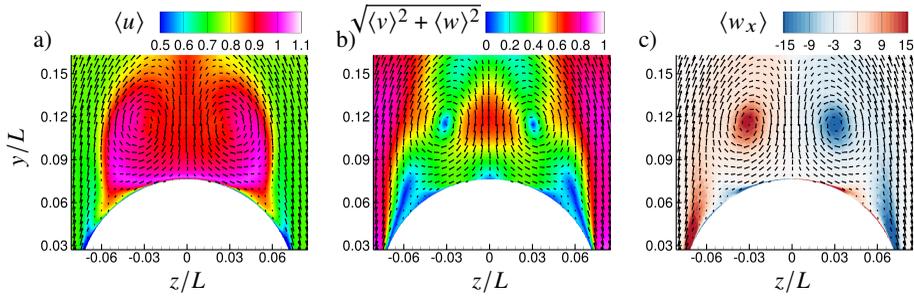
Figure \ref{fig: entrainment} shows the time--averaged axial velocity component, secondary velocity, and axial vorticity in and around the entrainment region. After the previously discussed deceleration on the outer side of the primary vortex, the flow is accelerated in the azimuthal direction to reach a maximum between the vortex pair on the leeside meridian plane. At this location, the velocity vector is aligned with the negative $y$ direction. From then on, the axial velocity increases to reach a maximum close to the wall on the same vertical plane. The flow is redirected on each side of the meridian plane to pass between the primary vortex and the secondary vortex before intersecting the separated shear layer at $z =\pm 0.05L$. The increase in turbulent kinetic energy and vorticity at this location is understood as a consequence of the interaction between the separated and entrained sheets. The vorticity in the entrained sheet, close to the separated sheet, is small and has a sign opposite to that of the primary vortex. This suggests that the fluid does not come from the separated layer but from the free--stream. \cite{jiang2016} noted a similar feature as a spiraling of two sheets of opposite sign into the vortex, coming from two different parts of the flow.
This entrainment is understood to be similar to the downwash created in the lee and wake of profiled geometries. 
Close to the wall, a boundary layer forms starting from $\phi = 180^\circ$ and travels in the opposite azimuthal direction as the primary boundary layer. This boundary layer separates below the primary vortex and forms a secondary vortex pair, visible as a region of opposite axial vorticity compared to the primary vortex of the corresponding side.
Four 2D stagnation points are observed in this flow: One for $\phi = 180^\circ$, above the primary vortex pair where the separated shear layers intersect; one at $\phi = 180^\circ$, where the entrainment sheet intersects the wall; one on each side where the entrainment flow intersects the separated shear layers.

\subsection{Recirculating wake}
\begin{figure}
\begin{centering}
\includegraphics[width=40mm,trim={1cm 1cm 1cm 1cm},clip]{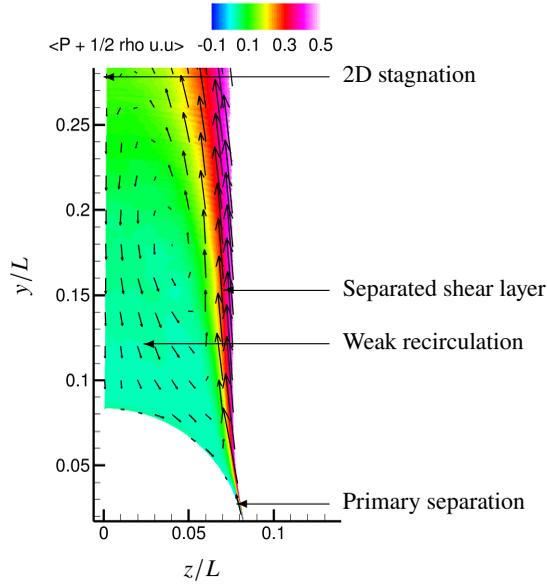}
\put(0,180){2D stagnation}
\put(-5,182){\vector(-1,0){85}}
\put(0,100){Separated shear layer}
\put(-5,102){\vector(-1,0){40}}
\put(0,80){Weak recirculation}
\put(-5,82){\vector(-1,0){70}}
\put(0,20){Primary separation}
\put(-5,22){\vector(-1,0){35}}
\put(-125, 100){\rotatebox{90}{$y/L$}}
\put(-60, -5){$z/L$}
\caption{Time--averaged stagnation pressure in a transverse slice for $(x/L, \alpha, Re) = (0.5, 90^\circ, 2M)$, showing the recirculating wake.}
\label{fig: wake overview}
\end{centering}
\end{figure}

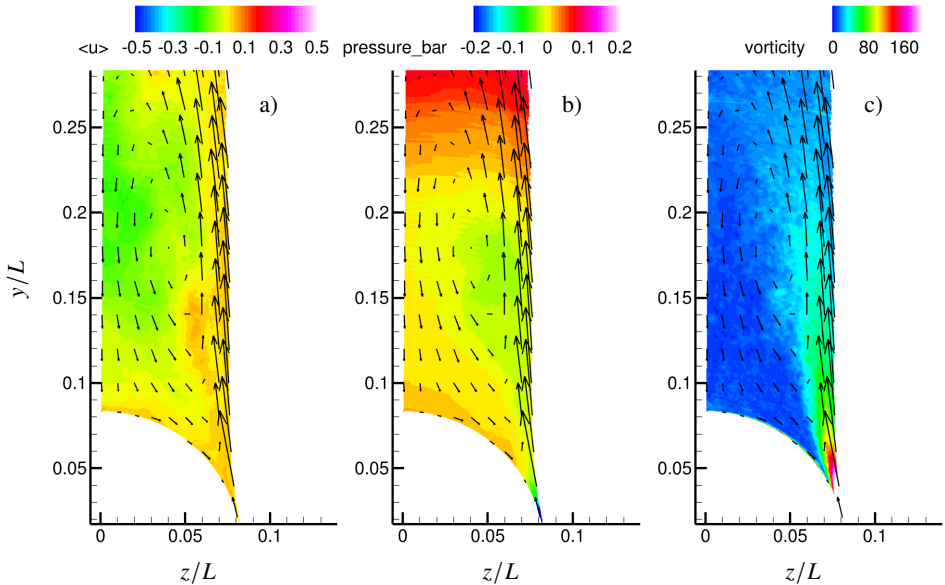
\begin{figure}
\begin{centering}
\includegraphics[width=40mm,trim={1cm 1cm 1cm 1cm},clip]{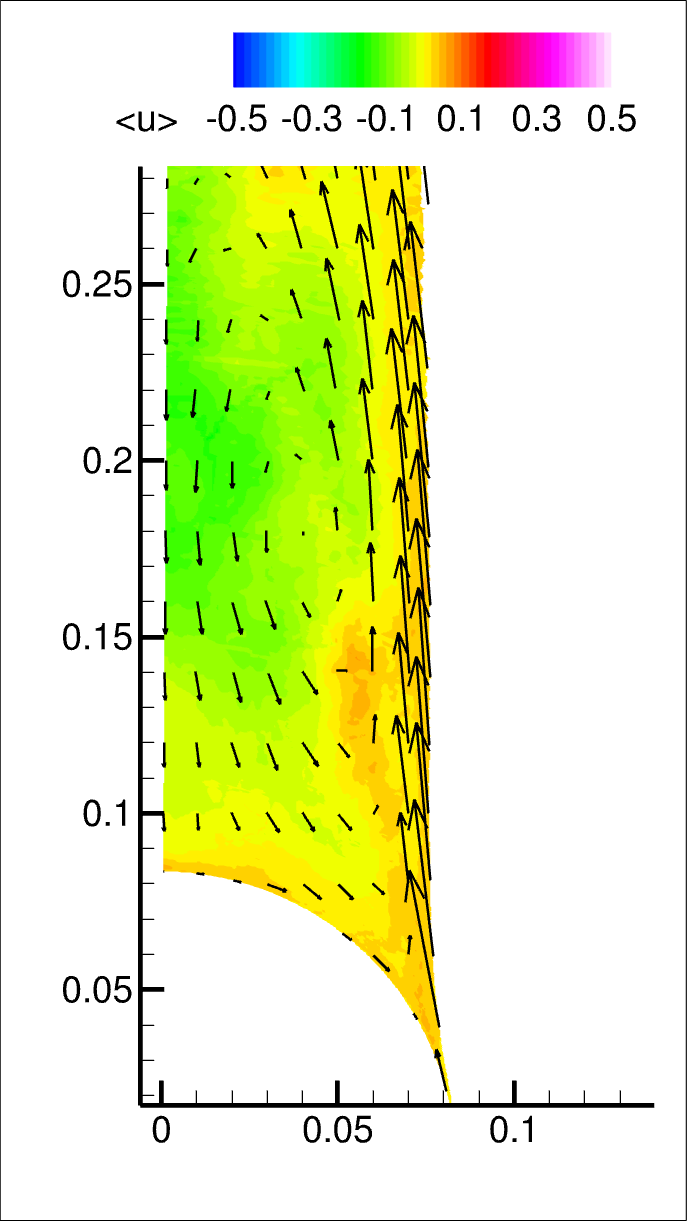}
\put(-30, 170){a)}
\put(-125, 100){\rotatebox{90}{$y/L$}}
\put(-60, -5){$z/L$}
\includegraphics[width=40mm,trim={1cm 1cm 1cm 1cm},clip]{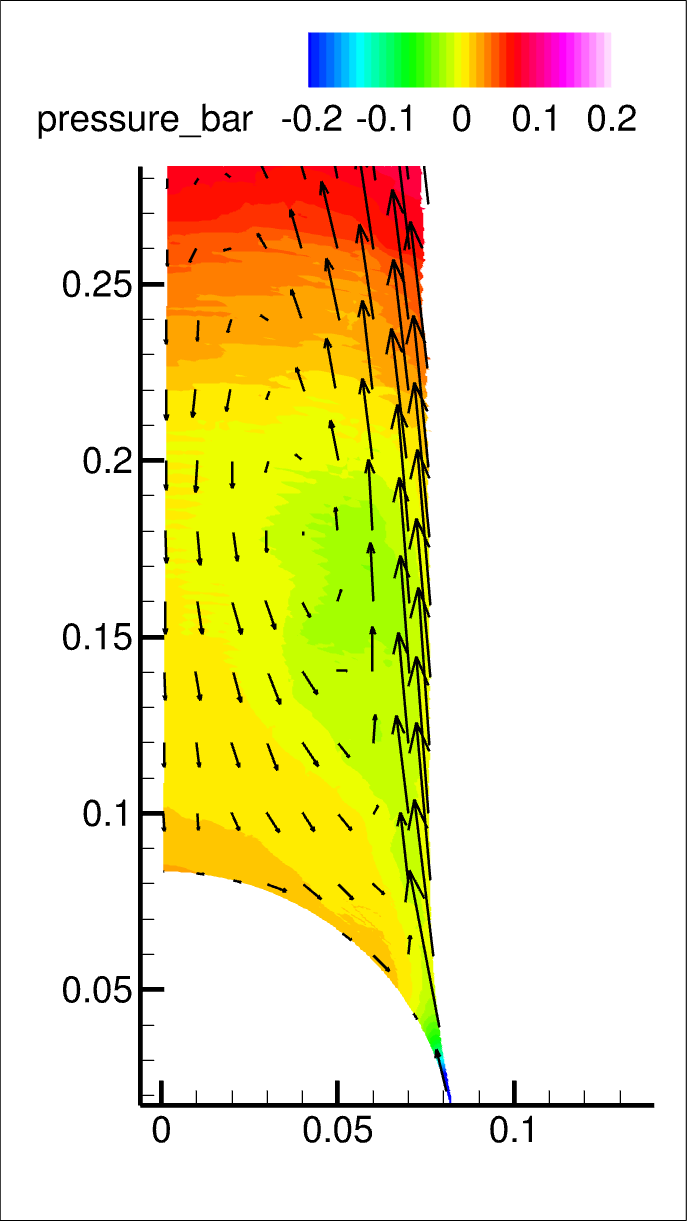}
\put(-30, 170){b)}
\put(-60, -5){$z/L$}
\includegraphics[width=40mm,trim={1cm 1cm 1cm 1cm},clip]{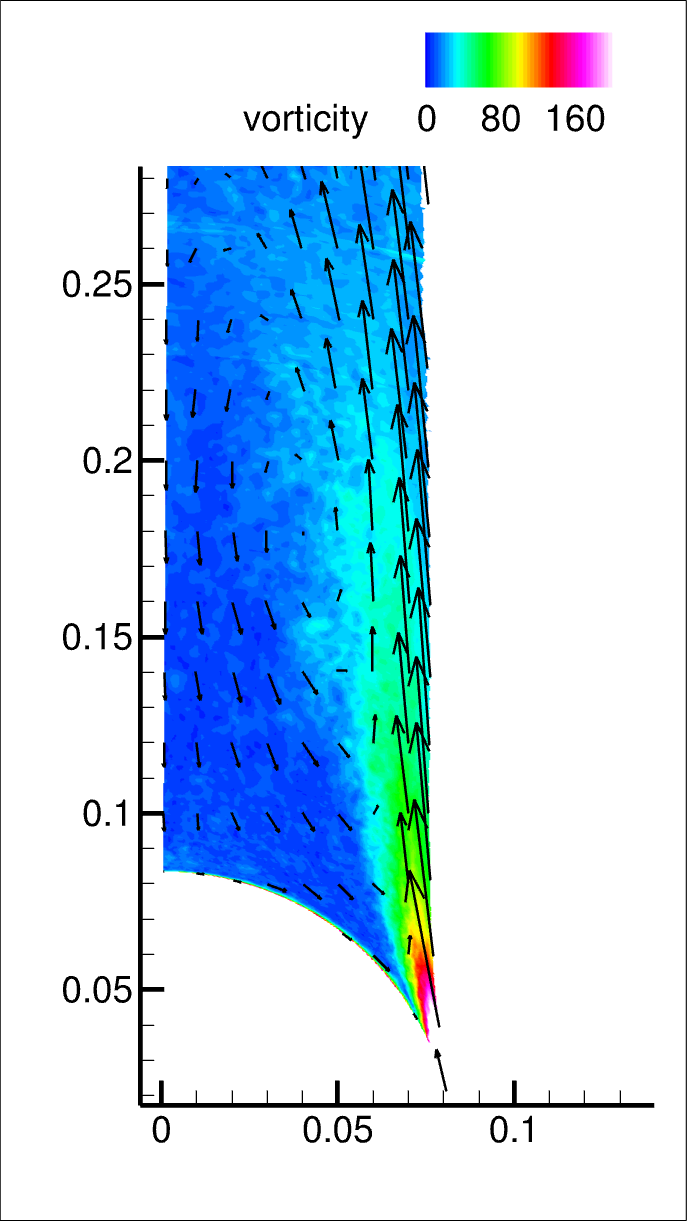}
\put(-30, 170){c)}
\put(-60, -5){$z/L$}
\caption{Time--averaged a) axial velocity, b) pressure and c) vorticity magnitude for $(x/L, \alpha, Re) = (0.5, 90^\circ, 2M)$.}
\label{fig: wake fields}
\end{centering}
\end{figure}

Figure \ref{fig: wake overview} shows the recirculating wake at $(x/L, \alpha, Re) = (0.5, 90^\circ, 2M)$. Downstream of separation, the separated shear layer does not form a coherent vortex. Instead, the projections of the streamlines in the transverse plane from the separation are unbounded. The vorticity carried by the shear layer slowly diffuses in the streamwise direction. The pair of separated layers then meet at a 2D stagnation point on the lee meridian plane. The distance between this 2D stagnation point and the wall is called the local separation length $L_s$ and is discussed in more detail in section \ref{sec: recirculation area}. The symmetric pair of shear layers forms a cavity where the pressure, stagnation pressure, and vorticity have small gradients and where the fluid slowly recirculates back toward the leeside wall of the spheroid.

\begin{figure}
\begin{centering}
\includegraphics[width=60mm]{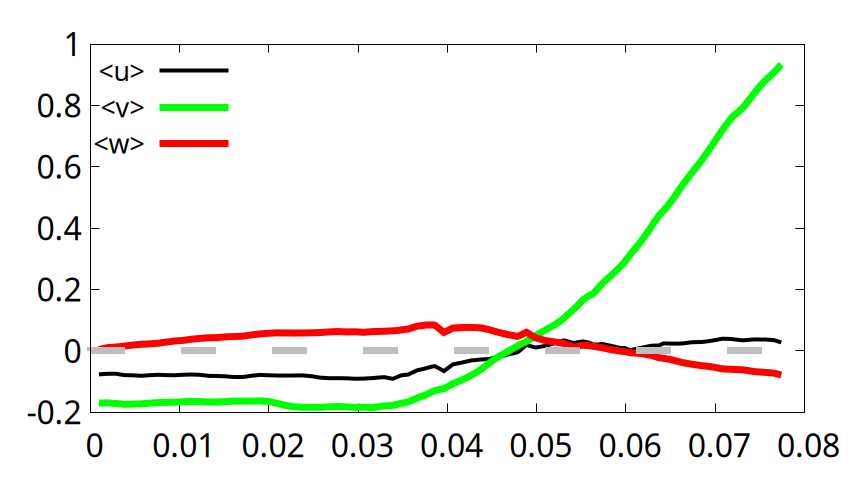}
\put(-90,-5){$z/L$}
\put(-180, 25){\rotatebox{90}{$\ave{u}$, $\ave{v}$, $\ave{w}$}}
\caption{Three components of time--averaged velocity in the body frame of reference along the span direction z for $(x/L, \alpha, Re) = (0.5, 90^\circ, 2M)$ at $y = 0.15L$ from the axis of the spheroid. The zero--velocity line is drawn in gray for ease of reading.}
\label{fig: velocity at 90deg}
\end{centering}
\end{figure}

Figure \ref{fig: velocity at 90deg} shows the components of velocity on a line that runs in the z--direction, at $y = 0.15L$. The vertical velocity is larger than the other two components. Two regions are visible:
for $z < 0.04L$, the axial and vertical components are negative, while the spanwise component is positive. All three components are almost constant; the spanwise component is $0$ at $z = 0$ by symmetry and increases slightly until $z = 0.04L$. This region corresponds to the cavity, as observed in figure \ref{fig: wake fields}. For $z > 0.04L$, all three components switch sign, the vertical velocity increases the fastest to match the freestream value. This transitional region corresponds to the separated shear layer.

\subsection{Evolution of recirculation area with axial location}
\label{sec: recirculation area}

\begin{figure}
\begin{centering}
\includegraphics[width = 120 mm]{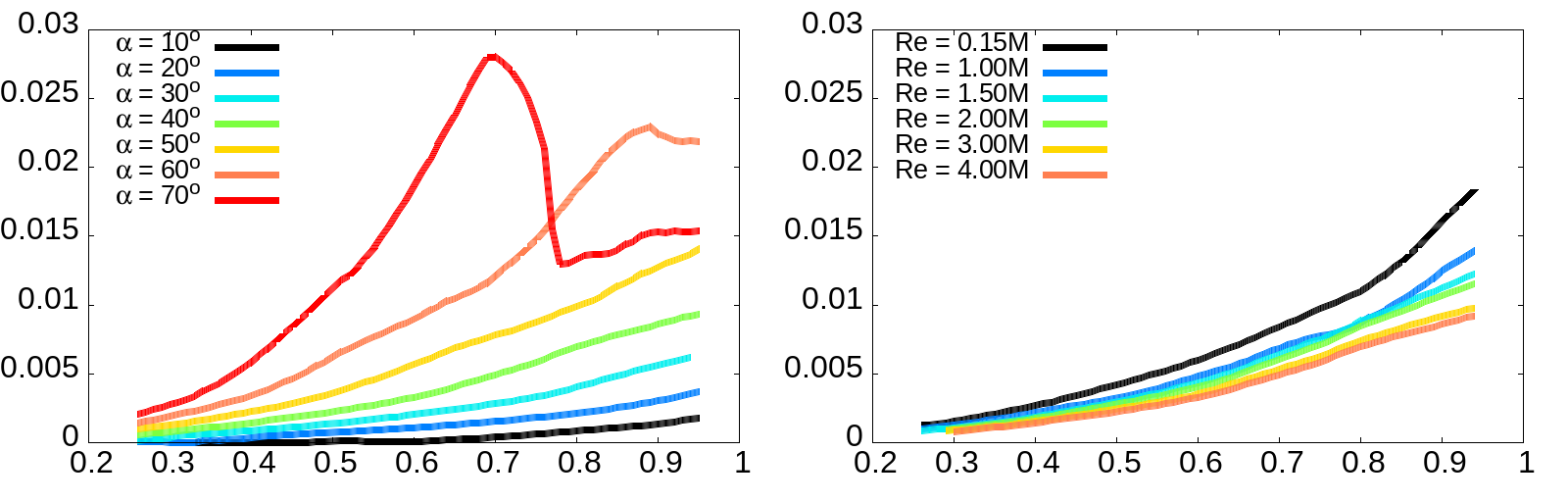}
\put(-260,-5){$x/L$}
\put(-90,-5){$x/L$}
\put(-355, 55){\rotatebox{90}{$A^t$}}
\put(-195,90){a)}
\put(-25,90){b)}
\caption{Total recirculation area versus $x/L$ for $\alpha \in [20^\circ, 70^\circ]$, $Re = 4M$ (a) and for $\alpha = 40^\circ$, $Re \in [0.15M, 4M]$ (b).}
\label{fig: total area vs x}
\end{centering}
\end{figure}
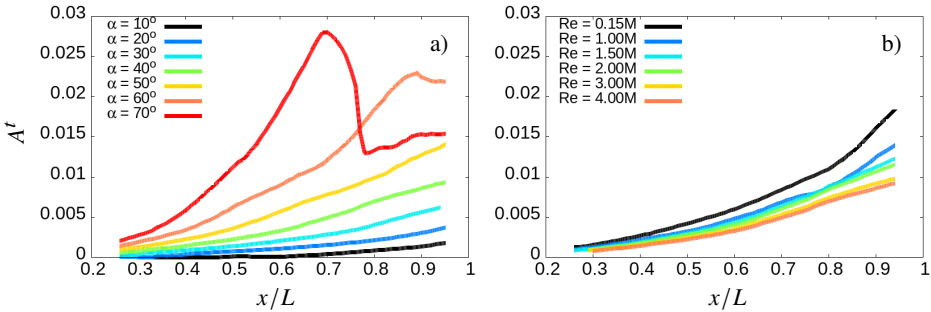

Figure \ref{fig: total area vs x} represents the area of recirculation defined in section \ref{sec: methodology}, versus the axial distance for $Re = 4M$. For $\alpha \leq 50^\circ$, the area increases monotonically. The slope of the curves increases with decreasing Reynolds numbers. For $\alpha \geq 70^\circ$, the area increases until $x/L \approx 0.7$ for $\alpha = 70^\circ$, at which point the area decreases. This decrease corresponds to a state change in the recirculation topology, dividing a regime in which the primary vortex is coherent from a regime in which the recirculation is decoherent.
The slope of the $A^t$ versus $x$ curve is increasingly steeper with increasing angle of attack. This is representative of the faster axial growth of the recirculation area at higher angles of attack and mirrors the linear rise of the radius of the vortex (see figure \ref{fig: vortex radius vs x}). A smaller albeit noticeable correlation is also observed between the slope of the curve and Re, with lower Reynolds numbers having higher slopes. This indicates an increased growth of the recirculation area with decreasing $Re$. This is consistent with the qualitative observation made in figure \ref{fig: overview wx}.

\begin{figure}
\begin{centering}
\includegraphics[width = 60 mm]{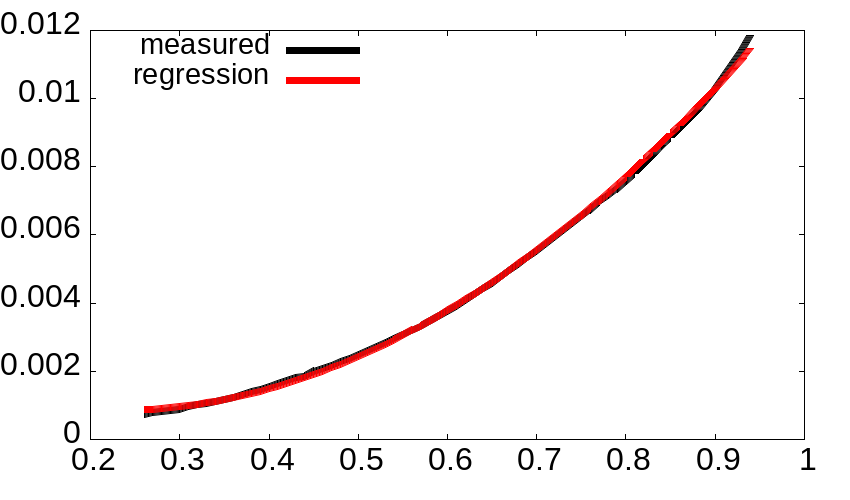}
\put(-90,-5){$x/L$}
\put(-185, 50){\rotatebox{90}{$A^t$}}
\caption{Total area of recirculation $A^t$ versus x/L (black) at $\alpha = 30$, $Re = 0.15M$, compared with a quadratic regression (red).}
\label{fig: total area vs x, AoA30 Re0p15M}
\end{centering}
\end{figure}
Figure \ref{fig: total area vs x, AoA30 Re0p15M} shows the total area of recirculation $A^t$ at $\alpha = 30^\circ$, $Re = 0.15M$.  It shows that the monotonic increase of area is quadratic, as evidenced by the comparison to the second--order regression. 

\begin{figure}
\begin{centering}
\includegraphics[width = 80 mm]{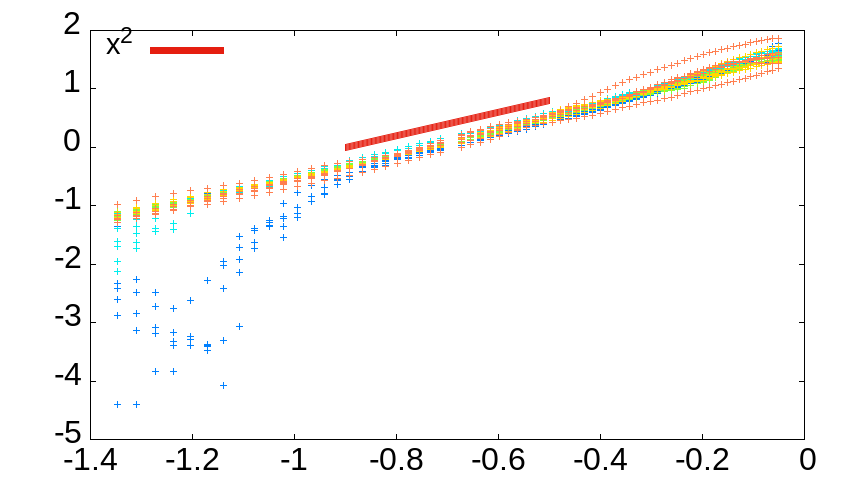}
\put(-120,-5){$log(x)$}
\put(-230, 50){\rotatebox{90}{$log(A^t)$}}
\caption{Logarithm of the recirculation area vs $log(x)$ for $\alpha \in [20^\circ, 60^\circ]$.}
\label{fig: log total area vs x}
\end{centering}
\end{figure}

Figure \ref{fig: log total area vs x} shows the log of the total area versus the log of the axial coordinate for $\alpha \in [20^\circ, 60^\circ]$. The slope of the curves is close to 2, showing that the quadratic scaling of $A^t$ with $x$ discussed in figure \ref{fig: total area vs x, AoA30 Re0p15M} can be extended to a large portion of the cases. Figure \ref{fig: log total area vs x} also shows that for $\alpha = 20^\circ$ and $\alpha = 30^\circ$, $A^t$ initially rises faster than $x^2$ when the area is small, but then converges toward a slope of $x^2$. In addition, the area of the recirculation only follows a quadratic curve when the recirculation is coherent. When a recirculating wake is formed, the vertical extent of the area with non--zero vorticity may become unbounded. In that case $A^t$ becomes undefined.

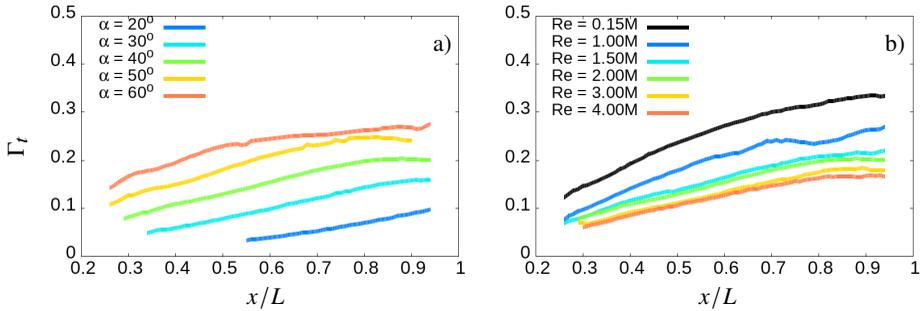
\begin{figure}
\begin{centering}
\includegraphics[width = 120 mm]{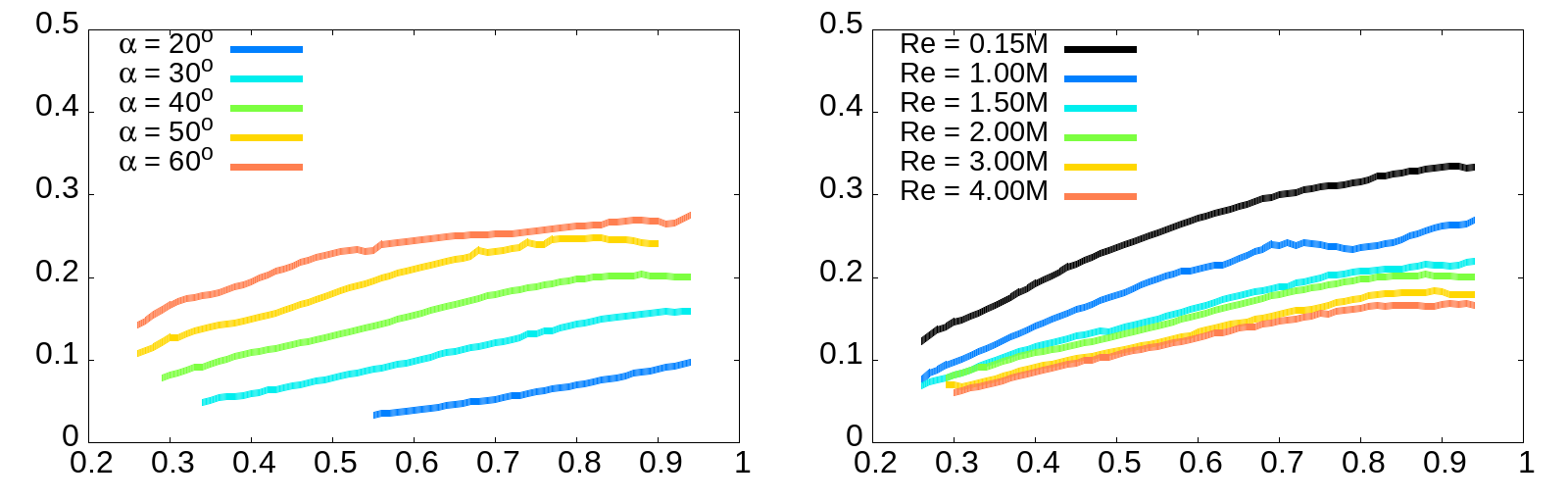}
\put(-260,-5){$x/L$}
\put(-90,-5){$x/L$}
\put(-350, 50){\rotatebox{90}{$\Gamma_t$}}
\put(-190, 90){a)}
\put(-20, 90){b)}
\caption{Recirculation circulation vs $x/L$ for a): $\alpha \in [20^\circ, 60^\circ]$, $Re = 4M$; b): $\alpha = 40^\circ$, $Re \in [0.15M, 4M]$.}
\label{fig: total circulation vs x}
\end{centering}
\end{figure}

Figure \ref{fig: total circulation vs x} shows the total circulation versus the axial direction. Unlike the area itself, which increased quadratically, the leeside circulation increases more slowly. This linear increase is related to an almost constant vorticity flow rate at separation. Linear increase in circulation coupled with quadratic increase in area translates into a lower average circulation with increasing $x$. It also suggests axial compression of the flow in the axial direction, and is consistent with the decrease of swirl observed in figure \ref{fig: swirl Re = 2M}. 

\begin{figure}
\begin{centering}
\includegraphics[width = 80 mm]{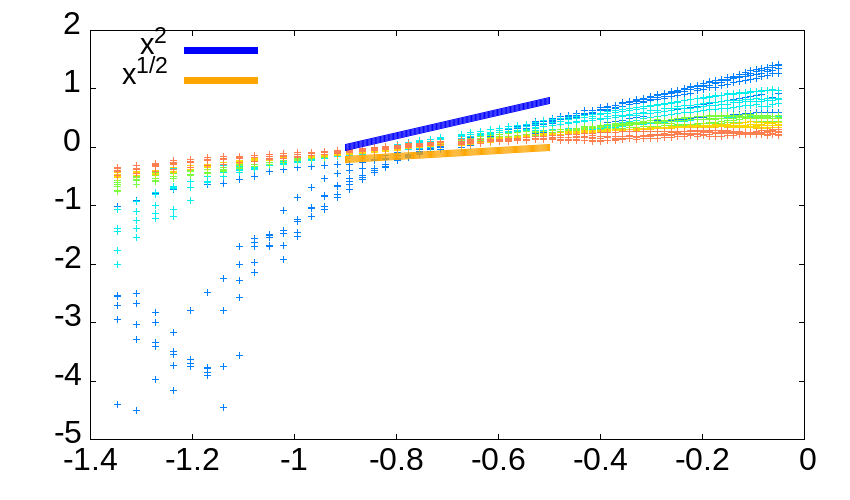}
\put(-120,-5){$log(x)$}
\put(-230, 50){\rotatebox{90}{$log(\Gamma_t)$}}
\caption{Logarithm of the total circulation vs $log(x)$ for $\alpha \in [20^\circ, 60^\circ]$.}
\label{fig: log total circ vs x}
\end{centering}
\end{figure}

Figure \ref{fig: log total circ vs x} shows the logarithm of total circulation versus the logarithm of $x$. Similarly to the total area (figure \ref{fig: log total area vs x}), the total circulation follows a power law where all the cases with the same incidence have the same slope regardless of Re, however, that slope changes with $\alpha$. The circulation in the cases with the lowest incidences follows a quadratic law, while the circulation scales with $x^{1/2}$ at $\alpha = 60^\circ$.

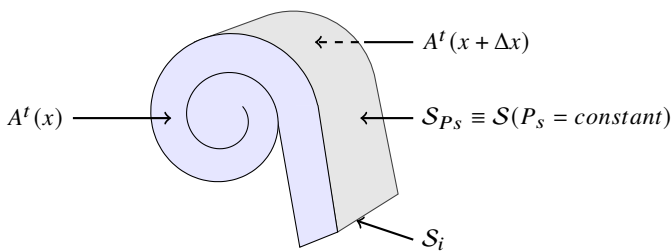
\begin{figure}
\begin{center}
\begin{tikzpicture}
\draw[fill=gray!20!white, draw=gray!50!black] (1.23,0.16) -- (1.5,-1.5) -- (2.3,-1.0) -- (2.0,0.5) arc (10:110:1.1) -- (-0.4,1.0);
\draw[fill=blue!10!white] (1.24,0.17) -- (1.5,-1.5) -- (1.0,-1.7) -- (0.67,0.17);
\draw[domain=12:25,variable=\t,smooth,samples=75,fill=blue!10!white]
        plot ({-\t r}: {0.002*\t*\t});
\draw[black, thick, ->] (-2,0) node[anchor=east]{$A^t(x)$} -- (-0.7,0);
\draw[black, thick, -] (2.5,1) node[anchor=west]{$A^t(x+\Delta x)$} -- (1.85,1);
\draw[black, thick, dashed, ->] (1.78,1) -- (1.2,1);
\draw[black, thick, ->] (2.5,-1.6) node[anchor=west]{$\mathcal{S}_i$} -- (1.75,-1.35);
\draw[black, thick, ->] (2.5,0) node[anchor=west]{$\mathcal{S}_{Ps}\equiv \mathcal{S}(P_s=constant)$} -- (1.8,0);
\end{tikzpicture}
\caption{Schematic of the control volume of the recirculation area, bounded by $\mathcal{S}_i$, across the separated shear layer between $x$ and $x+\Delta x$; $\mathcal{S}(P_s=constant)$, the surface of constant stagnation pressure surrounding recirculation area; $A^t(x)$ and $A^t(x+\Delta x)$, the area of recirculation area at $x$ and $x+\Delta x$ respectively.}
\label{fig: recirculation control volume}
\end{center}
\end{figure}

Figure \ref{fig: recirculation control volume} shows a schematic of the control volume surrounding the recirculation volume, which is used to better understand these scalings. The conservation of mass in this control volume can be written as
\[
\int_{\mathcal{S}_i} \vec{u}\cdot\hat{n}dS + \int_{\mathcal{S}_{Ps}} \vec{u}\cdot\hat{n}dS +
\int_{A^t(x)} \vec{u}\cdot\hat{n}dS +
\int_{A^t(x+\Delta x)} \vec{u}\cdot\hat{n} dS= 0
\]

In the inviscid limit, $\mathcal{S}_{Ps}$ has constant stagnation pressure and is therefore a material surface \citep{plasseraud2024}, and the mass flow rate across that surface is zero.$\int_{A^t(x)}$ and $\int_{A^t(x+\Delta x)}$ are normal to the axial direction, so $\vec{u}\cdot\hat{n} = u$.$\mathcal{S}_i$ is the only surface left to balance the growth of $A^t$:
\[
\int_{\mathcal{S}_i} \vec{u}\cdot\hat{n} dS+
\int_{A^t(x)} u dydz-
\int_{A^t(x+\Delta x)} u dydz = 0
\]
We write $\int_{\mathcal{S}_i} \vec{u}\cdot\hat{n} dS \equiv \mathcal{I} \Delta x$ where $\mathcal{I}$ is the net mass flow rate added to the recirculation per unit axial length. Note that the inflow may come from the primary separation and the entrainment region.
\[
\Leftrightarrow \mathcal{I}\Delta x+
\int_{A^t(x)} u dydz-
\int_{A^t(x+\Delta x)} u dydz = 0
\]
\[
\Leftrightarrow \frac{d}{dx}[\int_{A^t} u dydz]= \mathcal{I}
\]

This relation can be put into perspective with the linear increase in axial mass flow rate observed in the recirculation area, suggesting that the mass flow rate at separation is independent of $x$. Although the axial mass flow rate is linear, the area of recirculation is quadratic. This means that the mean axial velocity decreases and that the recirculation is squeezed in the axial direction. This is consistent with the smaller increase in total circulation (figure \ref{fig: log total circ vs x}), the increase in minimum pressure despite a constant stagnation pressure differential (figures \ref{fig: vortex pressure vs x} and \ref{fig: vortex dPs vs x}) and the decrease in swirl in the vortex (figure \ref{fig: swirl Re = 2M}).

\begin{figure}
\begin{center}
\begin{tikzpicture}
\filldraw[fill=gray!20!white, draw=gray!50!black] (0,0) -- (2,0) arc (0:90:2) -- (0,0);
\draw[blue, ultra thick, ->, rotate=15] (2,0) -- (2,3);
\draw[blue, ultra thick, -, rotate=15] (2,3.4) arc (15:75:1.2);
\draw[black, dotted, ultra thick, -, rotate=15] (0,0) -- (2,0.0) node[anchor=west]{$\phi_s$};
\draw (1.0,0.5) node{$r_x$};
\draw[black, ultra thick, <->] (0,2) -- (0,4.4);
\draw (-0.3,3.2) node{$L_s$};
\draw[black, ultra thick, <->] (0,3.35)node[anchor=south west]{$\propto (r_x + L_s \cdot cos(\phi_s))$} -- (1.1,3.35);
\draw (2.0,2.0) node{$\propto L_s$};
\draw[black, thick, ->, rotate=15] (1.85,2.4) -- (1.85,2.0);
\draw[black, thick, ->, rotate=15] (2.15,2) -- (2.15,2.4) node[anchor=north west]{$\tau_s$};
\draw[black, thick, ->] (0,4.9) -- (0,4.5);
\draw[black, thick, ->, rotate=-20] (-1,4.75)node[anchor=south west]{$p-p^{\infty}$} -- (-1,4.35);
\draw[black, thick, ->, rotate=-40] (-2,4.25) -- (-2,3.85);
\end{tikzpicture}
\caption{Schematic of the Riabouchinsky model for the prolate spheroid recirculation area.}
\label{fig: Riabouchinsky model}
\end{center}
\end{figure}
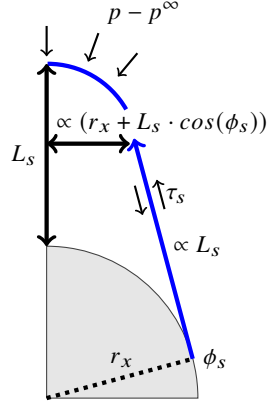

The quadratic increase for the area can also be understood by analogy to the cavity formed by a cylinder at $\alpha = 90^\circ$ as described by \cite{roshko1993} and \cite{williamson1995}. In that case, the cavity, also called the `mean recirculation region', extends to a point of maximal transverse fluctuation along the leeward meridian plane. The distance between that point and the wall is called the formation length and is similar to the previously defined local separation length $L_s$. A Riabouchinsky model can be made for such a cavity as shown in figure \ref{fig: Riabouchinsky model}, in which the viscous and turbulent stresses of the shear layers $\tau_s$ balance the linear pressure suction at the wall $f_y(x)$ \citep{riabouchinsky1921}:
\begin{equation}
\int_{\phi_s}^{180^\circ} p d\phi \equiv f_y(x) = \int_{\mathcal{S}}\tau_s \propto L_s \tau_s
\label{eq: Ria model}
\end{equation}

where $\mathcal{S}$ is the boundary of the cavity. This expression relates the force applied to the cavity and the separation length. In the case of the spheroid, the separated shear layer is tangential to the spheroid at separation $\phi_s$. If $\phi_s > 90^\circ$, the boundary of the cavity is angled toward the meridian plane and the cavity width decreases with the distance from the wall until $y = y_s = L_s + y_w$ where $y_w$ is the distance from the centerline of the spheroid to the wall. \cite{ElKhoury2012} measured this local separation length using DNS of the spheroid at low Reynolds number for $\alpha = 90^\circ$ and observed that $L_s$ increases linearly with the distance from the tips in the steady case and is maximum at $x/L = 0.5$. At higher $Re$, they also observed that $L_s$ flattens in the mid--section of the spheroid. 

\begin{figure}
\begin{centering}
\includegraphics[width = 80 mm]{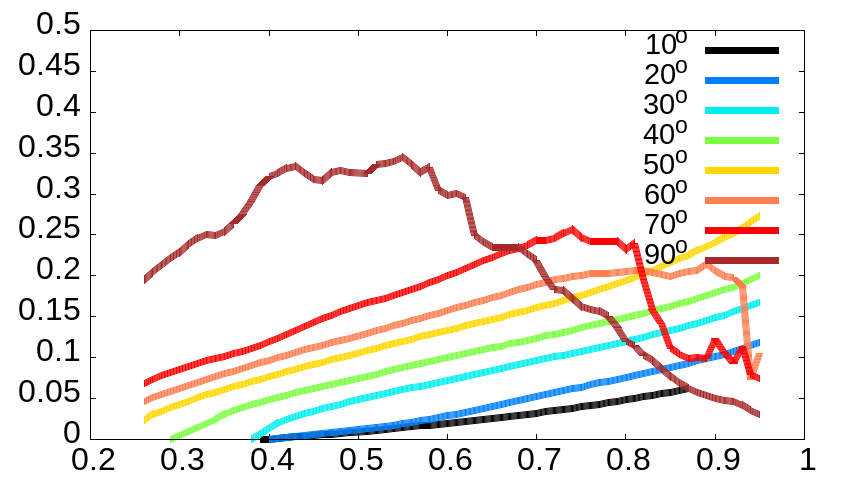}
\put(-120,-10){$x/L$}
\put(-240, 70){\rotatebox{90}{$L_s$}}
\caption{Local separation length $L_s$ vs $x/L$.}
\label{fig: Ls vs x}
\end{centering}
\end{figure}

Figure \ref{fig: Ls vs x} shows the local separation length for all the cases considered. For $\alpha < 60^\circ$, $L_s$ increases linearly with $x$. For $\alpha = 60^\circ$ and $\alpha = 70^\circ$, $L_s$ increases linearly, stagnates, and decreases more rapidly. At $\alpha = 90^\circ$, $L_s$ has a trapezoidal shape similar to the description of \cite{ElKhoury2012}. The linear increase of $L_s$ is related to the decrease in the azimuth of separation observed in figure \ref{fig: sep angle vs alpha}, which opens the angle of the separated shear layer. In turn, the opening of the angle of the separated shear layer increases the width of the recirculation, which can be scaled as $r_x + L_s cos(\phi_s)$, where $r_x$ is the radius of the spheroid for any given $x$. Thus, the area of recirculation can be scaled as $A^t \propto L_s \cdot (r_x + L_s cos(\phi_s))$. If $\phi_s \approx \pi/2$, $A^t \propto L_s \cdot (r_x + L_s \pi/2)$. Since $L_s \propto x$, $A^t \propto x^2$ in the leading order, as previously observed. This plateau observed at $\alpha = 90^\circ$ is understood to be similar to the previously commented vortex burst.


\subsection{Evolution of the skin friction with varying $Re$/$\alpha$}

\begin{figure}
\begin{centering}
\includegraphics[trim={50 10 80 30}, clip, width = 60 mm]{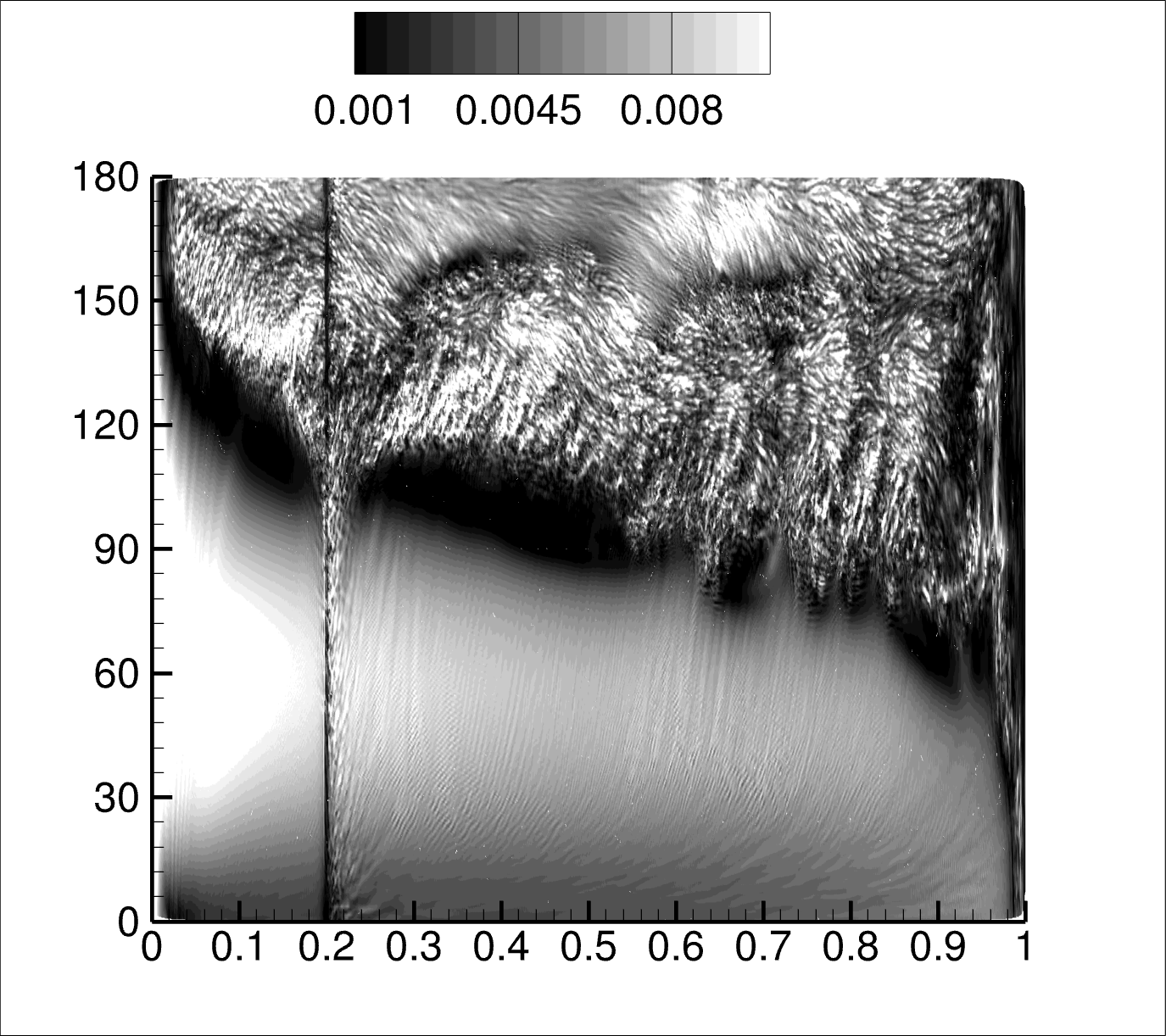}
\put(-160,155){$a)$}
\put(-180, 80){\rotatebox{90}{$\phi$}}
\put(-140,155){$c_f$}
\includegraphics[trim={50 10 80 30}, clip, width = 60 mm]{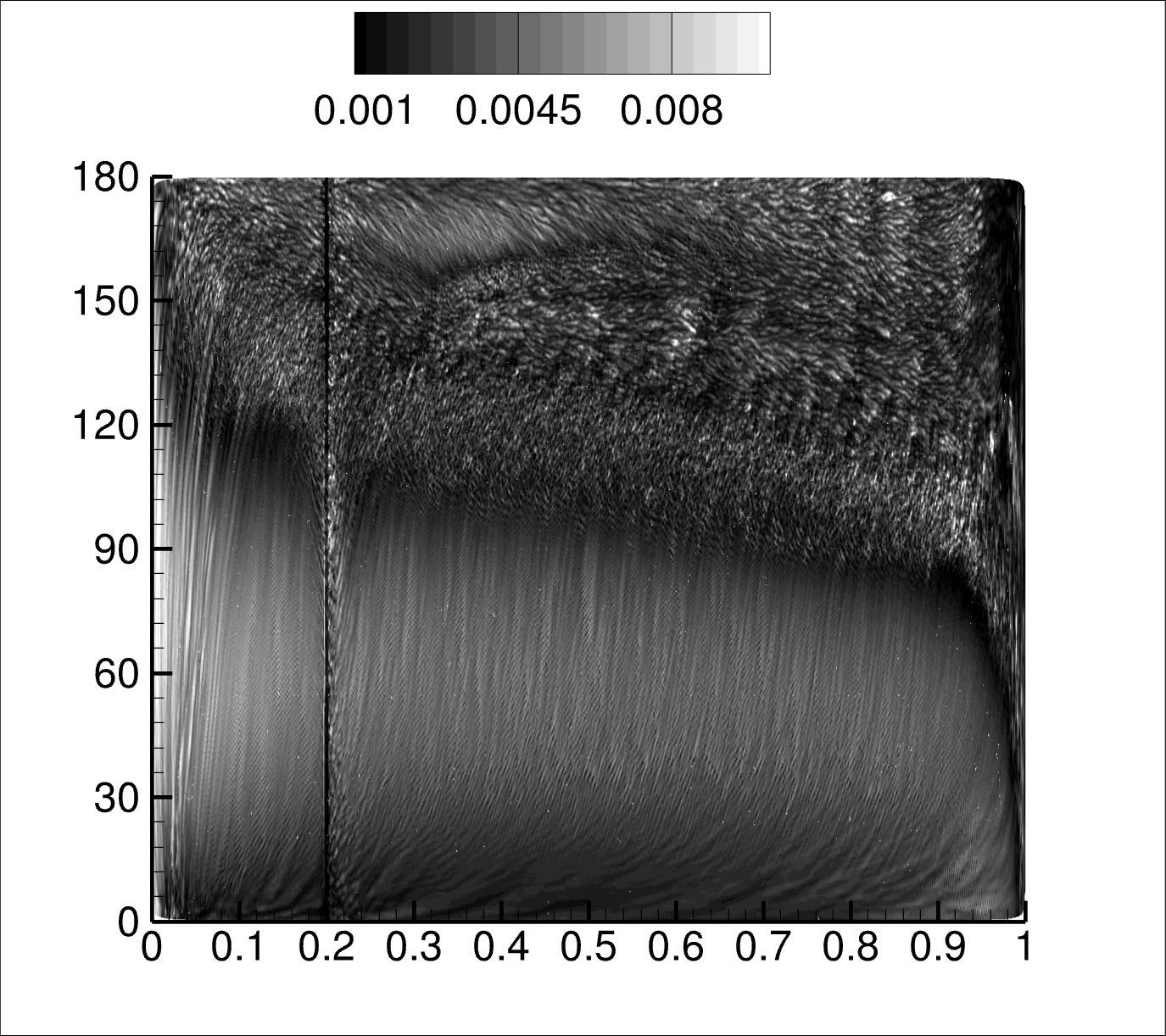}
\put(-160,155){$b)$}
\put(-140,155){$c_f$}

\includegraphics[trim={50 10 80 30}, clip, width = 60 mm]{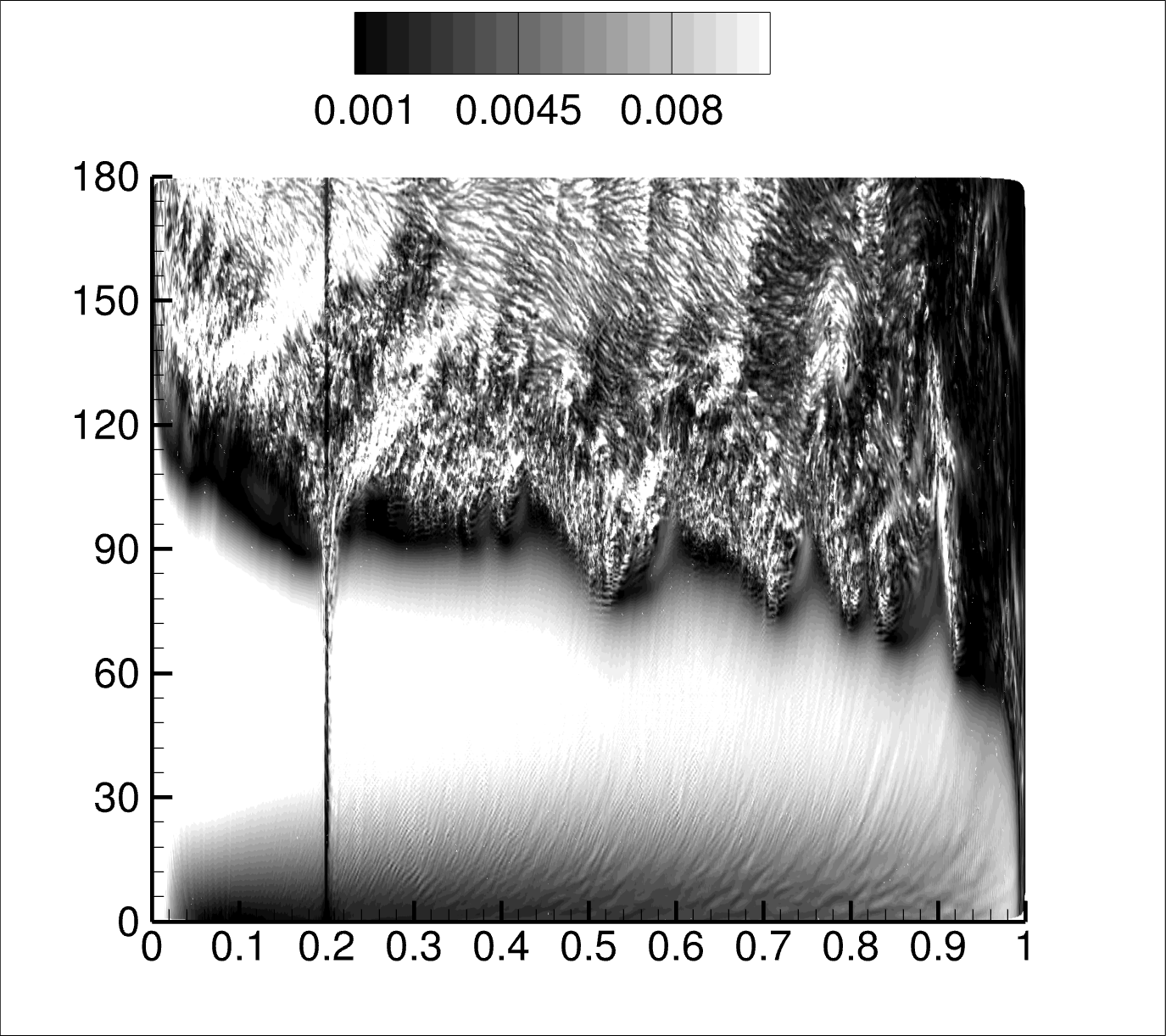}
\put(-160,155){$c)$}
\put(-180, 80){\rotatebox{90}{$\phi$}}
\put(-140,155){$c_f$}
\put(-90, -5){$x/L$}
\includegraphics[trim={50 10 80 30}, clip, width = 60 mm]{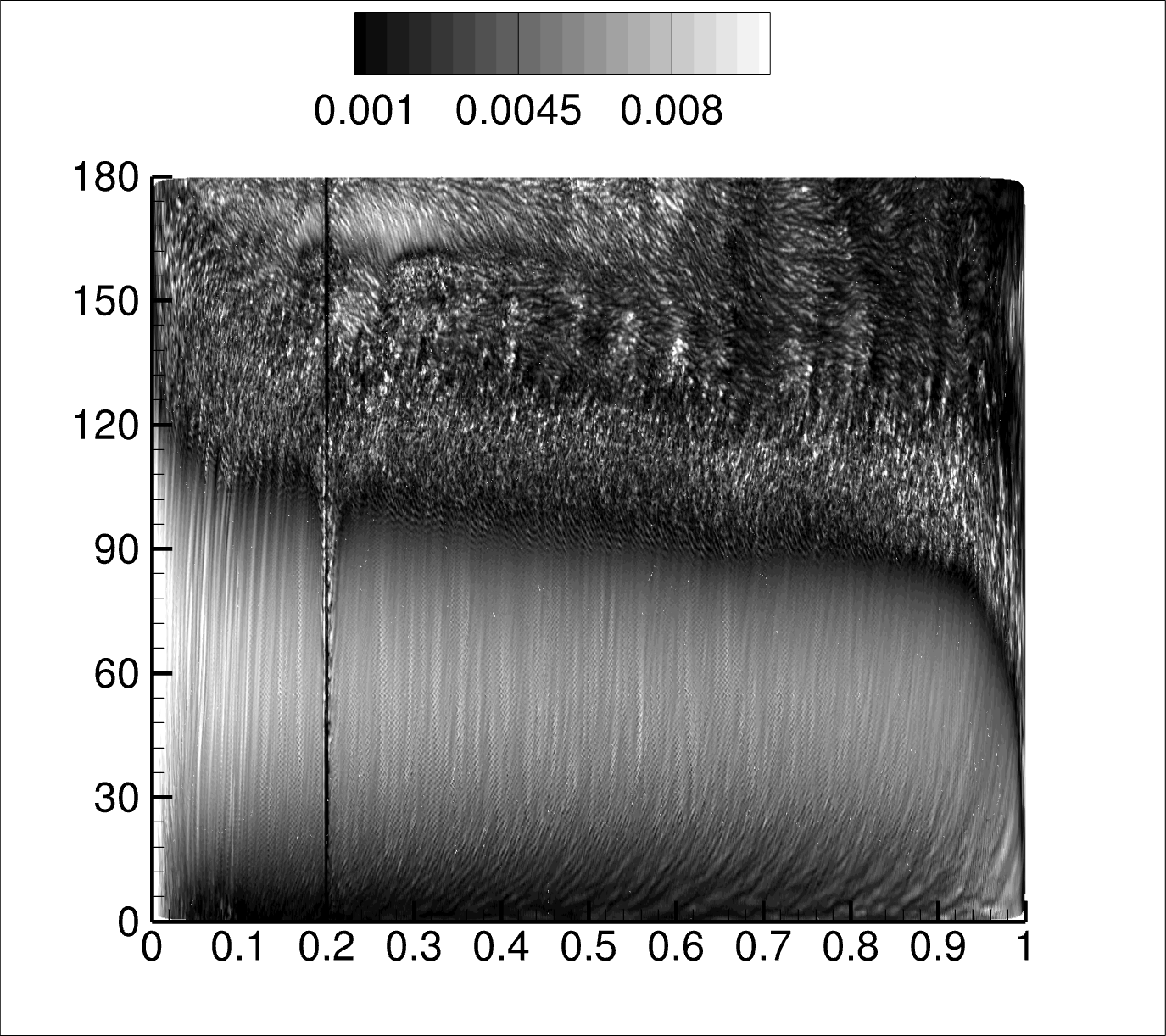}
\put(-160,155){$d)$}
\put(-140,155){$c_f$}
\put(-90, -5){$x/L$}
\caption{Instantaneous skin--friction coefficient versus $x/L$ and $\phi$ for a): $\alpha = 40^\circ$, $Re = 1M$; b) $\alpha = 40^\circ$, $Re = 4M$; c) $\alpha = 60^\circ$, $Re = 1M$; d) $\alpha = 60^\circ$, $Re = 4M$.}
\label{fig: cf}
\end{centering}
\end{figure}

\begin{figure}
\begin{centering}
\includegraphics[width=100 mm,trim={100 4 20 4},clip]{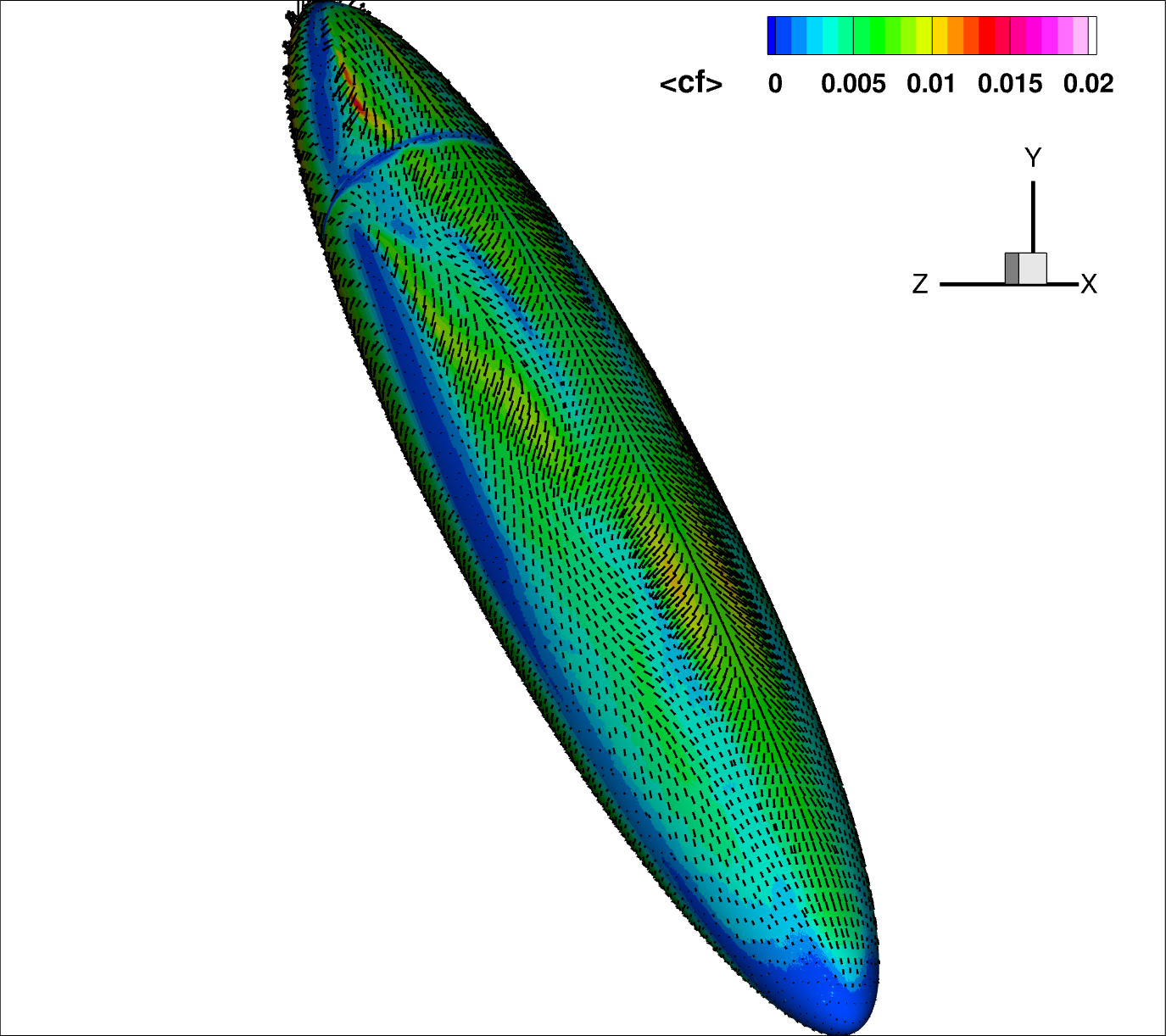}
\put(-70,80){Primary reattachment}
\put(-75,82){\vector(-1,0){25}}
\put(-70,100){Secondary separation}
\put(-75,102){\vector(-1,0){50}}
\put(-70,180){Primary separation (first vortex)}
\put(-75,182){\vector(-1,0){85}}
\put(-310,150){Primary separation}
\put(-238,152){\vector(1,0){50}}
\caption{Leeward side of the prolate spheroid showing the time--averaged skin friction coefficient magnitude and skin friction vector at $\alpha = 40^\circ$, $Re = 1M$.}
\label{fig: cf vectors}
\end{centering}
\end{figure}

Figure \ref{fig: cf} shows the instantaneous skin--friction coefficient $c_f$ versus $x/L$ and $\phi$, for $\alpha = 40^\circ$ (a and b), $\alpha = 60^\circ$ (c and d), $Re = 1M$ (a and c), $Re = 4M$ (b and d). The primary separation is visible as a downward band of increasing perturbations between $\phi = 90^\circ$ and $\phi = 120^\circ$, following a minimum of $c_f$. The slope of the primary separation is steeper at $40^\circ$ compared to $60^\circ$ as observed in figure \ref{fig: phi vs x}. The separation is wavy at $Re = 1M$, while it is straight with high--frequency perturbations at $Re = 4M$. In addition, secondary separation lines are visible in the four cases as minima of $c_f$ at $\phi \approx 160^\circ$, leading to the secondary vortex. Note that the minimum of separation at $\phi \approx 160^\circ$, $x/L \in [0.3, 0.5]$ is not the separation corresponding to a secondary vortex but the first component of a doublet primary vortex as observed in figure \ref{fig: AoA40_Re1M}. Figure \ref{fig: cf vectors} shows the time--average skin friction coefficient $\ave{c_f}$ and friction vectors for that particular case at $Re = 1M$ and $\alpha = 40^\circ$, helping to better visualize the behavior of the near--wall flow. 

On all $c_f$ maps of figure \ref{fig: cf}, the region between the leeward meridian $\phi = 180^\circ$ and the secondary separation is smoother, a fact previously observed in \cite{plasseraud2023}. This region is located below the entrainment zone where the primary vortex pair draws fluid from the freestream to the wall (previously described in \ref{sec: entrainment}). The region between the primary and secondary separations is located below the primary vortex and the separated lobe. Small azimuthal flow and large axial flow were previously observed in this region (see \ref{sec: 3d separation sheet}). Transverse perturbations are visible in this area, with higher frequencies for the higher Reynolds number. These perturbations have been observed to correlate with unsteady axial changes in the primary vortex.

\subsection{Loads}
\subsubsection{Effect of $Re$/$\alpha$ on the loads}

\begin{figure}
\begin{centering}
\includegraphics[width = 80 mm]{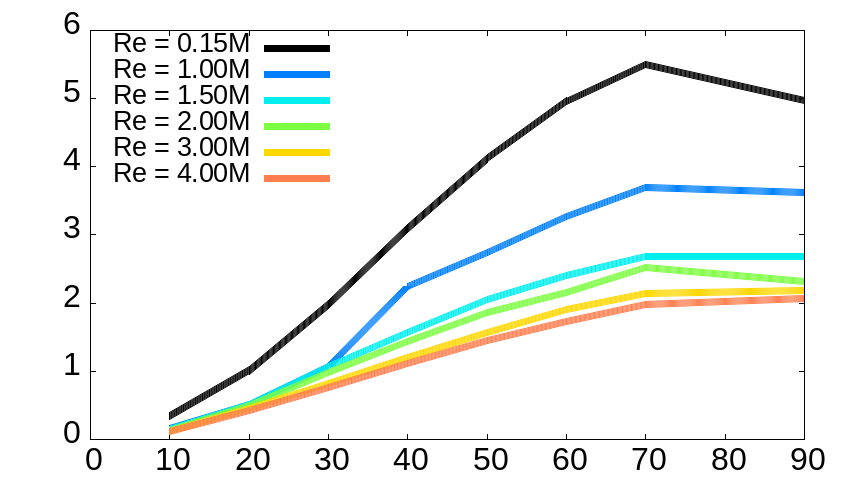}
\put(-110,-5){$\alpha$}
\put(-225,65){\rotatebox{90}{$F_y$}}
\caption{Normal force coefficient $F_y$ versus angle of attack $\alpha$ for all six Reynolds numbers.}
\label{fig: Fy vs alpha}
\end{centering}
\end{figure}

Figure \ref{fig: Fy vs alpha} shows the normal force versus the angle of attack. The force increases from $\alpha = 10^\circ$ to $70^\circ$ for all Reynolds numbers and is constant or decreasing for $\alpha > 70^\circ$. The force is larger at lower Reynolds numbers, particularly for the two lowest ones and for $\alpha > 40^\circ$. The effect of Reynolds number is smaller at the higher Reynolds numbers ($Re \geq 1.5M$). The rise in normal force and decay for $\alpha > 70^\circ$ correlates with the observed steepening of the slope of the vortex circulation versus $x$, as $\alpha$ increases. The plateau of the force between $\alpha = 70^\circ$ and $90^\circ$ corresponds to the incidences of the flow where the vortex breaks down. This corresponds to the flattening of $L_s$ previously observed in figure \ref{fig: Ls vs x}. This connection between the normal force and $L_s$ was previously detailed in equation \ref{eq: Ria model} using a Riabouchinsky model of the recirculation.

\begin{figure}
\begin{centering}
\includegraphics[width = 80 mm]{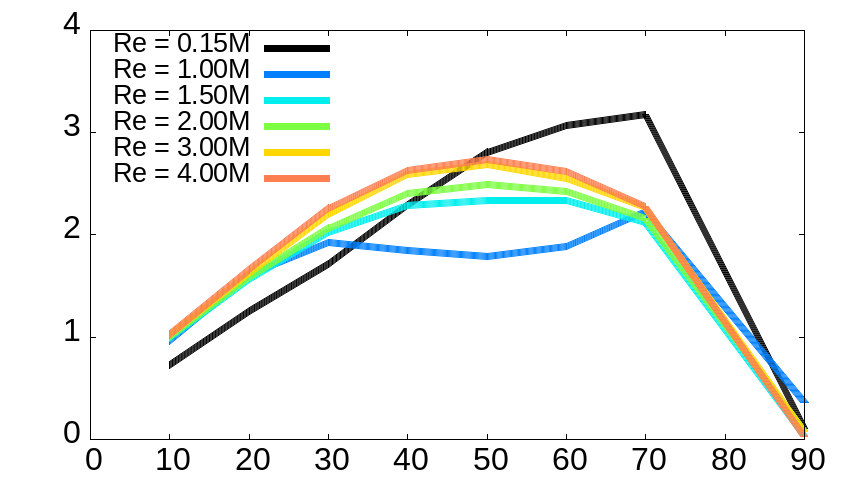}
\put(-110,-5){$\alpha$}
\put(-225,65){\rotatebox{90}{$M_z$}}
\caption{Pitching moment coefficient $M_z$ versus angle of attack $\alpha$ for all six Reynolds numbers.}
\label{fig: Mz vs alpha}
\end{centering}
\end{figure}

Figure \ref{fig: Mz vs alpha} shows the pitching moment $M_z$ versus the angle of attack. For $Re = 0.15M$, $M_z$ increases until $\alpha = 70^\circ$ and drops to 0 at $\alpha = 90^\circ$. For $Re = 1M$, the pitching moment increases until $\alpha = 30^\circ$, has a local minimum at $\alpha = 50^\circ$ and decreases again starting from $\alpha = 70^\circ$. For $Re \geq 1.5M$, has only one maximum $M_z$ between $\alpha = 40^\circ$ and $\alpha = 50^\circ$. The increasing pitching moment correlates with the increasing asymmetry of $L_s$ and $\Gamma_v$ in the axial direction (figures \ref{fig: Ls vs x} and \ref{fig: vortex circulation vs x}).

\subsubsection{Relation between loads and topology of recirculation}
As discussed in section \ref{sec: recirculation area}, a balance between the shear stress on the boundary of the cavity and the suction on the wall of the spheroid can be expressed as $f_y(x) = \int_{\mathcal{S}}\tau_s \propto L_s \tau_s$ where $f_y(x)$ is the linear force such that $F_y = \int_0^L f_y(x) dx$. That expression can be simplified as $\frac{1}{2}D(p_{wall}-p^\infty) = \tau_s L_s$ \citep{roshko1993} assuming that the fluid boundary is an inviscid streamline of constant velocity and pressure. This assumption is not valid in any cavity formed in the lee of the spheroid because constant velocity and pressure implies constant stagnation pressure, which is not the case in this study. In the proto--vortex and 3D vortex states, the stagnation pressure is constant at $P_s = 0.49 P_s^\infty$ along the separated shear layer but has higher values in the vicinity of the meridian plane. In the recirculating wake case, the 2D stagnation point is located in a region of lower stagnation pressure. However, the definition of the recirculation area and the vortex based on stagnation pressure allow for a similar statement \citep{plasseraud2024}:

At any axial location where a coherent vortex is formed, the pressure contribution of the in--plane vortex--induced force is zero by Crocco's equation:$f_y^{vortex}(x) = \int_{\mathcal{A}}\ave{\vec{u}\times\vec{\omega}} = \int_{\mathcal{A}} (\nabla P_s+\nabla \tau) = P_s^0\int_{\mathcal{S}} \hat{n} + \int_{\mathcal{S}} \tau_s \cdot \hat{n} = \int_{\mathcal{S}} \tau_s \cdot \hat{n}$ where $\mathcal{A}$ is the 2D area of the vortex at location $x$ bounded by $\mathcal{S}$ of normal $\hat{n}$ and constant saddle stagnation pressure $P_s^0$. In addition, the shear stress can be approximated as $\int_{\mathcal{S}} \tau_s \cdot \hat{n} \approx r_0 (\ave{v'w'}_r-\ave{v'w'}_l)$ where $_r$ and $_l$ refer to the left and right of the vortex boundary, respectively. Thus, the only contribution of vertical linear force from vortex comes from a left/right asymmetry in the turbulent stresses. If the vortex asymmetry is small, this term becomes small, and the force from the vortex is also small.

To assess the linear force from the total area of recirculation $A_t$, a similar argument can be made except that the isoline of stagnation pressure of that area is only constant in the fluid domain. On the wall, the value of stagnation pressure is equal to the pressure due to the no--slip condition, the pressure contribution to the vertical linear force can be reduced as: $f_y^{recirculation}(x) = \int_{\phi_s}^{\phi_{max}} (0.49P_s^\infty - p^{wall}) d\phi $ where $\phi_{max}$ is the maximum azimuth of the recirculation on the wall. From this expression, the suction from the recirculation scales with the width of the recirculation. Equivalently, the vortex force from a region $\mathcal{S}$ of the fluid can be written in the inviscid limit as the integral of the Lamb vector $F = \int_{\mathcal{S}} \ave{\vec{u} \times \vec{\omega}}d\mathcal{S} \approx \int_{\mathcal{S}} \nabla P_s d\mathcal{S}$ \citep{saffman1995}. Note that this expression is equivalent to the Kutta-Joukowski theorem in the case of a vortex aligned with $x$ and a force in the $y$ direction, $\ave{\vec{u} \times \vec{\omega}}_y = \ave{w \cdot \omega_x}$. Interestingly, this expression involves the spanwise component of velocity, which explains why the linear force is not greater when the recirculation is the largest (generally at the tail of the spheroid) but when the separation azimuth is delayed and the azimuthal velocity is highest (mid--body of the spheroid).



\section{Conclusion}
\label{sec: conclusion}
The flow around the prolate spheroid was studied for a wide range of Reynolds numbers and angles of attack. The separation of the boundary layer leads to a recirculation that is in one of three states: proto--vortex, coherent vortex, or recirculating wake. In the proto--vortex state, the recirculation on each side is strongly asymmetric, dominated by the axial component of velocity, and does not have a minimum of pressure. This state is prevalent at low angle of attack and in the early stage of flow separation. In the coherent vortex state, the recirculation forms a distinct vortex that turns the azimuthal component of velocity into axial velocity. The vortex was found to evolve in three stages: inception from proto--vortex, growth, and decay. During the growth stage, the circulation in the coherent vortex increases with $x$, fed by a constant flux of vorticity at separation. Despite an axial increase in circulation, the average vorticity and axial flow rate decrease. This decrease is driven by a quadratic increase in the area of recirculation, which is a consequence of the squeezing of the vortex. In turn, this compression leads to a decrease in the overall swirl and strength of the vortex. The decrease of swirl and the decay of the vortex leads to an increase in the leeward side pressure pressure and a topological change of the separation region as the flow becomes a recirculating wake. In this state, no coherent vortex is formed, instead, the cavity is bounded by a symmetric pair of decaying shear layer.
An increase in the angle of incidence was found to correlate with a higher azimuthal flux of mass and vorticity, leading to a faster axial increase in size and circulation of the recirculating flow. This faster increase is correlated with an increase in lift, up to $\alpha = 70^\circ$ where the primary vortex pair loses coherence and the mean suction decreases. An increase in Reynolds number leads to a delayed separation and a closing of the separated sheet. This closing leads to a smaller recirculation area, a smaller vortex, and a lower lift. For a fixed Reynolds number and angle of attack, the lift is higher for the first half of the spheroid and decreases toward the tail. This helps explain the overturning pitching moment created by the vortex despite the nose/tail symmetry of the geometry. Maximum suction was not found at the location of highest circulation or where the recirculation area is the largest, but in regions of high swirl and vortex stretching.
The current study helps to understand how the angle of attack and the Reynolds number affect boundary layer separation, recirculation, and loads on the prolate spheroid. These dynamics are commonly found on other canonical flows such as cylinders, cones, delta wings, and in more complex, practical applications such as aerial and underwater vehicles.

\backsection[Acknowledgements]{Soham Prajapati, Theo Leasca, and Dr Gary Wu for technical discussions, Tejas Kadambi and Swamenathan Ramesh for assistance in the meshing process. Preliminary results on the effects of Reynolds number and angle of attack are detailed in \cite{plasseraud2024snh}.}

\backsection[Funding]{This work is supported by the United States Office of Naval Research (ONR) under ONR Grant N00014-20-1-2717 with Dr. Peter Chang as technical monitor. Computational resources for this work were provided through a United States Department of Defense (DoD) Frontier project of the High Performance Computing Modernization Program (HPCMP) and the Engineer Research and Development Center (ERDC) of HPCMP}

\backsection[Declaration of interests]{The authors report no conflict of interest.}

\backsection[Author ORCIDs]{
M. Plasseraud, https://orcid.org/0000-0001-9990-0313
K. Mahesh, https://orcid.org/0000-0003-0927-5302
}
                     
\bibliographystyle{jfm}
\bibliography{jfm}

\end{document}